\documentclass[letterpaper, 11pt]{article}

\usepackage{mathtools}
\usepackage{array}
\usepackage{multirow}
\usepackage[T1]{fontenc}
\usepackage[utf8]{inputenc}
\usepackage{helvet}
\usepackage{enumitem}
\usepackage{amsmath,amsfonts,amssymb,bm}
\usepackage{titlesec}
\usepackage{pgfgantt,rotating}
\usepackage{float}
\usepackage{graphicx}
\usepackage{sidecap}
\usepackage{wrapfig}
\usepackage{bold-extra}
\usepackage{authblk}
\usepackage{appendix}
\usepackage{amsfonts}
\usepackage{amsmath}
\usepackage{upgreek}
\usepackage{newtxmath}
\usepackage[utf8]{inputenc}
\usepackage{color}
\usepackage{rotating}

\usepackage{cite}

\usepackage[margin=1in]{geometry}

\newcommand{\beq}{\begin{equation}}
\newcommand{\eeq}{\end{equation}}
\newcommand{\bgqar}{\begin{eqnarray}}
\newcommand{\enqar}{\end{eqnarray}}
\newcommand{\bgqarn}{\begin{eqnarray*}}
\newcommand{\enqarn}{\end{eqnarray*}}
\newcommand{\bgary}{\begin{array}}
\newcommand{\enary}{\end{array}}

\newcommand{\etal}{{\it et al. }}

\graphicspath{{figures/}}

\title{Statistical guided-waves-based SHM via stochastic non-parametric time series models}

\author{Ahmad Amer}
\author{Fotis Kopsaftopoulos\footnote{Corresponding author.}}

\affil{\small Intelligent Structural Systems Laboratory (ISSL) \\ Department of Mechanical, Aerospace and Nuclear Engineering \\ Rensselaer Polytechnic Institute, Troy, NY, USA \\ Email: \{amera2,kopsaf\}@rpi.edu}

\date{\today}

\begin{document}

\maketitle


\begin{abstract}

Damage detection in active-sensing, guided-waves-based Structural Health Monitoring (SHM) has evolved through multiple eras of development during the past decades. Nevertheless, there still exists a number of challenges facing the current state-of-the-art approaches, both in the industry as well as in research and development, including low damage sensitivity, lack of robustness to uncertainties, need for user-defined thresholds, and non-uniform response across a sensor network. In this work, a novel statistical framework is proposed for active-sensing SHM based on the use of ultrasonic guided waves. This framework is based on stochastic non-parametric time series models and their corresponding statistical properties in order to readily provide healthy confidence bounds and enable accurate and robust damage detection via the use of appropriate statistical decision making tests. Three such methods and corresponding statistical quantities (test statistics) along with decision making schemes are formulated and experimentally assessed via the use of three coupons with different levels of complexity: an Al plate with a growing notch, a Carbon fiber-reinforced plastic (CFRP) plate with added weights to simulate local damages, and the CFRP panel used in the Open Guided Waves project~\cite{OGW}, all fitted with piezoelectric transducers and a pitch-catch configuration. The performance of the proposed methods is compared to that of state-of-the-art time-domain damage indices (DIs). The results demonstrate the increased sensitivity and robustness of the proposed methods, with better tracking capability of damage evolution compared to conventional approaches, even for damage-non-intersecting actuator-sensor paths. In particular, the $Z$ statistic emerges as the best damage detection metric compared to conventional DIs, as well as the other proposed statistics. This is attributed to the incorporation of experimental uncertainty in defining the $Z$ statistic, which results in a both sensitive and robust approach for damage detection. Overall, the proposed statistical methods exhibit greater damage sensitivity across different components, with enhanced robustness to uncertainty, as well as user-friendly application.

\end{abstract}

\newpage\pagebreak 

\tableofcontents 

\section{Introduction} \label{sec:intro}

In the near future, Structural Health Monitoring (SHM) systems will be capable of implementing all four levels of SHM, namely: damage detection, localization, quantification and remaining useful life estimation (prognosis) \cite{Qiu-etal13,Romano-etal19,Janapati-etal16,Das-Saha18}, with sustainable levels of performance in complex components under varying operational and environmental conditions. In order to reach this milestone, a number of challenges facing current SHM techniques needs to be addressed. These challenges originate from the deterministic nature of the majority of the currently-employed approaches, i.e. they do not allow for the extraction of appropriate confidence intervals for damage detection, localization and quantification \cite{Farrar-Worden07,Amer-Kopsaftopoulos19a,Amer-Kopsaftopoulos19b}. This leads to the characterization of those techniques as inefficient in the face of uncertainty, stochastic time-variant and non-linear structural responses \cite{Kopsaftopoulos-etal-MSSP18, Spiridonakos-Fassois2014, Zhang07}, as well as incipient damage types and complex failure modes that can be easily masked by the effects of varying states \cite{Ahmed-Kopsaftopoulos19a,Ahmed-Kopsaftopoulos19b}. Thus, there lies a need for the development of SHM frameworks, where proper understanding, modeling, and analysis of stochastic structural responses under varying states and damage characteristics is achieved for clearing the road towards achieving the aforementioned ultimate goal of SHM systems. Towards this end, many researchers have proposed the use of statistical distributions of damage-related features in devising SHM metrics and corresponding probabilities (also known as \textit{statistical} and/or \textit{probabilistic SHM}); for instance see \cite{Zhao-etal07,2011-2,2012-5,2013-9,2014-15,Mujica-etal13,Amer-Kopsaftopoulos19a} for probabilistic damage detection and localization, as well as \cite{Zhao-etal07,2013-23,2016-9,2017-5,Amer-Kopsaftopoulos19b} for probabilistic damage quantification). These approaches promise wider applicability due to the direct extraction of confidence bounds for detection, localization, and quantification, as well as present an alternative approach for SHM reliability quantification without the need for further non-destructive testing, such as that required to obtain the Probability of Detection (POD) \cite{MIL-1823A,Polish-Paper,Janapati-etal16,Metis-IWSHM17,2017-8}.

In a more specific context, as the most fundamental level of SHM \cite{Farrar-Worden07,Zhao-etal07,Giurgiutiu05}, damage detection has received significant attention throughout the last two decades. Within the framework of active-sensing guided-waves-based SHM,  one of the most widely-used techniques for damage detection (and quantification) is the concept of the Damage/Health Index/Indicator (D/HI) \cite{Ihn-Chang04a,Ihn-Chang04b}, where some features of the signal for an unknown structural state is compared to that coming from the healthy structure \cite{Ihn-Chang08,Giurgiutiu11}. To this end, the most-widely used DI-based approaches are based on the time delay of specific mode wave packets in the acousto-ultrasound signal, the amplitude/magnitude of the signal, and the energy content of the signals, all used as the features to differentiate between a healthy and a damaged structure \cite{Jin-etal18,Xu-etal13,Ihn-Chang08,Janapati-etal16,Giurgiutiu11,Nasrollahi-etal18}. These approaches, thereon denoted as conventional DI-based methods, have been used extensively in the literature due to their simplicity (no experience needed in interpretation of results) and the allowance of a damage/no-damage paradigm, which may facilitate the decision-making stage \cite{Jin-etal18}. However, there exists a number of challenges facing these types of methods when it comes to damage detection. Namely, their deterministic nature and the time-varying and non-linear structural responses within a structure can limit the applicability of such methods \cite{Farrar-Worden07, Kopsaftopoulos-etal-MSSP18, Spiridonakos-Fassois2014, Zhang07}. In addition, the effect of complex damage types and their stochastic evolution can be masked under varying operational and environmental states, further inhibiting the effectiveness of such DIs in damage detection \cite{Farrar-Worden07,Ahmed-Kopsaftopoulos19a,Ahmed-Kopsaftopoulos19b}. Other issues related to the conventional DI-based approach include the need for user-defined damage thresholds for damage detection \cite{Qiu-etal16,Wang-etal18} and the phenomenon of saturation \cite{Castro14}. 

As such, the challenges facing the aforementioned methods have been tackled throughout the literature using different approaches, with most of them based on advanced variants of conventional DIs. Although these endeavors span many strategies, the most common approaches either enhance current time-domain DIs (see for instance \cite{Janapati-etal16,An-etal12}), use frequency-domain or mixed-domain DIs (see for instance \cite{Jin-etal18,Su-etal14,Qiu-etal16}), or use advanced signal processing/modelling as a preliminary step before calculating DIs (see \cite{Su-Ye04,Song-etal08,Tibaduiza-etal16}). Another family of techniques is based on baseline-free approaches (see \cite{Wang-etal18,Lize-etal19,Hua-etal20}), which themselves can be further categorized into a number of approaches as will be discussed shortly. Although the proposed enhancements exhibit better damage detection performance compared to conventional DIs, no single technique seems to collectively address the current drawbacks of conventional DIs. The following discussion briefly outlines selected studies from each family of approaches, highlighting the advantages and drawbacks of each family of techniques.

In the context of enhanced time-domain approaches, Janapati \etal \cite{Janapati-etal16} proposed a DI that depends solely on guided-wave propagation signal normalization and applied it to many identical coupons in order to pinpoint the source of variability between seemingly-identical SHM systems, and thus better understand the effect of uncertainties on time-domain DIs. Su and coworkers \cite{Su-etal14} compared three different DIs for fatigue damage characterization within an active-sensing acousto-ultrasonic SHM framework: the traditional time-of-flight delay and energy dissipation indices, and a novel index utilizing non-linear features of guided-wave signals, such as the second harmonic generation. They observed higher sensitivities, as well as better damage evolution-tracking using nonlinear features compared to the traditional approaches relying on linear ones. In addition, they concluded that analyzing the time-frequency domain instead of the time domain alone enhances damage detection ability, especially for early-stage cracks.   Building upon that mixed-domain approach, Jin \etal \cite{Jin-etal18} used DIs in both the time and the frequency domains. In addition, in order to address the uncertainties in each individual path DI due to noise and varying conditions, they proposed an arithmetic fusion algorithm, where DIs based on amplitude and energy, both generated in the time and the frequency domains (a total of four DIs), are each summed over all the actuator-sensor path signals coming from a steel plate in order to ``visualize'' fatigue crack growth. Although these endeavors were capable in addressing certain challenges facing DIs, they are still not probabilistic in nature, and are thus still prone to error due to uncertainties. 

In order to address this, and building upon the fact that the frequency domain may offer a different representation of the signal dynamics, many researchers studied the effect of damage on the energy of wavelets (coming from some type of wavelet transformation of the signals) by using the concept of entropy \cite{Ibanez-etal15}. Basically, the Shannon entropy \cite{Shannon48} is calculated for windows/parts of the signal and the observed changes are related to damage, as appropriate. As this method is essentially based on the probability distribution of the energy of each wavelet, it is more effective in capturing uncertainties compared to conventional approaches. However, defining damage/healthy thresholds, as well as the analysis of changes in entropy, are not straight forward \cite{Rojas-etal15}, and user expertise is required in order to properly detect damage. Another approach was adopted by Qiu and coworkers \cite{Qiu-etal16}, where time-domain (as-received signal amplitude) and frequency-domain (frequency-response function) variants of the DI were used as features to develop a Gaussian mixture model for guided-wave signals in the healthy case. Then, upon the acquisition of new feature values, the model is migrated and a statistical technique is used to measure the differences between the baseline and the migrated models. Applying this approach to a real-life fuselage component with a developing fatigue crack, they observed the enhanced sensitivity and better damage evolution tracking. Most importantly, they also concluded the superiority of this technique owing to the lack of requiring user-experience in defining detection thresholds. One drawback to this approach, however, is the complexity in defining the original Gaussian mixture model, which requires many data sets, and multiple steps, including k-means clustering and expectation maximization algorithms.

In another attempt to enhance the detection capability of DIs under uncertainty, several researchers proposed baseline-free techniques, which do not require the presence of pre-sampled reference/healthy signals to compare to. In many of these techniques, either an instantaneous baseline is acquired from an identical path to the one being investigated, or from the reciprocal of that path, or the signal is reversed and analyzed for mode conversions and deviations from the original one, amongst other methods \cite{Lize-etal19}. Although these approaches appear to be more robust to varying conditions owing to the lack of pre-sampled baseline signals, most of them require knowledge of dispersion curves for the components being monitored, and dictate sophisticated actuation strategies. In addition, it has been shown that, depending on the actuator-sensor path from which the signal is coming, the evolution of the DIs can proceed in a manner uncorrelated with damage evolution \cite{Amer-Kopsaftopoulos19b,Jin-etal18}, which clearly limits the applicability of many of these techniques to specific sensor network designs and simple boundary conditions. Although there are recent studies on averting this latter drawback \cite{Qiu-etal19,Hua-etal20}, these approaches still require experience in defining detection thresholds.

Although the highlighted approaches show promise for enhancing the damage detection capability of DIs, from the discussion above it can be concluded that there exist no approaches capable of overcoming the aforementioned challenges facing DIs with a user-friendly way that can be widely applicable to different sensor networks and component designs. Namely, there still lies the need to develop damage detection techniques that overcome the following challenges, which are faced by the current DI-based approaches: 
\begin{itemize}
\item Complexity in defining/calculating the damage detection metric.
\item Lack of a straightforward approach in defining appropriate statistical damage thresholds, i.e. the need for historical data and user experience with specific damage cases.
\item Poor performance in complex damage cases and non-linear structural responses (e.g. complex composite parts).
\item Lack of robustness towards uncertainties originating from operational or environmental sources.
\item Failure to follow damage evolution in some cases, either entirely (e.g. damage non-intersecting signals \cite{Amer-Kopsaftopoulos19b}), or starting from a certain damage size (e.g. saturation phenomenon \cite{Castro14}).
\end{itemize}

In order to overcome these challenges, the use of non-parametric time series (NP-TS) models within a statistical framework is proposed herein in order to tackle damage detection under uncertainty. NP-TS models have been widely used in damage detection via vibration-based SHM \cite{Kopsaftopoulos-Fassois10,Kopsaftopoulos-Fassois13,Kopsaftopoulos-Fassois16} due to their stochastic nature, which inherently accounts for uncertainty and allows for the extraction of theoretical and experimental confidence intervals, avoiding the need for user-defined thresholds, based on statistical decision making schemes. Also, NP-TS frequency-domain representations of system dynamics can exhibit increased sensitivity to damage and entertain simplicity of application requiring little-to-no user experience \cite[pp. 212]{Kopsaftopoulos-Fassois10}. Finally, as will be shown herein, NP-TS models can prove superior to conventional DIs in following damage evolution. Thus, the use of stochastic NP-TS representations for damage detection has the potential to enhance the detection performance with straight-forward applicability due to the ability of extracting ``inherent thresholds'' from the SHM metrics themselves, as well as simplicity of application.

In a previous preliminary study \cite{Amer-Kopsaftopoulos19a}, the authors applied NP-TS representations and statistical hypothesis testing (SHT) to an Al coupon and a stiffened panel showing the extraction of estimation confidence intervals from the metrics being used, as well as the enhanced detection capability with these stochastic models compared to DI-based approaches. In the current study, work on NP-TS models for damage detection in active-sensing acousto-ultrasound SHM is significantly expanded, and their performance in detecting damage over three different structural coupons is compared and experimentally assessed with that of two state-of-the-art DIs from the literature \cite{Janapati-etal16,Qiu-etal16}. To the author's best of knowledge, the NP-TS-based damage detection metrics used herein have not been proposed previously within the framework of active-sensing guided-waves-based SHM. The main novel aspects of this study include:
\begin{itemize}
\item[(a)] The introduction of a novel data-based statistical damage detection framework based on NP-TS models in active-sensing guided-waves-based SHM, as well as the expansion of two previously proposed methods by the authors and co-workers \cite{Fassois-Kopsaftopoulos-encyclopedia,Kopsaftopoulos-Fassois10}.
\item[(b)] The application of the proposed methods in two composite panels with different types of simulated damage, as well as in a notched Al coupon.
\item[(c)] The extraction of statistical confidence intervals and the detection of damage via appropriate statistical hypothesis testing schemes, negating the requirement of user-defined thresholds.
\item[(d)] The proposal of a straight-forward damage detection method in active-sensing guided-waves-based SHM, with the advantage of enhanced detection capability over conventional DI-based approaches, without sacrificing simplicity.
\end{itemize}

The remainder of this paper is organized as follows: Section \ref{sec:SDDF} introduces the development of the statistical framework (Subsection \ref{sec:SHT}), the theory of the utilized stochastic NP-TS representation (Subsection \ref{sec:NP-TS}), the statistics used in this study for damage detection (Subsections \ref{sec:F}, \ref{sec:Fm} and \ref{sec:Z}), as well as briefly presents the literature-based DIs used for comparison in this study (Subsection \ref{sec:DIs}). Then, the experimental setup, the results and the discussions are presented for every coupon consecutively in Section \ref{sec:results}. Finally, Section \ref{sec:conc} concludes this study and proposes extra steps for enhancement of damage detection within active-sensing guided-wave SHM systems.

\section{The Statistical Damage Detection Framework}\label{sec:SDDF}

\subsection{The General Framework} \label{sec:SHT}

\begin{figure}[t!]
\hspace{-0.4cm}\includegraphics[scale=0.75]{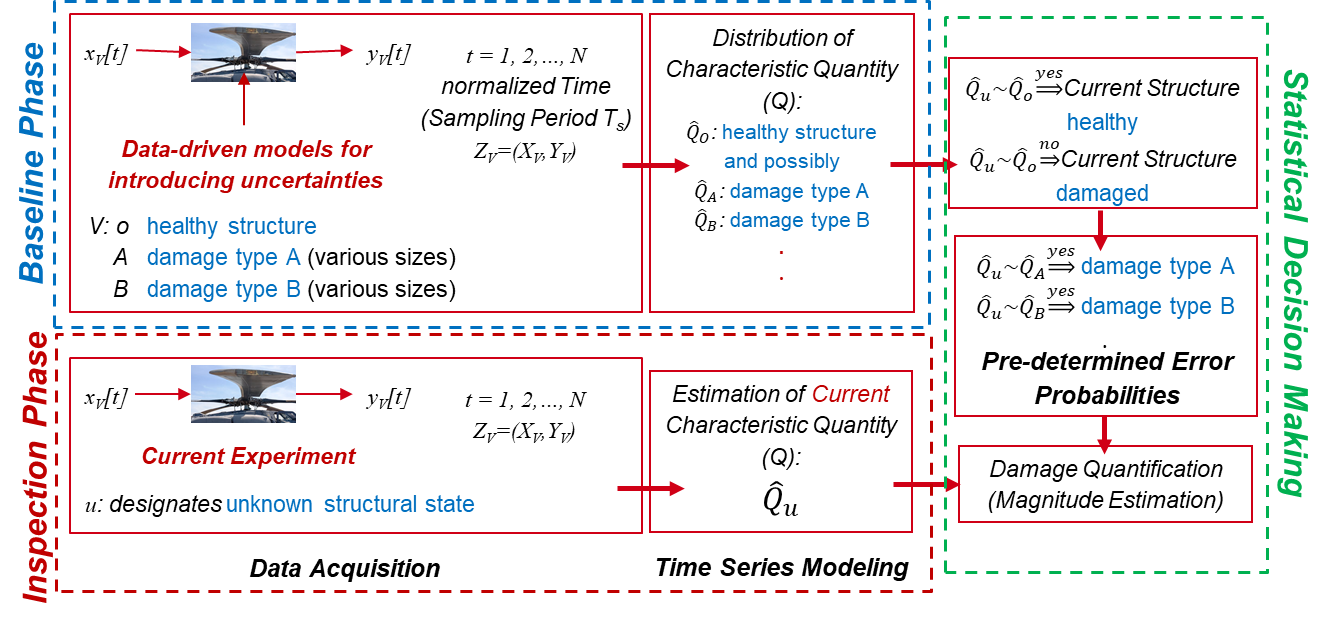} 
\caption{Framework for statistical time series methods for structural health monitoring  \cite{Amer-Kopsaftopoulos19a,Fassois-Kopsaftopoulos-encyclopedia,Kopsaftopoulos-Fassois10}.} 
\label{fig:SHT}
\end{figure}

The use of statistical methods for damage detection and identification has been previously reported for vibration-based SHM \cite{Kopsaftopoulos-Fassois08,Kopsaftopoulos-Fassois10ewshm,Kopsaftopoulos-Fassois10}, and only recently for active-sensing guided-wave SHM \cite{Amer-Kopsaftopoulos19a}. A typical statistical framework for damage detection, localization and quantification is shown in Figure \ref{fig:SHT} \cite{Amer-Kopsaftopoulos19a,Fassois-Kopsaftopoulos-encyclopedia}. In this framework, $x[t]$ and $y[t]$ are the individual actuation and response signals, respectively, for every structural case, respectively, indexed with discrete time $t$ ($t=1, 2, \ldots, N$), which can be converted to continuous time through the transformation $(t-1)T_s$, where $T_s$ is the sampling time for the recorded signals. The subscripts ($o, A, B, \ldots,$ and $u$) indicate the healthy, damage $A, B, \ldots,$ and unknown cases, respectively. In this context, the damage cases labeled as ($A, B, \ldots$) can resemble different types, sizes or locations of damage. For each structural case, all actuation ($X$) and response ($Y$) signals can be presented as $Z=(X,Y)$, with $Z_o, Z_A, Z_B,...,$ and $Z_u$ indicating the different cases as before. 

As shown in Figure \ref{fig:SHT}, the statistical time series framework consists of two phases, namely the baseline and inspection phase. In the baseline phase, NP-TS models, each producing a characteristic quantity $\widehat{Q}$, are identified and properly validated for the healthy ($\widehat{Q}_0$) time series signal, as well as, if available, different predefined damage cases ($\widehat{Q}_A$, $\widehat{Q}_B,\ldots$). Then, during the inspection phase, the same NP-TS models are identified for the unknown  ($\widehat{Q}_u$) state of the system. Next, damage detection is achieved through applying appropriate binary statistical hypothesis tests to assess the statistical deviation of the unknown quantity $\widehat{Q}_u$ from $\widehat{Q}_0$ corresponding to the healthy signal (damage detection). Based on data availability, the statistical similarity to one of the damage characteristic quantities $\widehat{Q}_A$, $\widehat{Q}_B, \ldots$ with the baseline quantity $\widehat{Q}_0$ can enable statistical damage identification/classification. In the present study, this framework is only used for damage detection. 

\subsection{Overview of Non-parametric Time Series Representations}\label{sec:NP-TS}

Stochastic NP-TS representations utilize time-domain Auto-/Cross-Covariance Functions (A/CCF) and/or frequency-domain Power-/Cross-Spectral Densities (P/CSD) in order to model a dynamic stationary signal \cite[Chapter 2, pp. 39]{Box-etal94}. As discussed above, frequency-domain models are used in this study. In this context, several estimators have been developed for the PSD (also referred to as Auto Spectral Density) of a sensor excitation and/or response signal, including the periodogram, the Thompson, the Blackman-Tukey, and the Bartlett-Welch (or simply Welch) estimators \cite[Chapter 5, pp. 235]{Ingle05}. As such estimators are random variables that represent the true PSD of a system, their corresponding statistical properties, such as the mean and variance, allow for the extraction of estimation confidence intervals that can be subsequently used to represent statistical damage thresholds. In this study, the Welch PSD estimate, which is a modified periodogram estimator using a series of overlapping windows \cite[Chapter 4, pp. 76]{Kay88} is used for damage detection. For a time series signal $x[t]$, the frequency-domain ($\omega$) Welch PSD ($\widehat{S}_{xx}(\omega)$) is based on the averaging of multiple-windowed periodograms using properly-selected sample windows $w[t]$ with 50\% overlap, and is calculated as follows \cite[Chapter 8, pp. 418]{Hayes96} (the hat indicates an estimated variable):
\begin{equation}
\widehat{S}_{xx}(\omega) = \frac{1}{KLUT} \sum_{i=0}^{K-1} \Bigl| T\sum_{t=0}^{L-1} w[t] \cdot \widehat{x}[t+iD]^{(-j2\pi \omega tT)} \Bigr|^2 
\end{equation}
with
\begin{equation}
U=\frac{1}{L} \sum_{t=0}^{L-1} w^2[t], \quad \widehat{x}[t] = x[t] - \widehat{\mu}_x, \quad N=L+D(K-1)
\end{equation}
and $N$, $L$, $K$, $D$, and $T$ being the total number of signal samples, the size of each window, the number of utilized windows, the number of overlapping data points in each window, and the time period of the signal, respectively. $\widehat{\mu}_x$ represents the mean of the time series. The estimation statistics, that is the mean and variance, of the Welch PSD can be described as follows in case the Bartlett window is used \cite[Chapter 8, pp. 419]{Hayes96}:
\begin{equation}
E\{\widehat{S}_{xx}(\omega)\}= \frac{1}{2\pi LU} {S}_{xx}(\omega) |W(\omega)|^2 
\end{equation}
\begin{equation}
Var\{\widehat{S}_{xx}(\omega)\} \approx \frac{9}{16} \frac{L}{N} {S}^2_{xx}(\omega) 
\end{equation}
where $W(\omega)$ designates the Fourier transform of the window function. One of the main reasons behind the wide use of the Welch PSD estimator is that it is asymptotically unbiased and consistent \cite{Ingle05}. In this study, the Welch PSD estimate is used in developing appropriate statistical quantities, also referred to as test statistics, and corresponding statistical hypothesis tests for damage detection, as described in the following section.

\subsection{The Single-set $F$ Statistic Method}\label{sec:F}

Based on the PSD-based NP-TS method and the corresponding statistical hypothesis testing setup presented in \cite{Kopsaftopoulos-Fassois10,Fassois-Kopsaftopoulos-encyclopedia}, damage can be detected by assessing changes in the Welch PSD of properly-determined wave packets/modes from an acousto-ultrasound time series signal. Thus, the characteristic quantity in this study is $Q={S}_{xx}(\omega)={S}(\omega)$. The main idea is based on the comparison of the Welch PSD of the response of the structure in an unknown state, $S_u(\omega)$, to that of the structure in its healthy state, $S_o(\omega)$. Damage detection can thus be tackled using the following SHT problem \cite{Amer-Kopsaftopoulos19a,Kopsaftopoulos-Fassois10}:
\begin{equation}
    \begin{array}{llll}
         H_o & : & S_u(\omega) = S_o(\omega) & \text{(null hypothesis -- healthy structure )} \\
         H_1 & : & S_u(\omega) \neq S_o(\omega) & \text{(alternative hypothesis -- damaged structure)}
    \end{array}
    \label{eq:hypothesis}
\end{equation}
Again, due to the finite nature of the experimental time series, the true PSD values are unknown, and thus corresponding estimated quantities are utilized instead ($\widehat{S}$). It can be shown that the Welch PSD estimate will have the following property \cite[Chapter 3, pp. 46]{Kay88}:
\begin{equation}
2K\widehat{S}(\omega)/{S}(\omega) \sim \chi^2(2K) \label{eq:estimator_prop}\end{equation}
In the above expression, the factor of 2 comes from the fact that every periodogram used in averaging the Welch PSD has a real and a complex component. Consequently, a damage detection statistic following the $\mathcal{F}$ distribution with $(2K,2K)$ degrees of freedom can be developed as follows:
\begin{equation}
F = \frac{\widehat{S}_o(\omega)/S_o(\omega)}{\widehat{S}_u(\omega)/S_u(\omega)} \, \sim \, \mathcal{F}(2K,2K)
\end{equation}
In the case of a healthy structure (null hypothesis), $S_u(\omega)$ and $S_o(\omega)$ coincide, thus:
\begin{equation}
\text{Under} \; H_o: \quad F = \frac{\widehat{S}_o(\omega)}{\widehat{S}_u(\omega)} \, \sim \, \mathcal{F}(2K,2K)
\end{equation}
Thus, the above SHT decision-making process can be modified as follows:

\begin{equation}
    \begin{array}{ccl}
     \hspace{-0.4cm} f_{\frac{\alpha}{2}}(2K,2K) \leq F=\frac{\widehat{S}_o(\omega)}{\widehat{S}_u(\omega)} \leq f_{1-\frac{\alpha}{2}}(2K,2K)\quad (\forall \; \omega) & \hspace{-0.2cm} \Longrightarrow & \hspace{-0.2cm} H_o \; \text{is accepted (healthy structure)} \\  
     \text{Else} & \hspace{-0.2cm} \Longrightarrow & \hspace{-0.2cm} H_1 \; \text{is accepted (damaged structure)} \\  
    \end{array} 
\end{equation} 
where $\alpha$ is the Type I error (\textit{false alarm}) probability, $f_{\frac{\alpha}{2}}$, $f_{1-\frac{\alpha}{2}}$ designate the $\mathcal{F}$ distribution’s $\frac{\alpha}{2}$ and $1-\frac{\alpha}{2}$ critical points, respectively ($f_{\alpha}$ is defined such that Prob$(F \leq f_{\alpha}) = \alpha$).

\subsection{The Multiple-set Modified $F_m$ Statistic Method} \label{sec:Fm}

In many realistic cases, a single baseline signal may not be representative of the healthy structure, and the average of many signal realizations might be more meaningful. In that case, multiple closely-spaced (time-wise) response signal realizations available for each state of the component being monitored under nominally-constant environmental/operational conditions can be used to entail some experimental statistics in the estimation of the SHM metric being used. Towards this end, the sample expectation, that is,
\begin{equation}
E\{\widehat{S}_o(\omega)\} = \frac{1}{M} \sum_{h=1}^{M} \widehat{S}_o(\omega) \; 
\label{eq:mean}
\end{equation}
can be used in order to ``expand'' the baseline/healthy estimates for the structure being monitored. In the above expression, $M$ is the number of healthy data sets used in the estimation of the metric being used. Then, following the aforementioned property of PSD estimates in equation (\ref{eq:estimator_prop}), the following expression can be developed:
%
%
%
%
\begin{equation}
2KME\{\widehat{S}_o(\omega)\}/{S}_o(\omega) \sim \chi^2(2KM) 
\end{equation}
As such, a modified $F$ statistic can be developed by replacing the Welch PSD estimate with the mean of all PSD estimates of $M$ number of time-series signals taken for the system under the baseline/healthy state:
\begin{equation}
F_m = \frac{ E\{\widehat{S}_o(\omega)\}/S_o(\omega)}{\widehat{S}_u(\omega)/S_u(\omega)} \; \sim \; \mathcal{F}(2KM,2K)
\end{equation}	
Under the null hypothesis in equation (\ref{eq:hypothesis}), the $S_o(\omega)$ and $S_u(\omega)$ coincide: 
\begin{equation}
\text{Under} \; H_o: \quad F_m = \frac{ E\{\widehat{S}_o(\omega)\}}{\widehat{S}_u(\omega)} \; \sim \; \mathcal{F}(2KM,2K)
\end{equation}	
thus, the modified decision making scheme with the appropriate confidence levels, can be expressed as follows:
\begin{equation}
    \begin{array}{ccl}
\hspace{-0.4cm} f_{\frac{\alpha}{2}}(2KM,2K) \leq F_m = \frac{E\{\widehat{S}_o(\omega)\}}{\widehat{S}_u(\omega)} \leq f_{1-\frac{\alpha}{2}}(2KM,2K)\quad (\forall \; \omega) & \hspace{-0.2cm} \Longrightarrow & \hspace{-0.2cm} H_o \; \text{is accepted (healthy)} \\  
     \text{Else} & \hspace{-0.2cm} \Longrightarrow & \hspace{-0.2cm} H_1 \; \text{is accepted (damaged)} \\  
    \end{array} 
\end{equation} 
with $f_{\frac{\alpha}{2}}$, $f_{1-\frac{\alpha}{2}}$ designating the $\mathcal{F}$ distribution’s $\frac{\alpha}{2}$ and $1-\frac{\alpha}{2}$ critical points, respectively ($f_{\alpha}$ is defined such that Prob$(F_m \leq f_{\alpha}) = \alpha$).

\subsection{The Multiple-set $Z$ Statistic Method} \label{sec:Z}

With the availability of a sufficiently-large number of data sets, that is a large $M$ in equation (\ref{eq:mean}), $E\{\widehat{S}_o(\omega)\}$ would approach the true PSD, and according to the Central Limit Theorem (CLT) \cite[Chapter 3, pp. 62]{Bendat-Piersol00}, $E\{\widehat{S}_o(\omega)\}$ would also follow a normal distribution with the true PSD being the mean and $\sigma_0$ the variance. Utilizing these statistical phenomena, and based on the $Z$ statistic developed by Fassois and coworkers \cite{Kopsaftopoulos-Fassois10,Fassois-Kopsaftopoulos-encyclopedia} for the Frequency Response Function (FRF) of vibration-based SHM signals, a novel $Z$ statistic is proposed herein utilizing the Welch PSD estimate for many baseline active-sensing acousto-ultrasound SHM signals. The following SHT problem is posed for damage detection in this case:
\begin{equation}
    \begin{array}{llll}
         H_o & : & S_o(\omega) - S_u(\omega) = 0       & \text{(null hypothesis -- healthy structure)} \\
         H_1 & : & S_o(\omega) - S_u(\omega) \neq 0    & \text{(alternative hypothesis -- damaged structure)}
    \end{array}
\end{equation}
where both terms in the hypothesis above are the true values of the respective PSDs. As mentioned above, under the assumption of a large $M$ in equation (\ref{eq:mean}) (many baseline signals used for expectation estimation), the first term in the hypothesis test above ($S_o(\omega)$) can be replaced by the expectation, which would be normally distributed as aforementioned. Additionally, assuming the availability of many data points (that is, a large $N$) used for PSD estimation, the second term in the formulation of the hypothesis test ($S_u(\omega)$) can be replaced by an estimate \cite[Chapter 3, pp. 45]{Kay88}, which will also follow a normal distribution due to the asymptotic properties of $\chi^2$-distributed estimates, as also dictated by the central limit theorem (CLT) \cite[Chapter 3, pp. 62]{Bendat-Piersol00}. Under the null hypothesis, both of these terms would follow the same distribution since both would be coming from a healthy structural case. Thus, under the null hypothesis:
%
%
\begin{equation}
    \begin{array}{llll}
          \text{Under} \; H_o & : & E\{\widehat{S}_o(\omega)\} - \widehat{S}_u(\omega) \; \sim \; \mathcal{N}(0,2\sigma_o^2(\omega)) & \mbox{(null hypothesis -- healthy structure)} \\
    \end{array}
\end{equation}
where $\sigma_o^2(\omega)$ can be estimated from the baseline phase, and can be assumed to have negligible variability if a large number of signals is used in estimating the value of the PSDs \cite{Kopsaftopoulos-Fassois10,Fassois-Kopsaftopoulos-encyclopedia}. Thus, by defining an appropriate type I error, or false alarm, probability ($\alpha$), the Welch PSD-based $Z$ statistic can be expressed as follows:
\begin{equation}
    \begin{array}{ccl}
     Z = \frac{\mid E\{\widehat{S}_o(\omega)\} - \widehat{S}_u(\omega) \mid}{\sqrt{2\widehat{\sigma}_o^2(\omega)}} \leq Z_{1-\frac{\alpha}{2}}\quad (\forall \; \omega) & \Longrightarrow & H_o \; \text{is accepted (healthy structure)} \\  
     \text{Else} &  \Longrightarrow & H_1 \; \text{is accepted (damaged structure)} \\  
    \end{array} 
\end{equation} 
with $Z_{1-\frac{\alpha}{2}}$ designating the standard Normal distribution's $1-\frac{\alpha}{2}$ critical point. Table \ref{tab:statistics} summarizes all three statistics used in this study for damage detection.

\begin{table}[b]
\centering
\caption{Summary of the different damage detection statistics utilized in this study.}\label{tab:statistics}
\renewcommand{\arraystretch}{1.2}
{\footnotesize
\begin{tabular}{|lccc|} 
\hline
Quantity & $F$ Statistic & $F_m$ Statistic & $Z$ Statistic \\
\hline
Property & $2K\widehat{S}(\omega)/{S}(\omega) \sim \chi^2(2K)$ & $2KME\{\widehat{S}(\omega)\}/{S}_o(\omega) \sim \chi^2(2KM)$ & $E\{\widehat{S}(\omega)\} - S(\omega) \; \sim \; \mathcal{N}(0,2\sigma_o^2(\omega))$ \\
Test Statistic & $F=\frac{\widehat{S}_o(\omega)}{\widehat{S}_u(\omega)}$ & $F_m = \frac{ E\{\widehat{S}_o(\omega)\}}{\widehat{S}_u(\omega)}$ &  $Z = \frac{\mid E\{\widehat{S}_o(\omega)\} - \widehat{S}_u(\omega) \mid}{\sqrt{2\sigma_o^2(\omega)}}$ \\
 & \multicolumn{3}{c|}{$K$: Number of non-overlapping segments} \\
Comment  & \multicolumn{3}{c|}{$M$: Number of available baseline data sets} \\
 & \multicolumn{3}{c|}{$\widehat{S}(\omega)$: Welch PSD estimate; $\omega \; \epsilon[0,2\pi/T_s]$: frequency in radians per second ($T_s$ is the sampling time). } \\
\hline
\end{tabular}} 
\end{table}

It is worth noting here that, although there lies a difference between the statistical assumptions behind the formulations of the $F_m$ and the $Z$ statistics, both are applied to the same data sets in this study, and their results are compared with respect to which statistic achieves better detection capabilities.

\subsection{Reference State-of-the-Art Damage Indices}\label{sec:DIs}
In this work, two time-domain damage indices are utilized as reference in order to compare between the performance of DIs and the performance of the NP-TS models proposed herein. The first DI was adopted from the work of Janapati \etal \cite{Janapati-etal16}, which is characterized by high sensitivity to damage size and orientation, and low sensitivity to other variations such as adhesive thickness and the material properties of the structure, sensors, and adhesive. Given a baseline $y_0[t]$ and an unknown $y_u[t]$ signal indexed with normalized discrete time $t$ ($t=1,2,3,\ldots, N$ where $N$ is the number of data samples considered in the calculation of the DIs, which depends on the studied coupon as will be shown in Section \ref{sec:results}), the formulation of that DI is as follows:
\begin{equation}
Y_{u}^n[t] =\frac{y_u[t]}{\sqrt{\sum_{t=1}^{N}{y^2_u[t]}}},
\quad Y_{0}^n[t]=\frac{\sum_{t=1}^{N}{(y_0[t]\cdot Y_{u}^n[t])}}{y_0[t]\cdot \sum_{t=1}^{N}{y_0^2[t]}},
\quad DI=\sum_{t=1}^{N}{(Y^n_{u}[t] - Y^n_{0}[t])}
\end{equation}
In this notation, $Y^n_u[t]$ and $Y^n_0[t]$ are normalized unknown (inspection) and baseline signals, respectively. The second DI used in this study is the time-domain DI presented by Qiu \etal \cite{Qiu-etal16} and used in training their Gaussian mixture models due to its sensitivity to changes in wave form and time of flight. The formulation of that DI is as follows:
\begin{equation}
DI=1-\sqrt{\frac{(\sum_{t=1}^{N}{y_0[t]\cdot y_u[t])^2}}{\sum_{t=1}^{N}{y_0^2[t]} \cdot \sum_{t=1}^{N}{y^2_u[t]}}}
\end{equation}
%
\section{Results and Discussion}\label{sec:results}

In this work, the comparison between state-of-the-art DIs and the proposed NP-TS approaches in damage detection was carried out over three components with different damage cases: a notched Al plate, a Carbon Fiber-Reinforced Plastic (CFRP) coupon with weights taped on the surface to simulate a crack, and the open-source data sets available on the Open Guided-Waves project's website \cite{OGW}.

\subsection{Test Case I: Damage Detection in an Aluminum Plate}

\subsubsection{Test Setup, Damage Types and Data Acquisition} 

\begin{figure}[t!]
\centering
\includegraphics[scale=0.38]{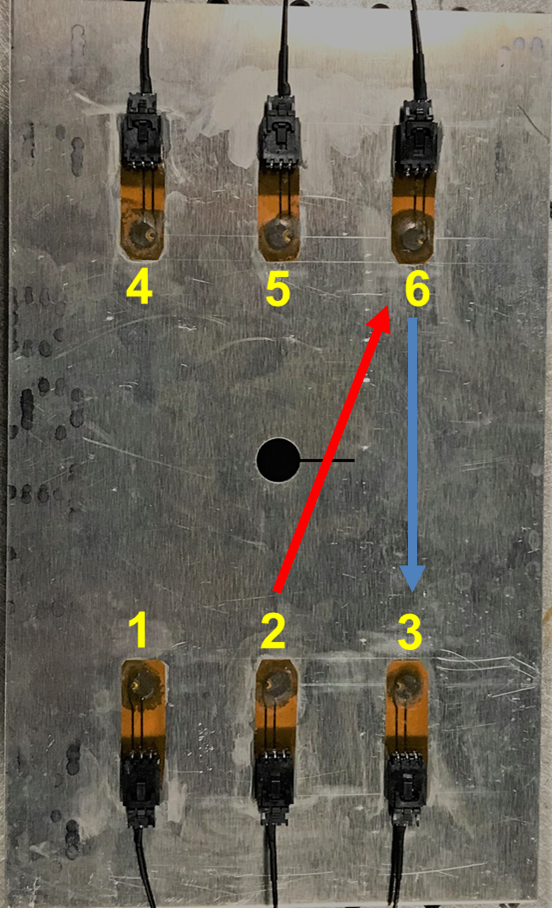} 
\caption{The Al coupon used in this study shown here with a 20-mm notch (the largest damage size of this test case). The arrows indicate the paths used in the analysis presented herein.}  
\label{fig:Al_coupon}
\end{figure}

This first coupon was a 6061 Aluminum $152.4 \times  254$ mm ($6 \times 10$ in) coupon ($2.36$ mm/$0.093$ in thick) (McMaster Carr) with a 12-mm (0.5-in) diameter hole in the middle, as shown in Figure \ref{fig:Al_coupon}. Using Hysol EA 9394 adhesive, the coupon was fitted with six single-PZT (Lead Zirconate Titanate) SMART Layers type PZT-5A (Acellent Technologies, Inc) as shown in Figure \ref{fig:Al_coupon}. The PZT sensors are $0.2$ mm ($0.00787$ in) in thickness and $3.175$ mm ($1/8$ in) in diameter. To simulate damage, using an end-mill and a $0.8128$-mm ($0.032$-in) hand saw, a notch was generated extending from the middle hole of the coupon with length varying between $2$ and $20$ mm, in $2$-mm increments.

Actuation signals in the form of 5-peak tone bursts (5-cycle Hamming-filtered sine wave) having an amplitude of 90 V peak-to-peak and various center frequencies were generated in a pitch-catch configuration over each sensor consecutively. With a sampling rate of 24 MHz, data was collected using a ScanGenie III data acquisition system (Acellent Technologies, Inc). Preliminary analysis was conducted, and a center frequency of 250 kHz was chosen for the complete analysis presented in this study based upon the best separation between the first two wave packets in various signal paths. All data sets were exported to MATLAB for analysis.\footnote{Matlab version R2018a; function \textit{pwelch.m} (window size: 100 for single wave packet analyses and 500 for the full signal/two-wave packet analyses; NFFT: 2000; Overlap: 50\%).} Table \ref{tab:Al_exp_info} summarizes the relevant experimental details for this coupon.

\begin{table}[b!]
\centering
\caption{Summary of experimental details for the Al coupon.}\label{tab:Al_exp_info}
\renewcommand{\arraystretch}{1.2}
{\footnotesize
\begin{tabular}{|cc|}
\hline 
Structural State & Number of Data Sets \\
\hline
Healthy & 20$^{\dagger}$ \\
2-mm notch & 20 \\
4-mm notch & 20 \\
6-mm notch & 20 \\
8-mm notch & 20 \\
10-mm notch & 20 \\
12-mm notch & 20 \\
14-mm notch & 20 \\
16-mm notch & 20 \\
18-mm notch & 20 \\
20-mm notch & 20 \\
\hline
\multicolumn{2}{l}{{\bf Sampling Frequency:} $f_s=24$ MHz. Center frequency range: [$50:50:750$] kHz} \\
\multicolumn{2}{l}{{\bf Number of samples} per data set $N= 8000$.}\\
\multicolumn{2}{l}{$^\dagger$M=20 in equation (\ref{eq:mean}).}\\
\end{tabular}} 
\end{table}

\subsubsection{Damage Detection Results}

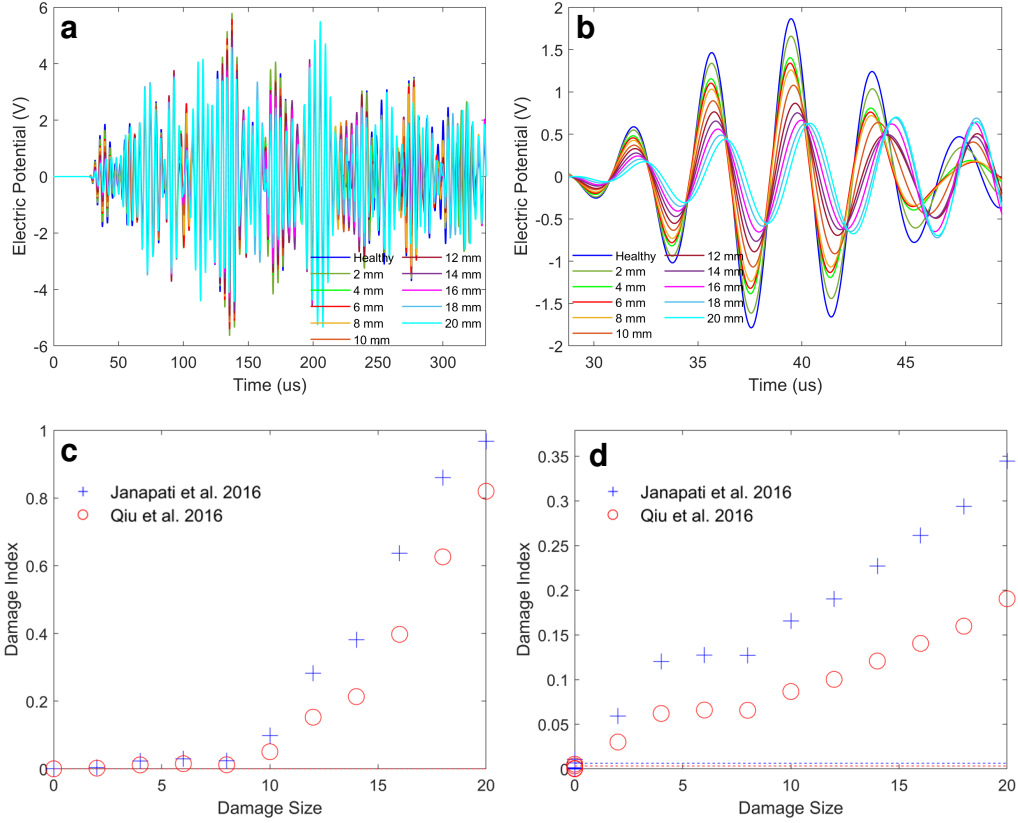
\begin{figure}[t!]
\centering
\begin{picture}(400,300)
\put(0,160){\includegraphics[scale=0.5]{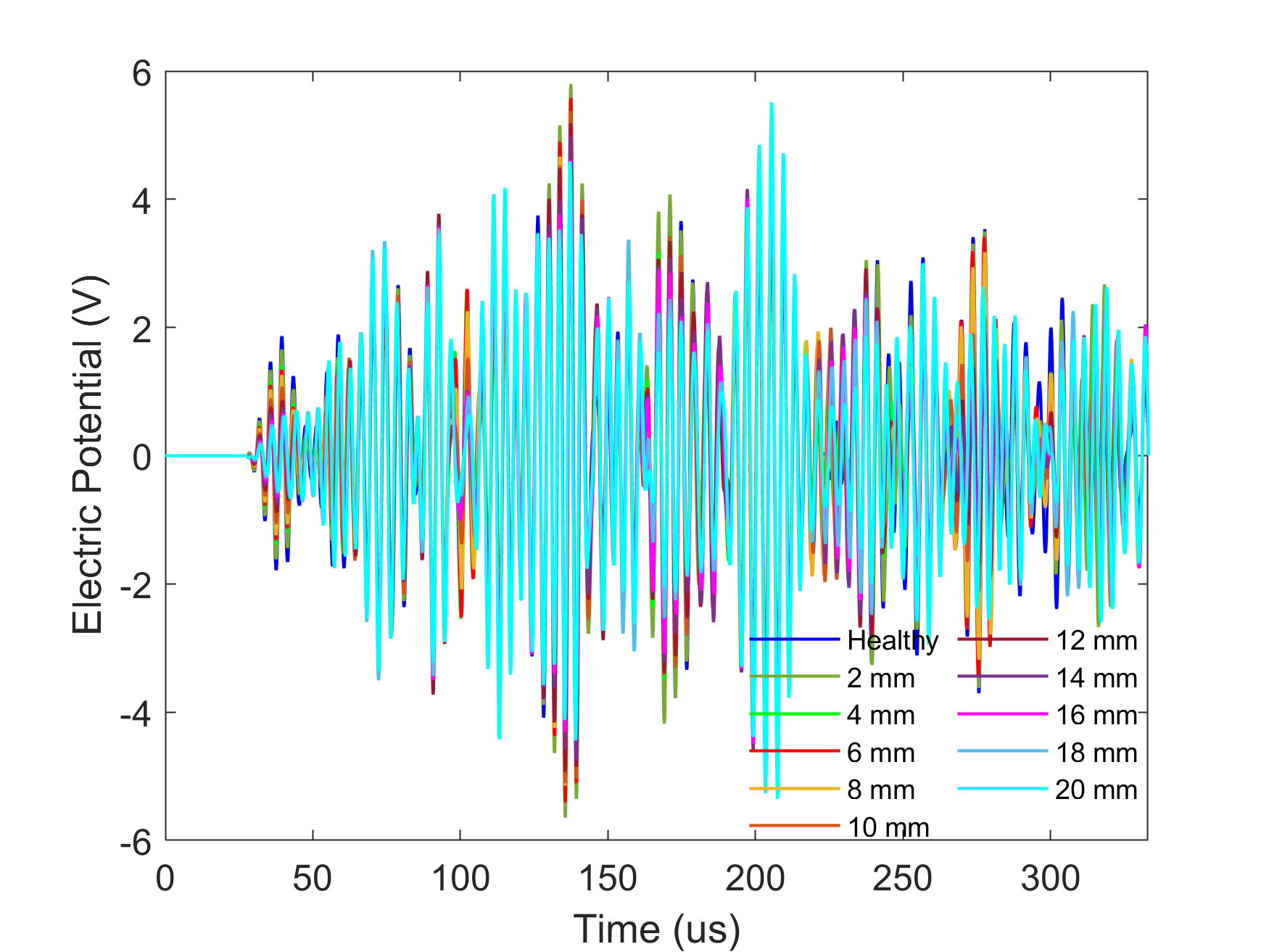}}
\put(195,160){\includegraphics[scale=0.5]{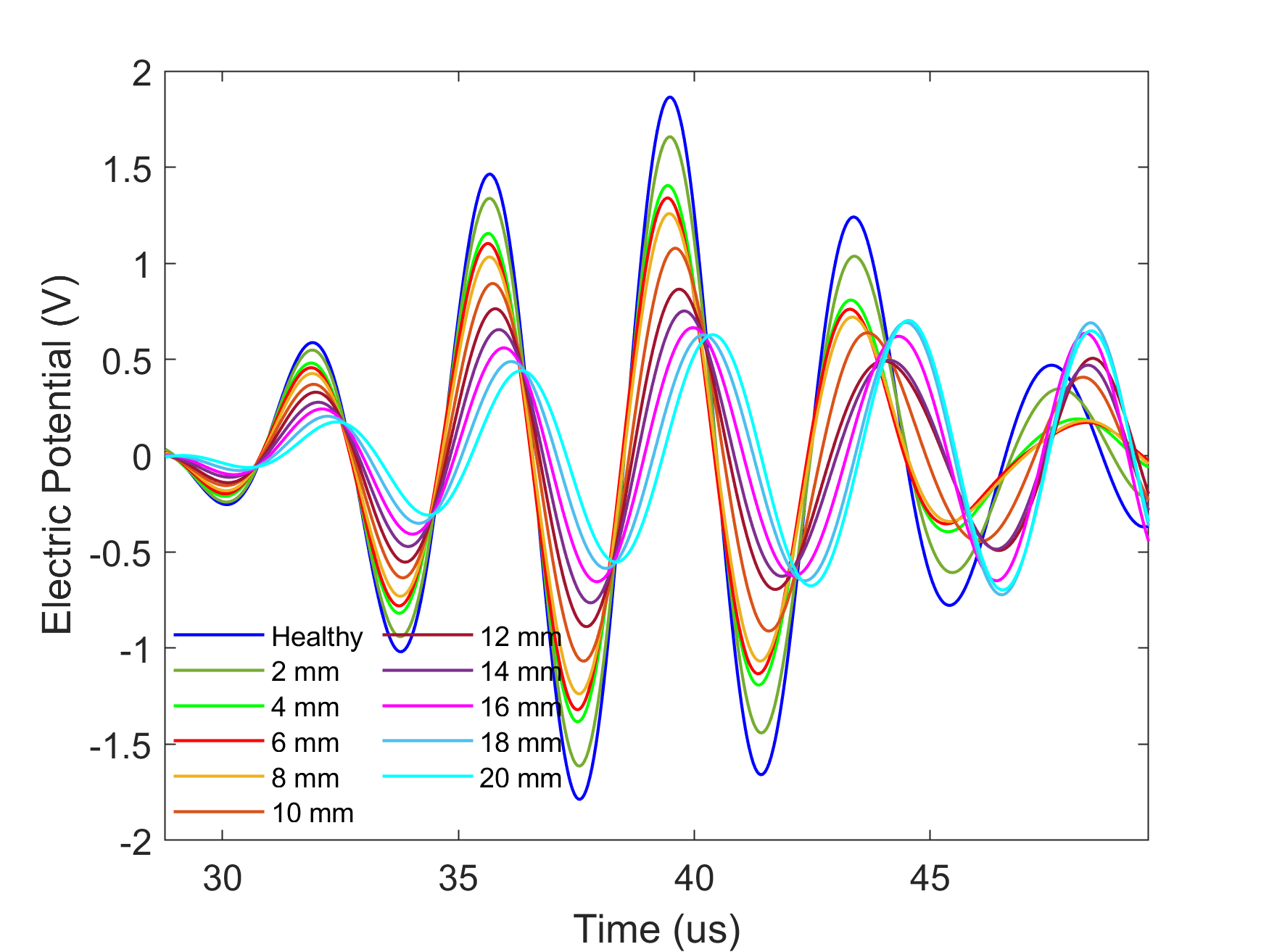}}
\put(0,0){\includegraphics[scale=0.5]{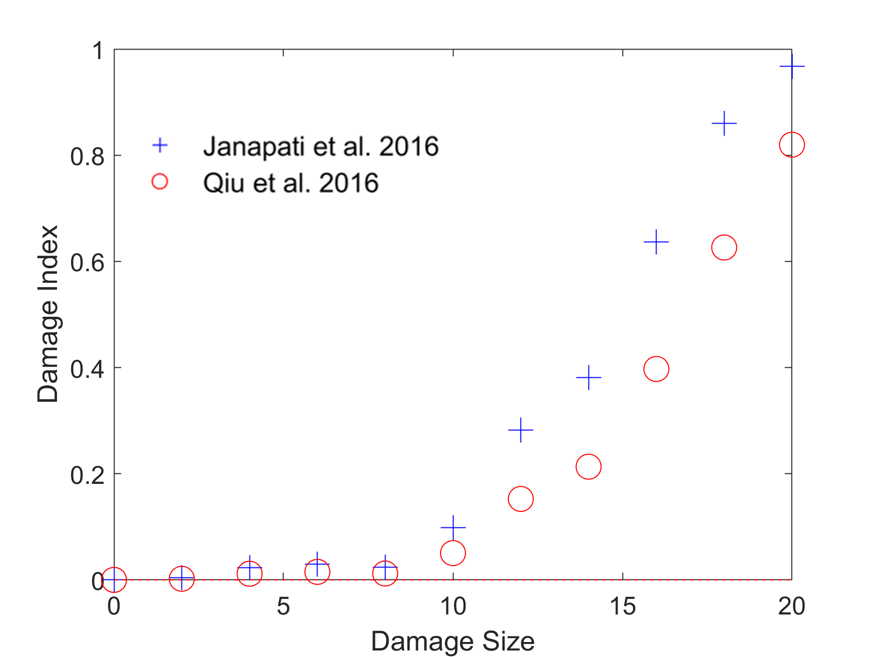}}
\put(197,0){\includegraphics[scale=0.5]{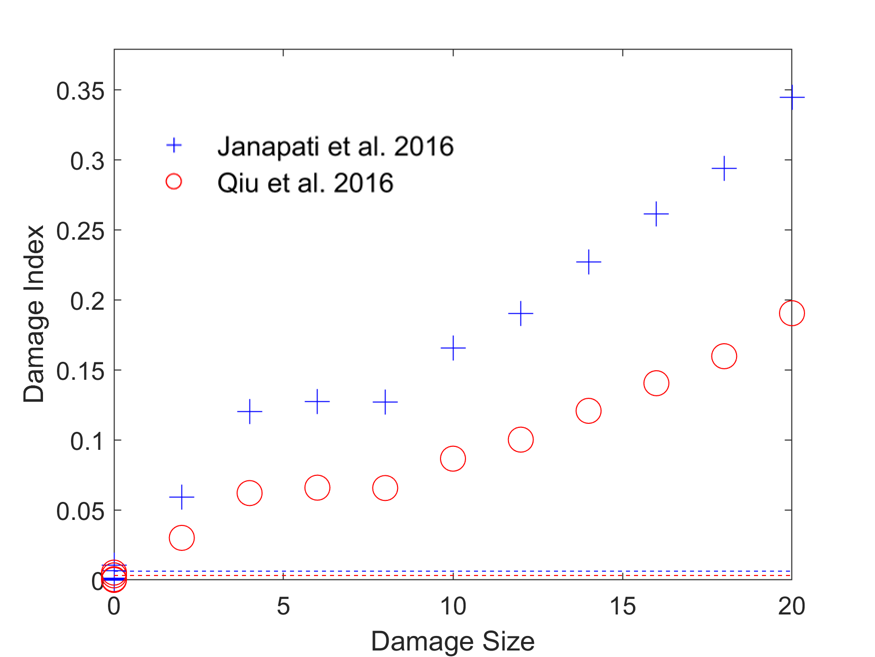}}
\put(30,295){\color{black} \large {\fontfamily{phv}\selectfont \textbf{a}}}
\put(225,295){\color{black} \large {\fontfamily{phv}\selectfont \textbf{b}}}
\put(30,135){\color{black} \large {\fontfamily{phv}\selectfont \textbf{c}}} 
\put(230,133){\color{black} \large {\fontfamily{phv}\selectfont \textbf{d}}}
\end{picture} 
\caption{Indicative signal from the Al coupon for signal path 2-6 (damage-intersecting) under different notch sizes: (a) full signal; (b) first-arrival wave packet; (c) single wave packet DIs -- the dashed lines designate the upper and lower $95\%$ confidence bounds for the Janapati \etal (blue) and Qiu \etal (red) DIs; (d) two wave packet DIs.}  
\label{fig:Al_2-6_sig}
\end{figure}

\begin{figure}[t!]
\centering
\begin{picture}(400,300)
\put(-2,160){\includegraphics[scale=0.5]{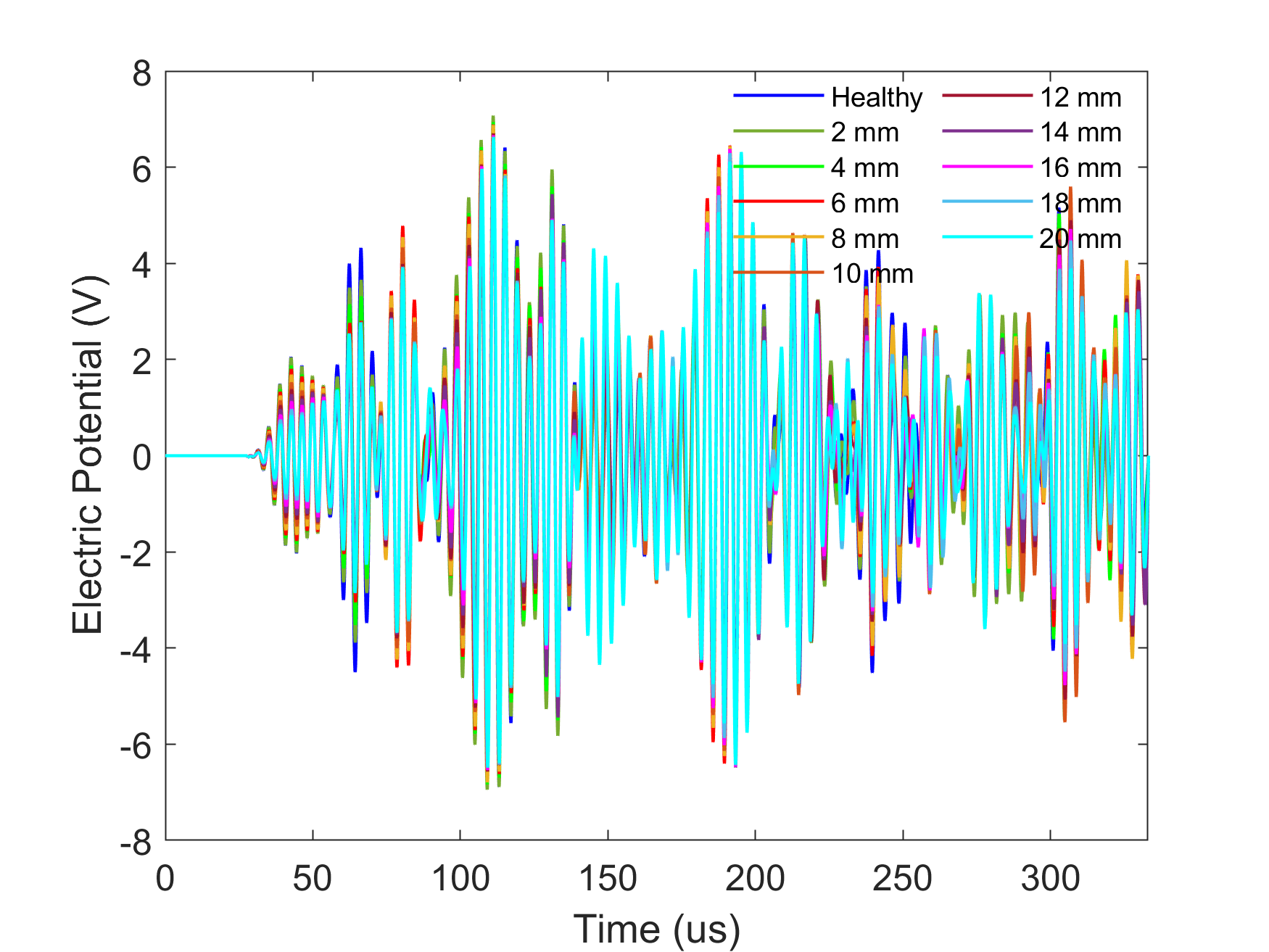}}
\put(191,160){\includegraphics[scale=0.5]{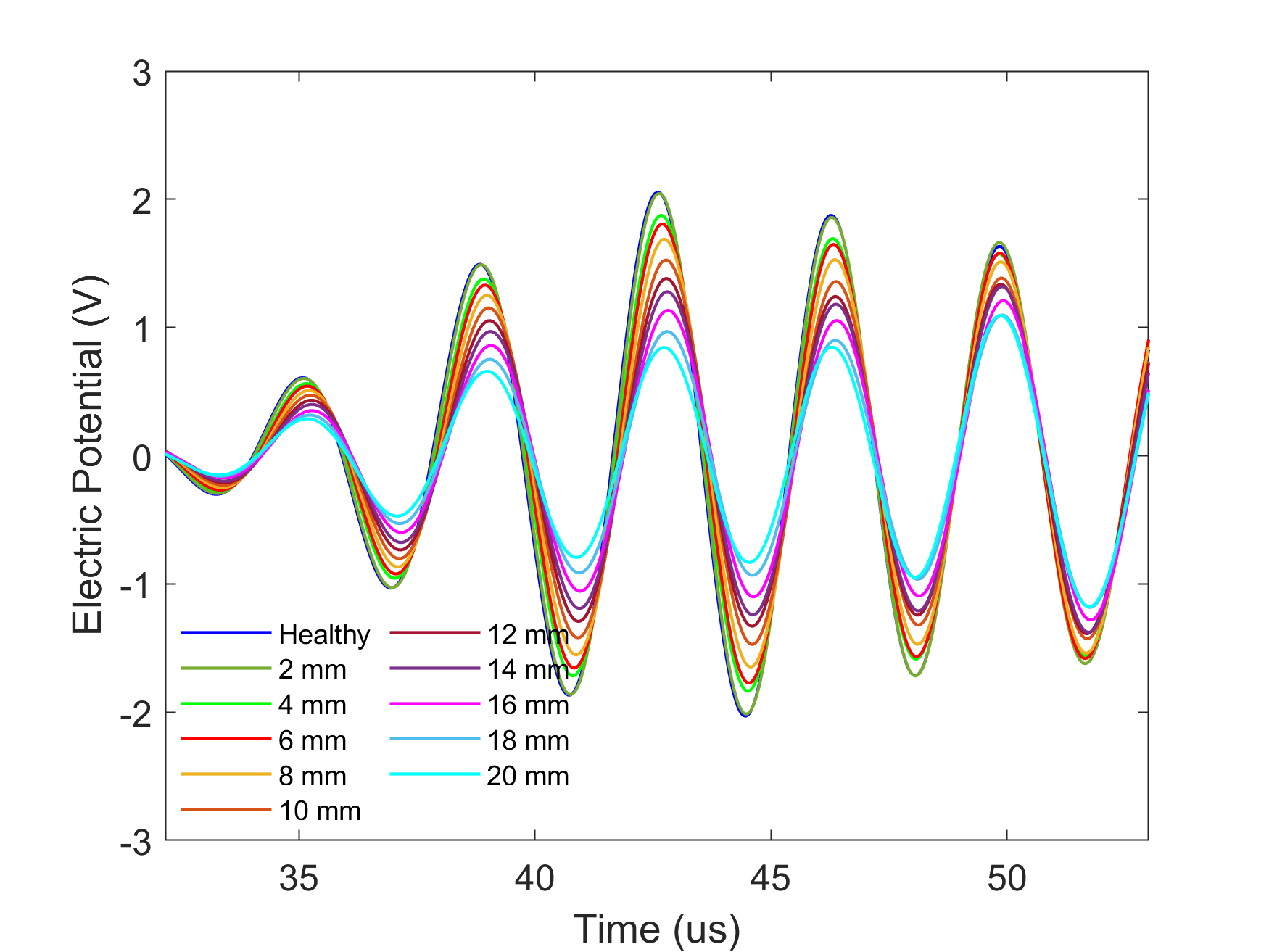}}
\put(-1,0){\includegraphics[scale=0.5]{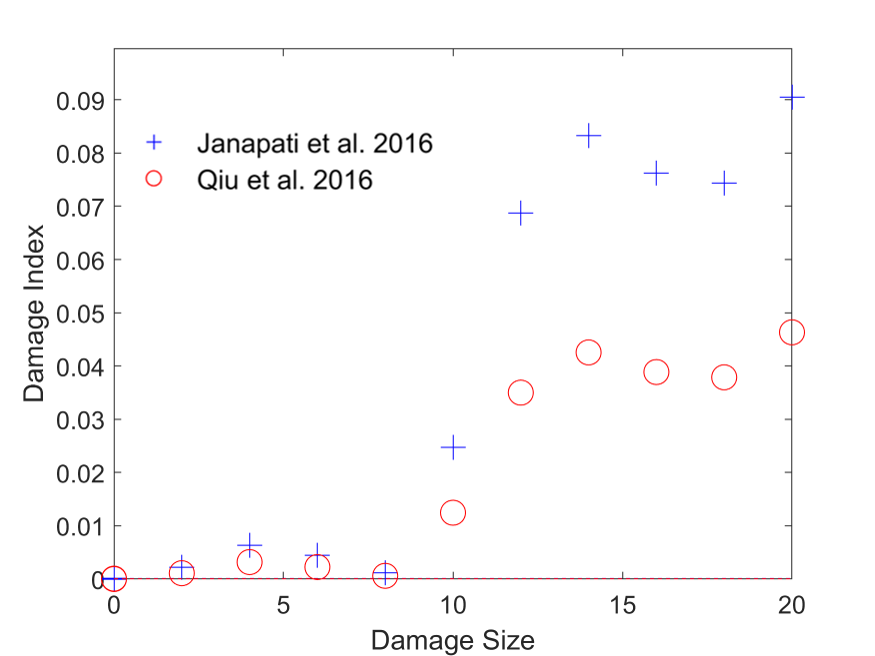}}
\put(192,0){\includegraphics[scale=0.5]{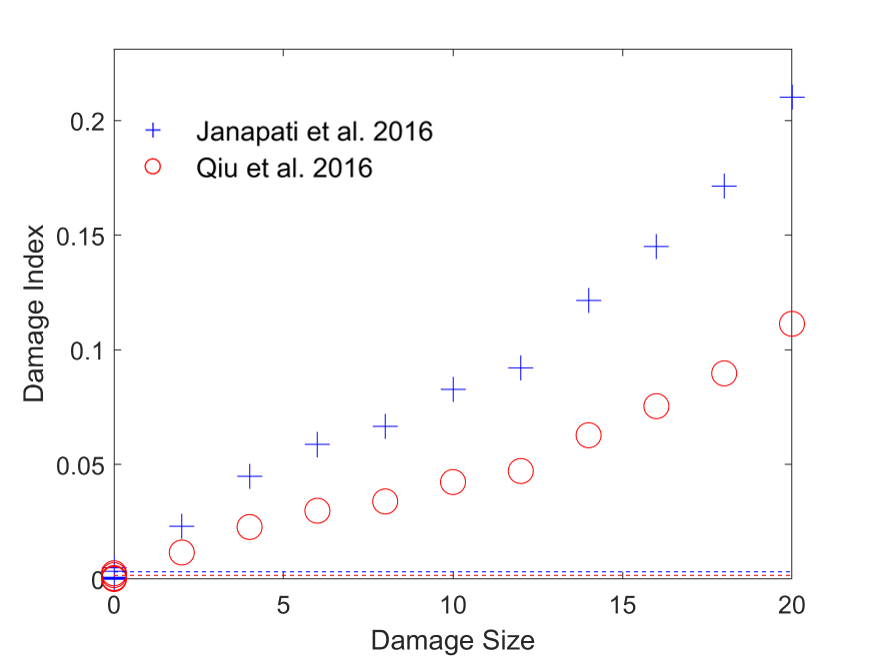}}
\put(28,295){\color{black} \large {\fontfamily{phv}\selectfont \textbf{a}}}
\put(221,295){\color{black} \large {\fontfamily{phv}\selectfont \textbf{b}}}
\put(30,135){\color{black} \large {\fontfamily{phv}\selectfont \textbf{c}}} 
\put(223,135){\color{black} \large {\fontfamily{phv}\selectfont \textbf{d}}}
\end{picture} 
\caption{Indicative signal from the Al coupon for signal path 6-3 (damage-non-intersecting) under different notch sizes: (a) full signal; (b) a single wave packet; (c) single-wave packet DIs -- the dashed lines designate the upper and lower $95\%$ confidence bounds for the Janapati \etal (blue) and Qiu \etal (red) DIs; (d) two-wave packet DIs.}  
\label{fig:Al_6-3_sig}
\end{figure}

In order to assess the performance of the proposed approach, a simple isotropic Al coupon was initially used. Figure \ref{fig:Al_2-6_sig} panels a and b, respectively, show one indicative full response signal and its corresponding first-arrival wave packet off of sensor 6 when sensor 2 was actuated (refer to Figure \ref{fig:Al_coupon} for sensor numbering) under different notch sizes. Because this is a damage-intersecting path, a gradual decrease in signal amplitude, with a slight delay, can be observed with increasing notch size. This is expected since the notch scatters the wave, decreasing the amount of energy going through to sensor 6 as scattering increases \cite{Amer-Kopsaftopoulos19b}. Figure \ref{fig:Al_2-6_sig}c shows the evolution of the two chosen state-of-the-art DIs for the first-arrival wave packet. As shown, although the DIs closely follow damage for notch sizes more than 8 mm, it might be difficult to detect damages up to 8 mm in size, given the proximity of the DI values for the healthy case and the damaged cases. Without prior experience with these types of materials/components, assigning a threshold between a healthy component and a damaged one might be challenging in that range of damages. As the length of the analyzed signal increases, the DIs become more sensitive to small damages as shown in Figure \ref{fig:Al_2-6_sig}d. However, it can be observed that the DIs do not follow the increase in notch size uniformly even for a damage-intersecting path like path 2-6. Exploring a damage-non-intersecting path (Figure \ref{fig:Al_6-3_sig}), the DIs fail to follow damage evolution to a greater extent, with fluctuations being observed as notch size increases. Such fluctuation in the DIs can be mistaken for a change in conditions surrounding the component, which would make the task of damage detection and threshold identification even more challenging. 

\begin{figure}[t!]
\centering
\begin{picture}(400,300)
\put(-3,160){\includegraphics[scale=0.5]{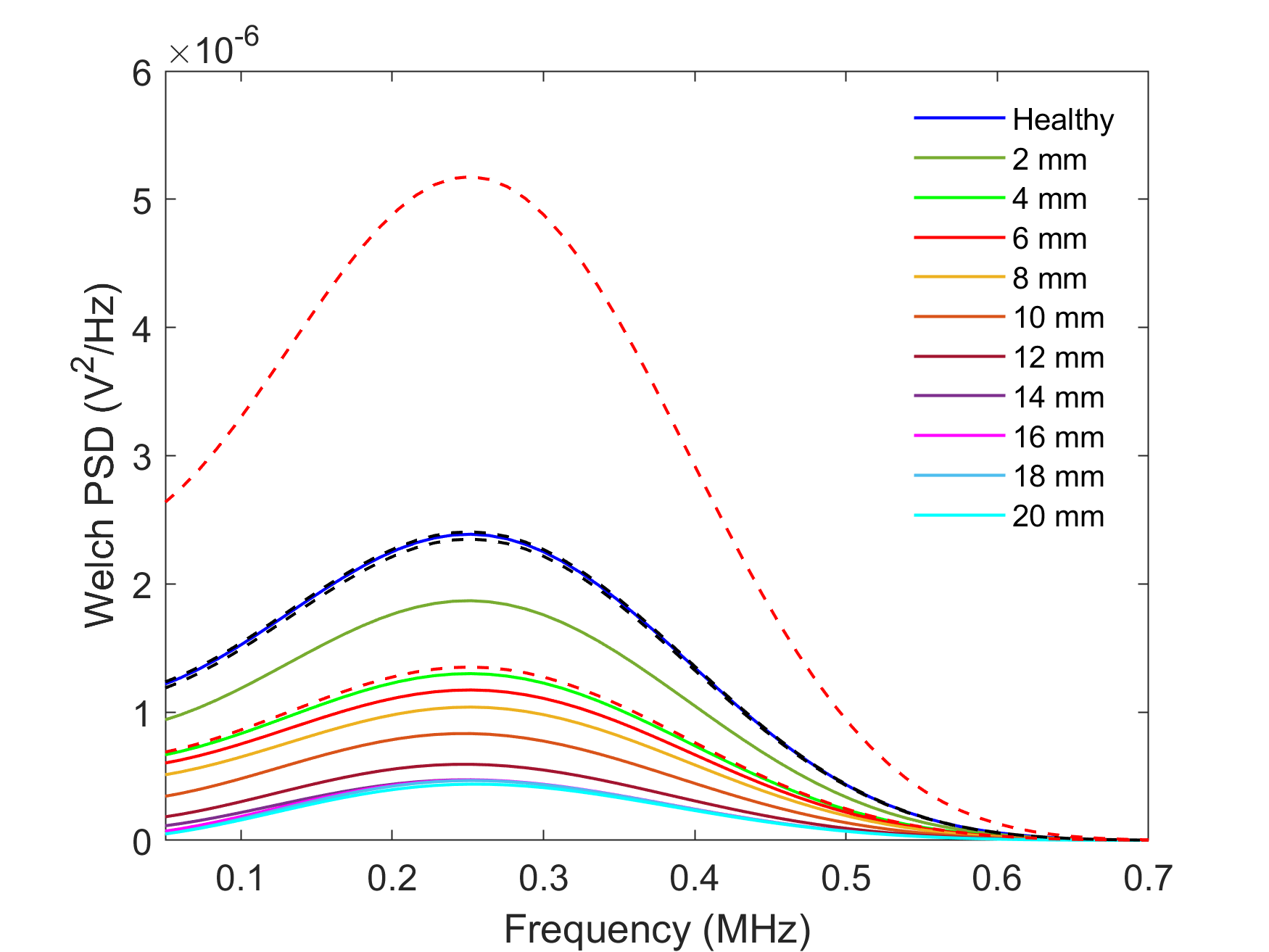}}
\put(192,160){\includegraphics[scale=0.5]{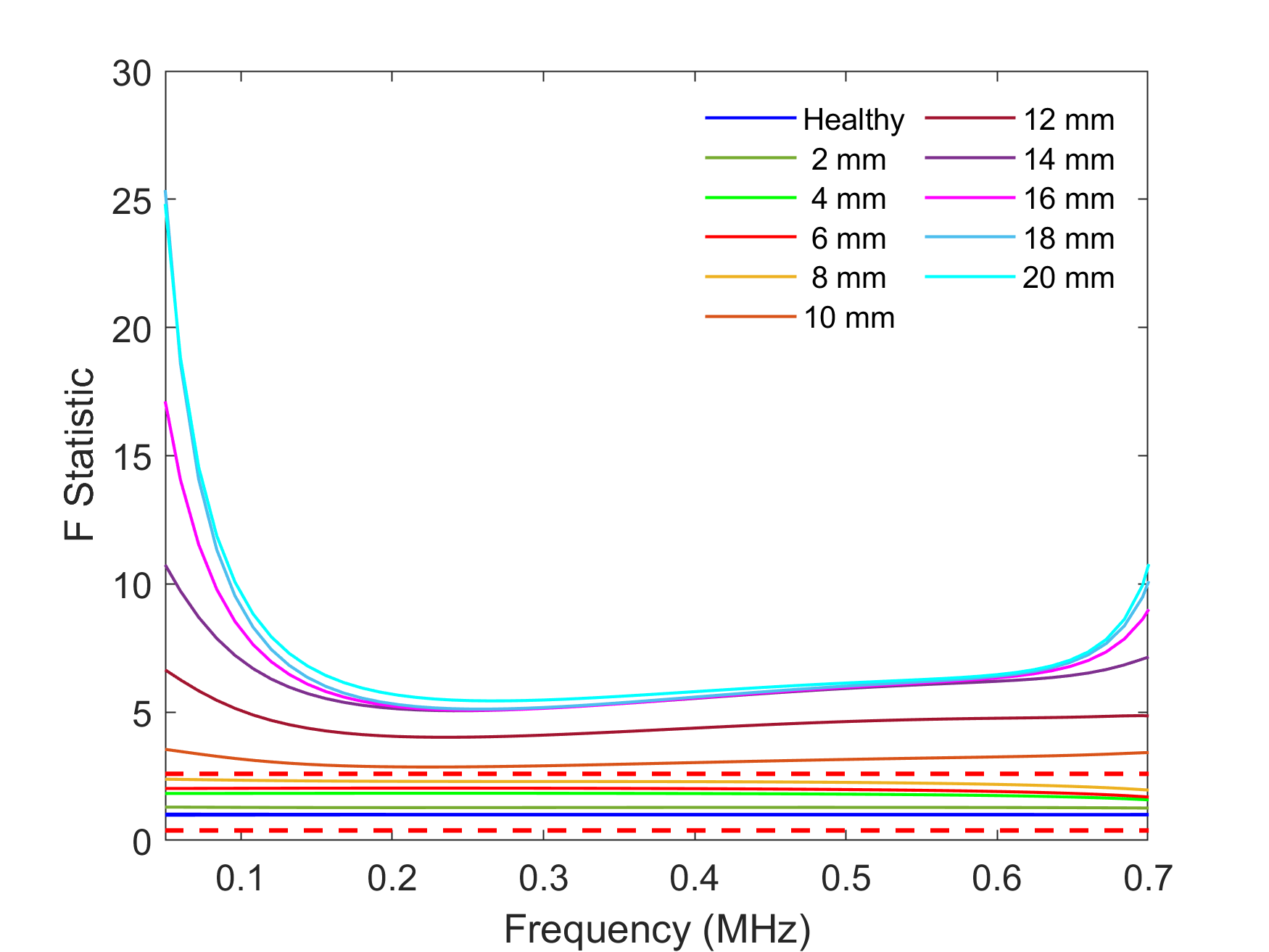}}
\put(-3,0){\includegraphics[scale=0.5]{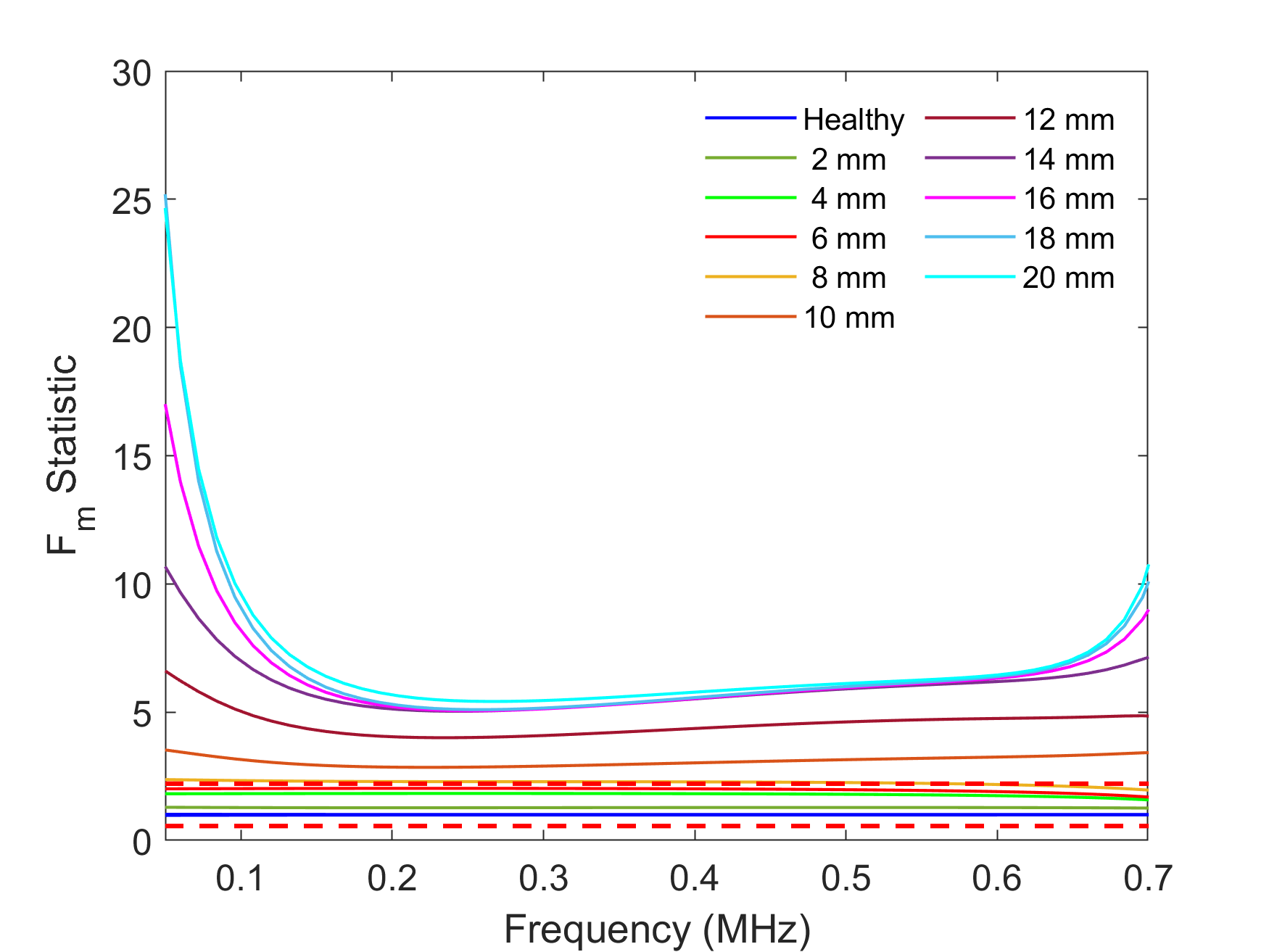}}
\put(192,0){\includegraphics[scale=0.5]{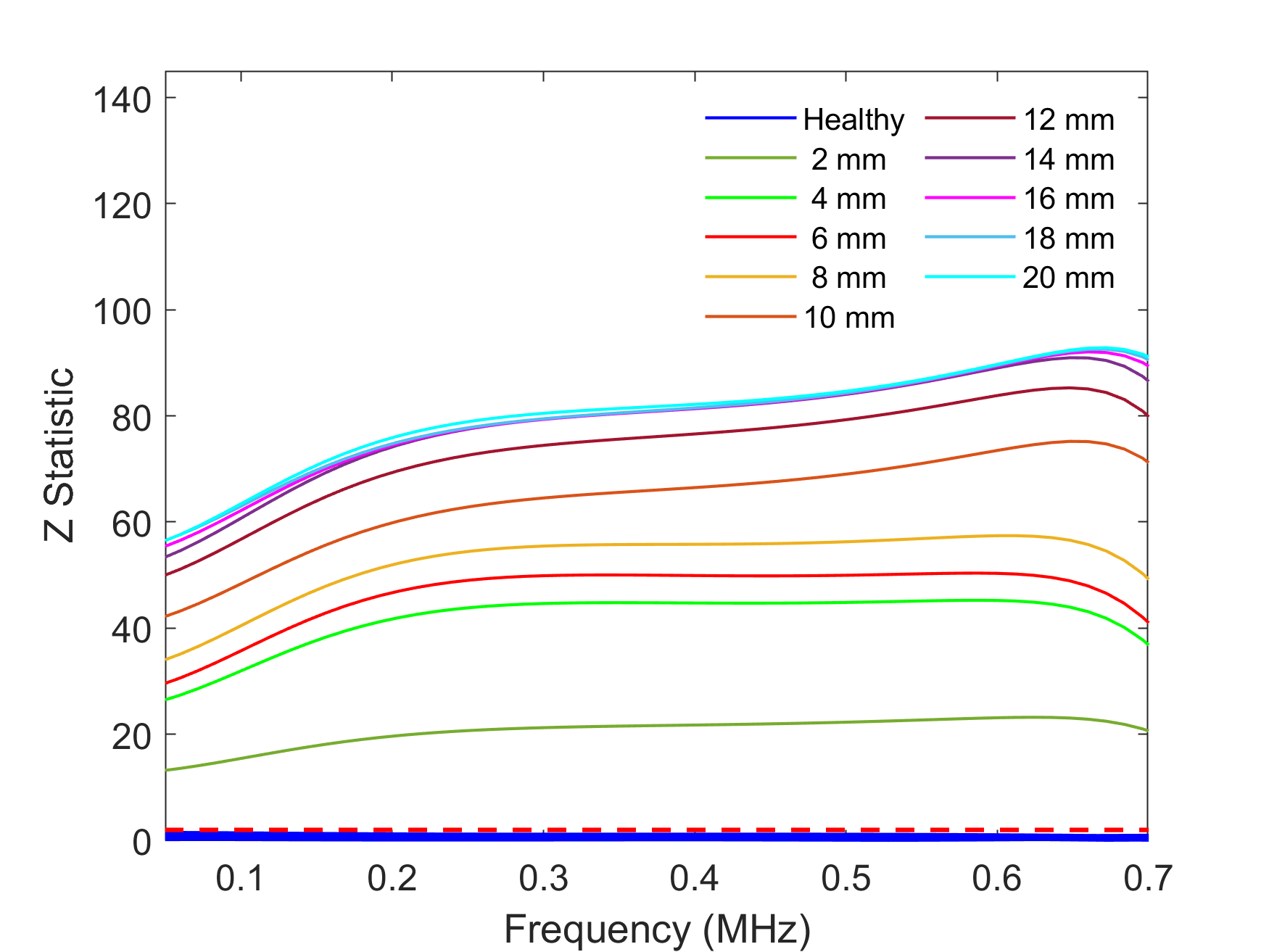}}
\put(30,295){\color{black} \large {\fontfamily{phv}\selectfont \textbf{a}}}
\put(225,295){\color{black} \large {\fontfamily{phv}\selectfont \textbf{b}}}
\put(30,135){\color{black} \large {\fontfamily{phv}\selectfont \textbf{c}}} 
\put(225,135){\color{black} \large {\fontfamily{phv}\selectfont \textbf{d}}}
\end{picture} 
\caption{Indicative results from applying the proposed NP-TS approach to the first arrival wave packet from path 2-6 in the Al coupon under different damage sizes: (a) Welch PSD -- the red and the black dashed lines indicate the theoretical (estimation uncertainty) and the experimental 95\% confidence bounds of the healthy PSD, respectively; (b) $F$ statistic; (c) $F_m$ statistic; (d) $Z$ statistic.}
\label{fig:Al_2-6_npts1}
\end{figure}

\begin{figure}[t!]
\centering
 \begin{picture}(400,300)
\put(-3,160){\includegraphics[scale=0.5]{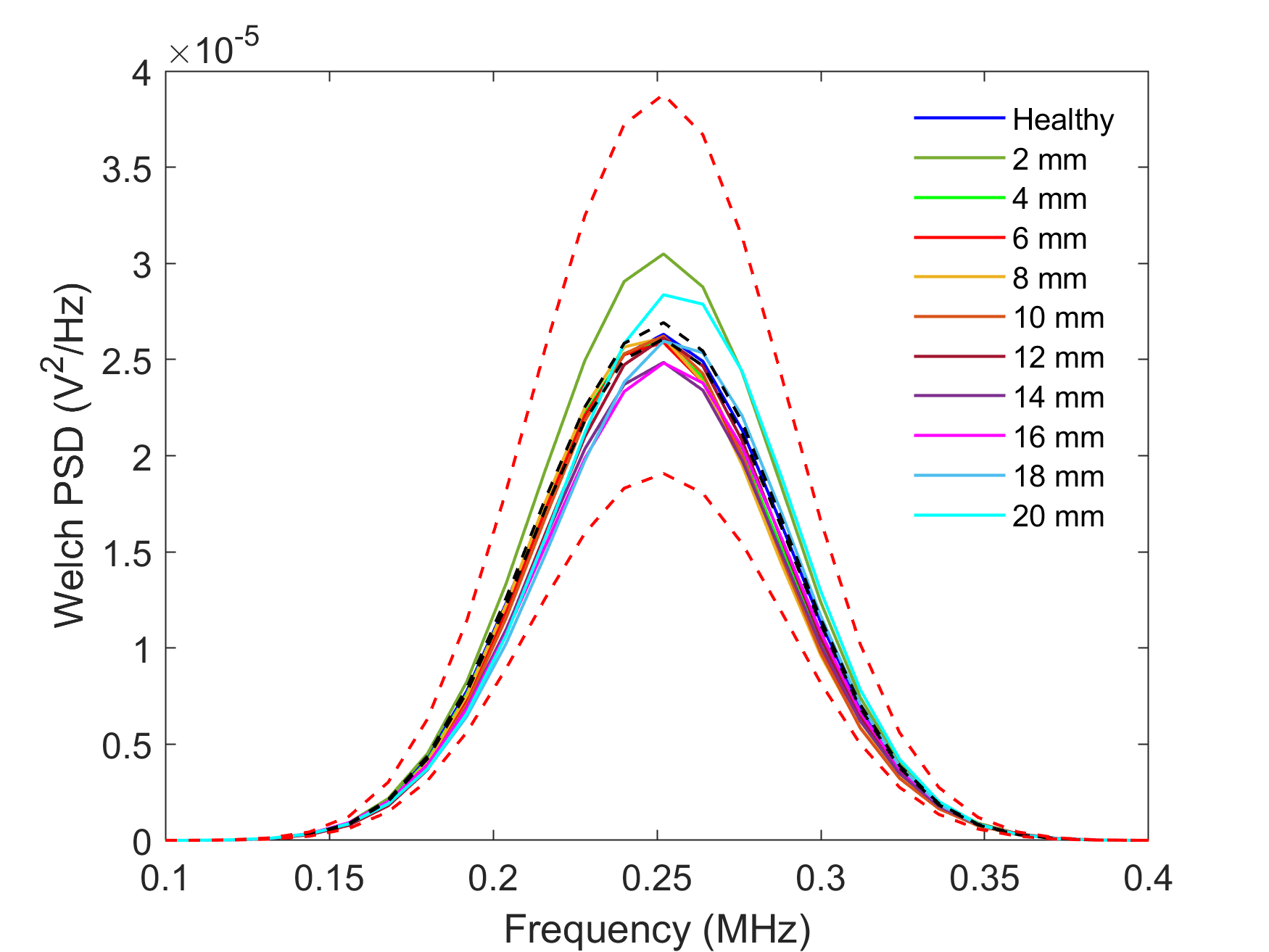}}
\put(192,160){\includegraphics[scale=0.5]{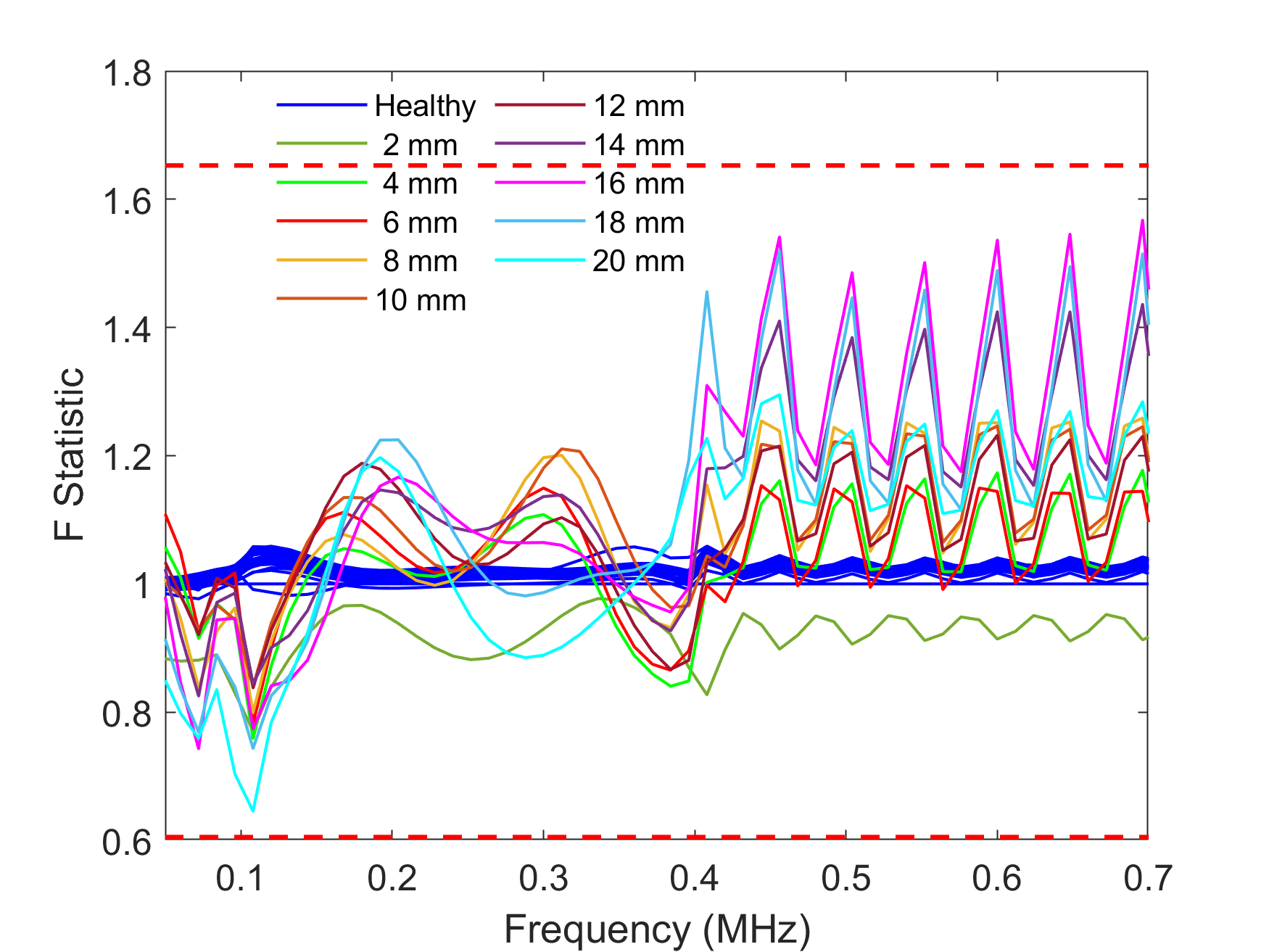}}
\put(-3,0){\includegraphics[scale=0.5]{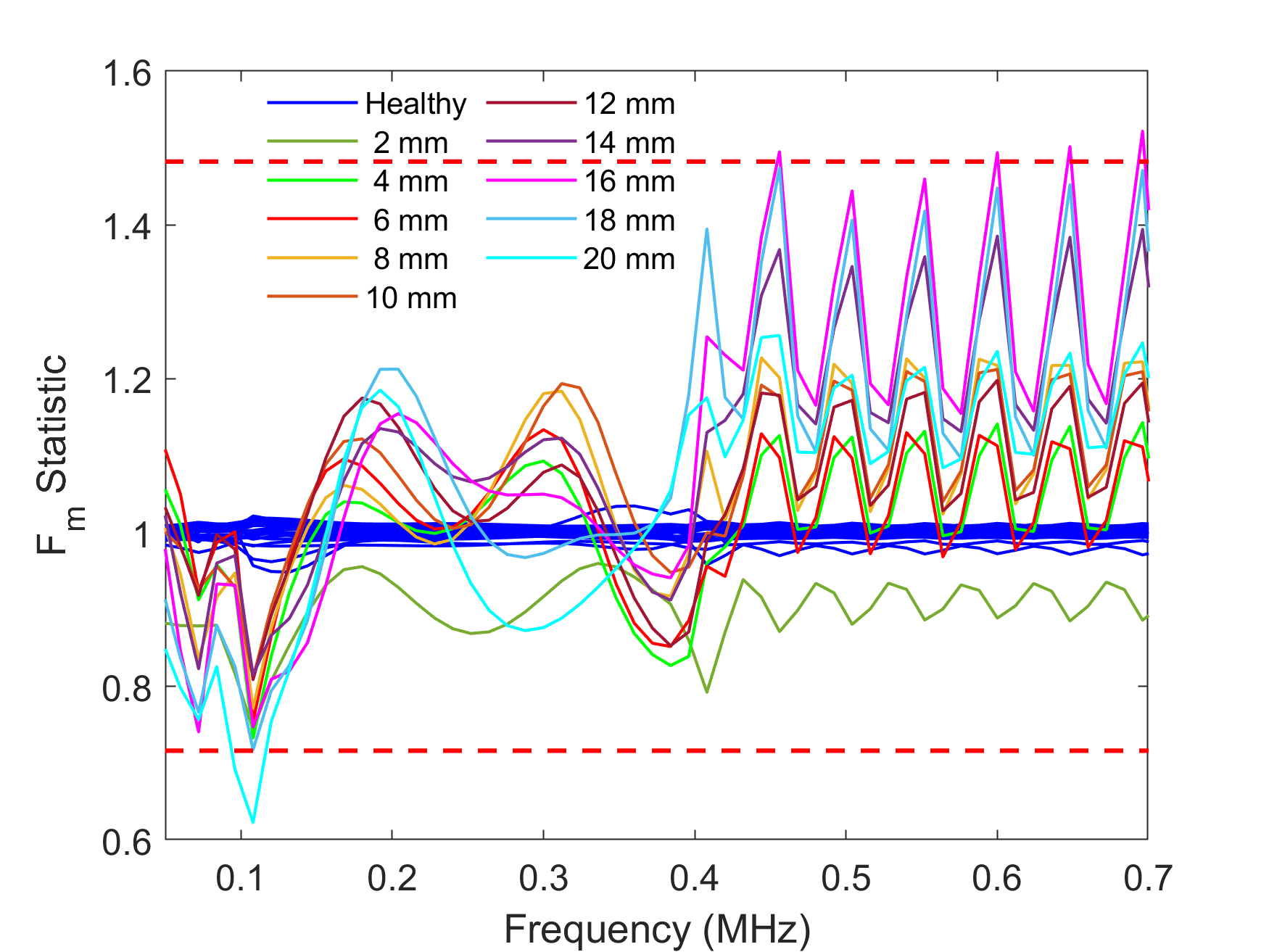}}
\put(192,0){\includegraphics[scale=0.5]{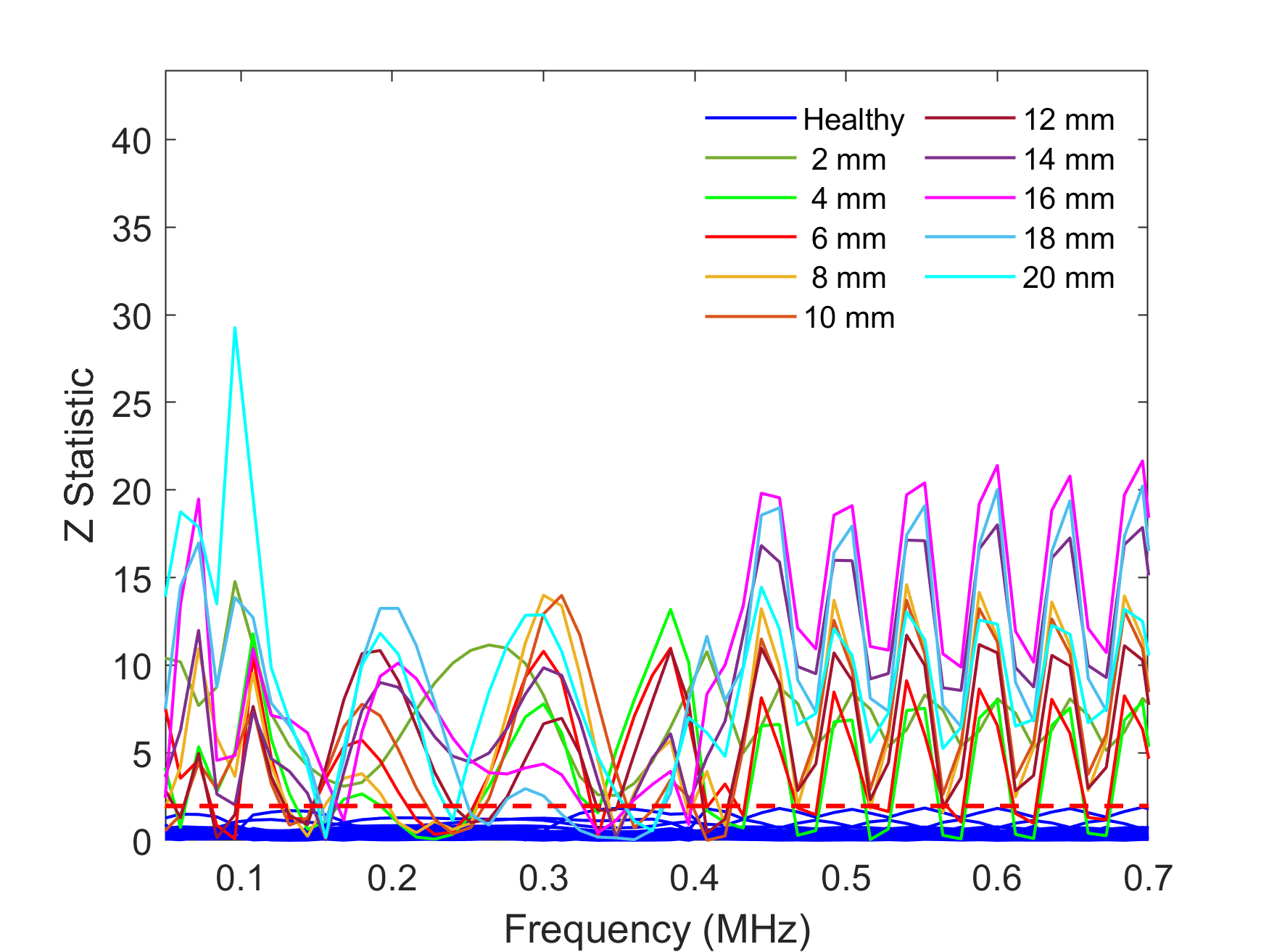}}
\put(30,295){\color{black} \large {\fontfamily{phv}\selectfont \textbf{a}}}
\put(225,295){\color{black} \large {\fontfamily{phv}\selectfont \textbf{b}}}
\put(30,135){\color{black} \large {\fontfamily{phv}\selectfont \textbf{c}}} 
\put(225,135){\color{black} \large {\fontfamily{phv}\selectfont \textbf{d}}}
\end{picture} 
\caption{Indicative results from applying the proposed NP-TS approach to the full signal from path 2-6 in the Al coupon under different damage sizes: (a) Welch PSD -- the red and the black dashed lines indicate the theoretical (estimation uncertainty) and the experimental 95\% confidence bounds of the healthy PSD, respectively; (b) $F$ statistic; (c) $F_m$ statistic; (d) $Z$ statistic.}
\label{fig:Al_2-6_npts2}
\end{figure}

Figure \ref{fig:Al_2-6_npts1} presents indicative results of applying the proposed framework to the response signal from path 2-6 in the Al coupon (see Table \ref{tab:Al_2-6_FA/MF} in the Appendix for summary results). Figure \ref{fig:Al_2-6_npts1}a shows the evolution of the Welch PSD of the signals as notch size increases, with the red and the black dashed lines indicating the theoretical (estimation uncertainty) and the experimental 95\% confidence intervals of the healthy PSD, respectively. The first thing to be observed in this figure is that, using the theoretical estimation confidence intervals, notch sizes more than 2 mm can be detected with 95\% confidence, and all damage sizes are detected when the experimental 95\% confidence levels are considered. Although the latter result is expected due to the nominally-controlled lab environment significantly inhibiting change in the Welch PSD over multiple healthy signals, the former observation shows the enhanced detection capability of the frequency-domain PSD compared to time-domain DIs for damages of this type in Al. Furthermore, in contrast to the DIs in Figure \ref{fig:Al_2-6_sig}, the PSDs evolve uniformly with damage, which hints on the enhanced damage quantification capability of these techniques. Thus, the Welch PSD emerges as a better metric when it comes to damage detection and quantification for the case at hand. Applying the SHT frameworks developed in Section \ref{sec:SHT} (Figure \ref{fig:Al_2-6_npts1} panels b-d), one can assess the difference between both approaches in a statistical way. As shown in Figure \ref{fig:Al_2-6_npts1}b, the $F$ statistic is capable of only detecting the last three damage cases (10-18 mm) with 95\% confidence. Although this performance is somewhat similar to that of the DIs, an advantage in the proposed approach is the extraction of confidence intervals directly from the SHM metric being used, without the need for user experience for defining damage thresholds. The $F_m$ statistic (Figure \ref{fig:Al_2-6_npts1}c) does a slightly better job by detecting the 8-mm damage with 95\% confidence, which is attributed to the inclusion of some experimental statistics into the definition of this metric. Examining the $Z$ statistic (Figure \ref{fig:Al_2-6_npts1}d), one can observe that all damage cases are detected with $95\%$ confidence. Furthermore, the effect of damage on the $Z$ statistic is again uniform, indicating the superior performance of this statistic in damage quantification compared to the conventional time-domain DI approach. Figure \ref{fig:Al_2-6_npts2}a shows indicative Welch PSD estimates for the first two wave packets (using a window size equal to the width of a single wave packet). As shown, although the peak amplitude of the PSD at the actuation frequency decreases, the detection performance remains the same as for a single wave packet. Exploring the three statistics proposed in this study (Figure \ref{fig:Al_2-6_npts2} panels b-d), it can also be observed that the detection performance stays the same, with all damages being detected by the $Z$ statistic with 95\% confidence.

\begin{table}[b!]
\centering
\caption{The different parameters used in estimating the Welch PSD for the Al coupon data sets.}\label{tab:Al_npts_info}
\renewcommand{\arraystretch}{1.2}
{\footnotesize
\begin{tabular}{|ll|}
\hline
Segment Length & $100$ \\
Window Type & Hamming \\
Frequency Resolution & $\Delta f=12$ kHz \\
Sampling Frequency & $24$ MHz \\
\hline
\multicolumn{2}{|c|}{Single Wave Packet} \\
\hline
Data Length & $N=500$ samples ($\sim 20$ $\mu s$) \\
No of non-overlapping segments & $9$ \\
\hline
\multicolumn{2}{|c|}{Full Signal Length} \\
\hline
Data Length & $N=8000$ samples ($\sim 330$ $\mu s$) \\
No of non-overlapping segments & $159$ \\
\hline
\hline
\end{tabular}} 
\end{table}

\begin{figure}[t!]
\centering
 \begin{picture}(400,300)
\put(-3,160){\includegraphics[scale=0.5]{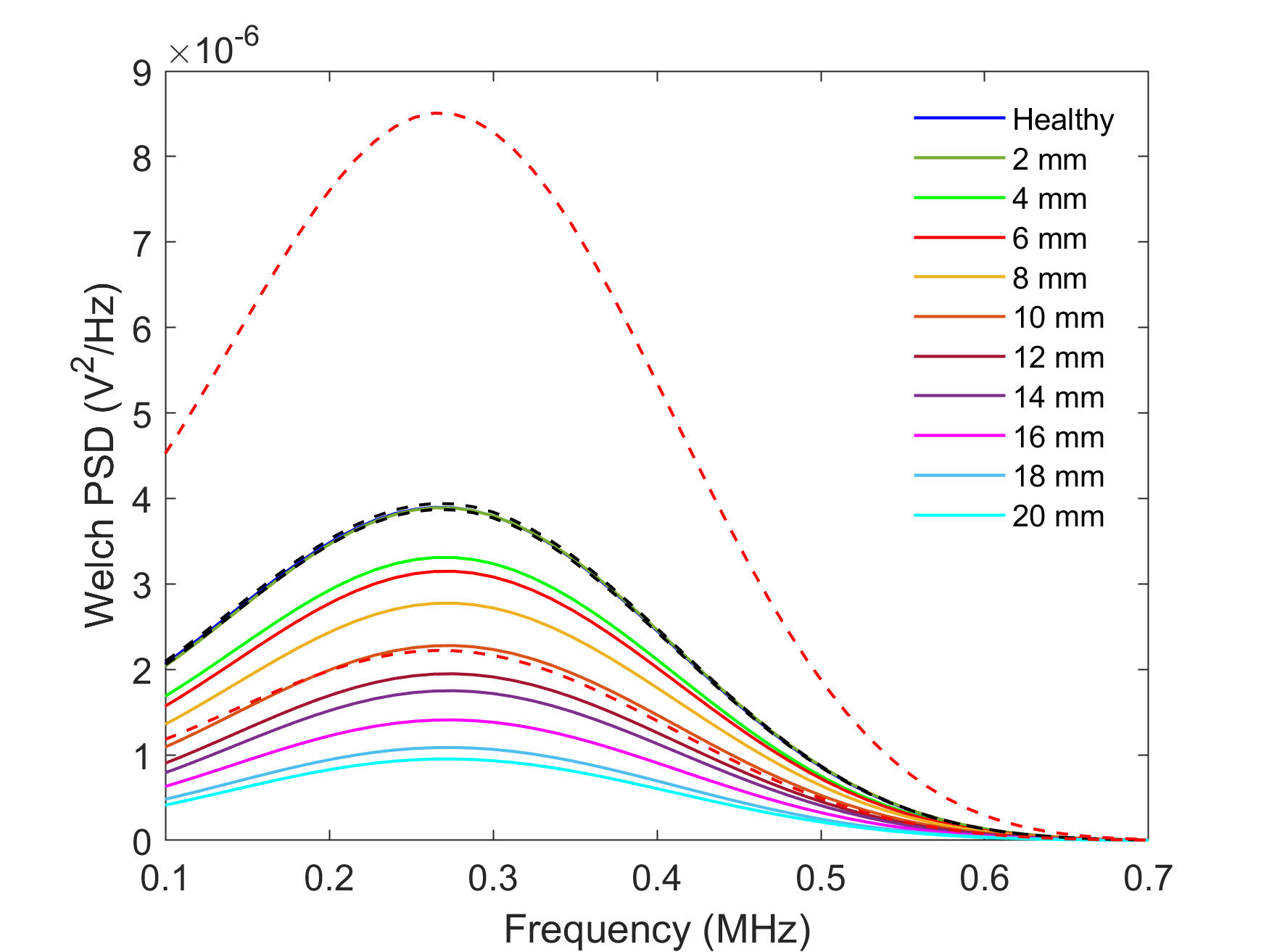}}
\put(192,160){\includegraphics[scale=0.5]{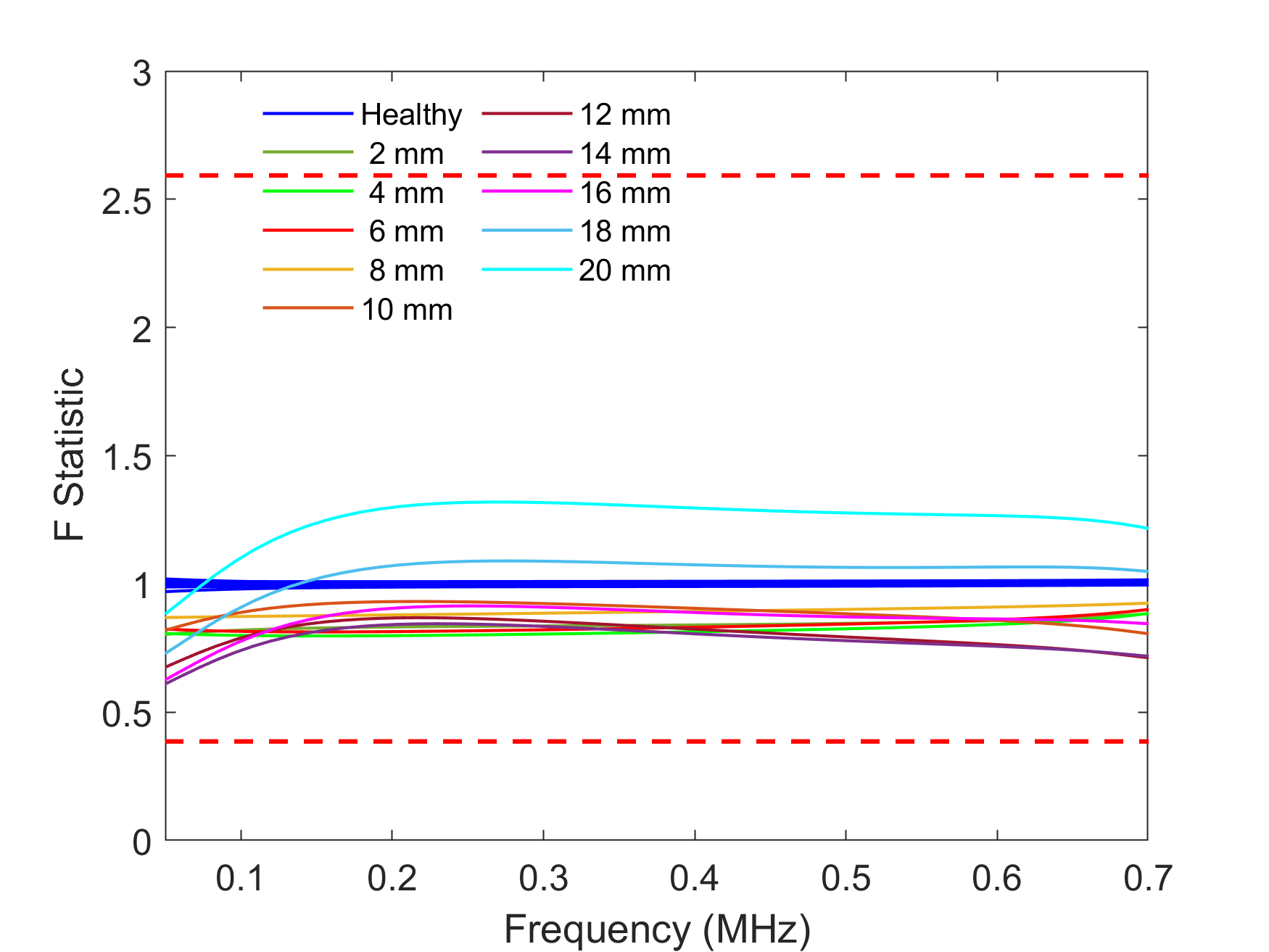}}
\put(-3,0){\includegraphics[scale=0.5]{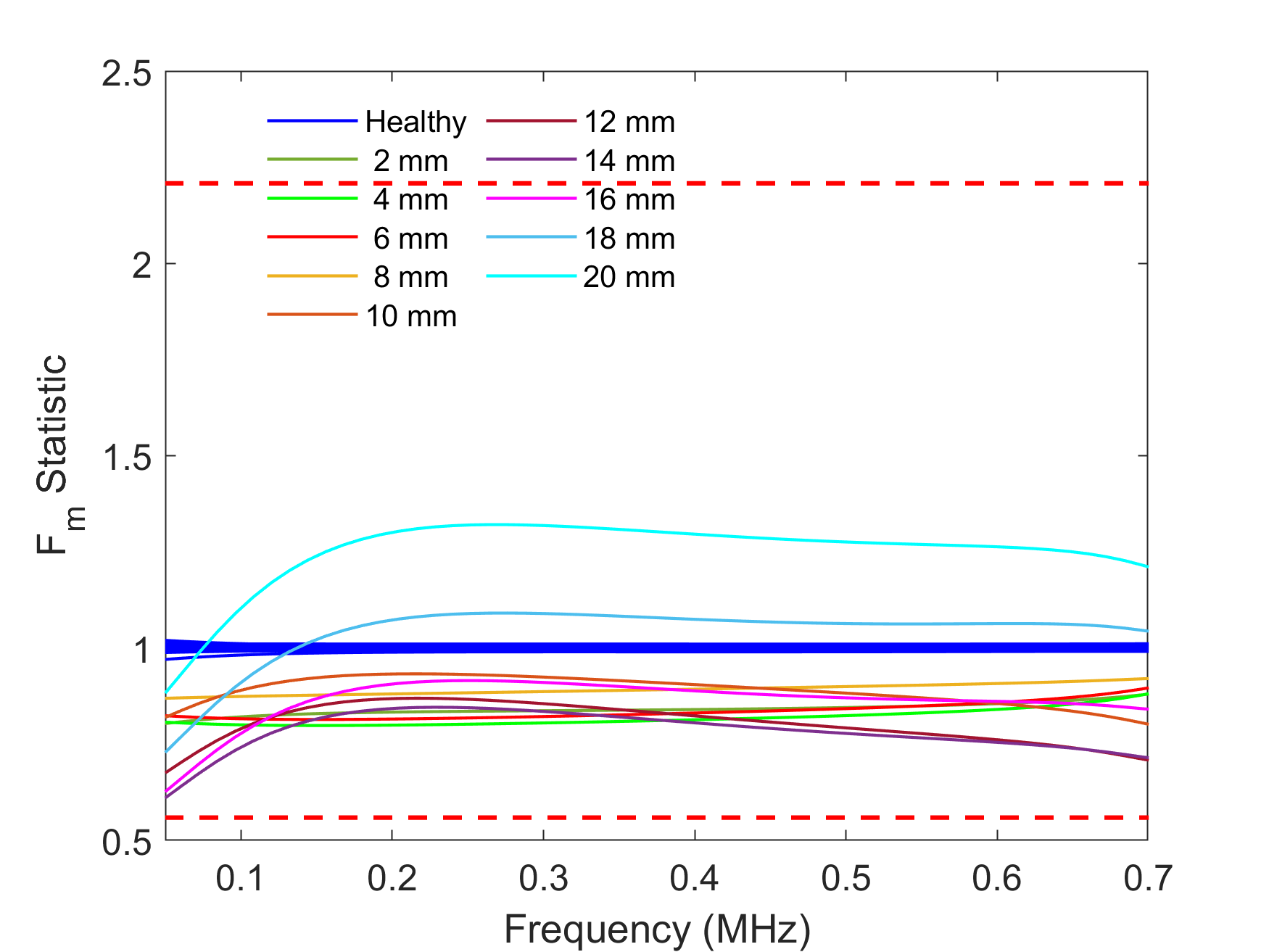}}
\put(192,0){\includegraphics[scale=0.5]{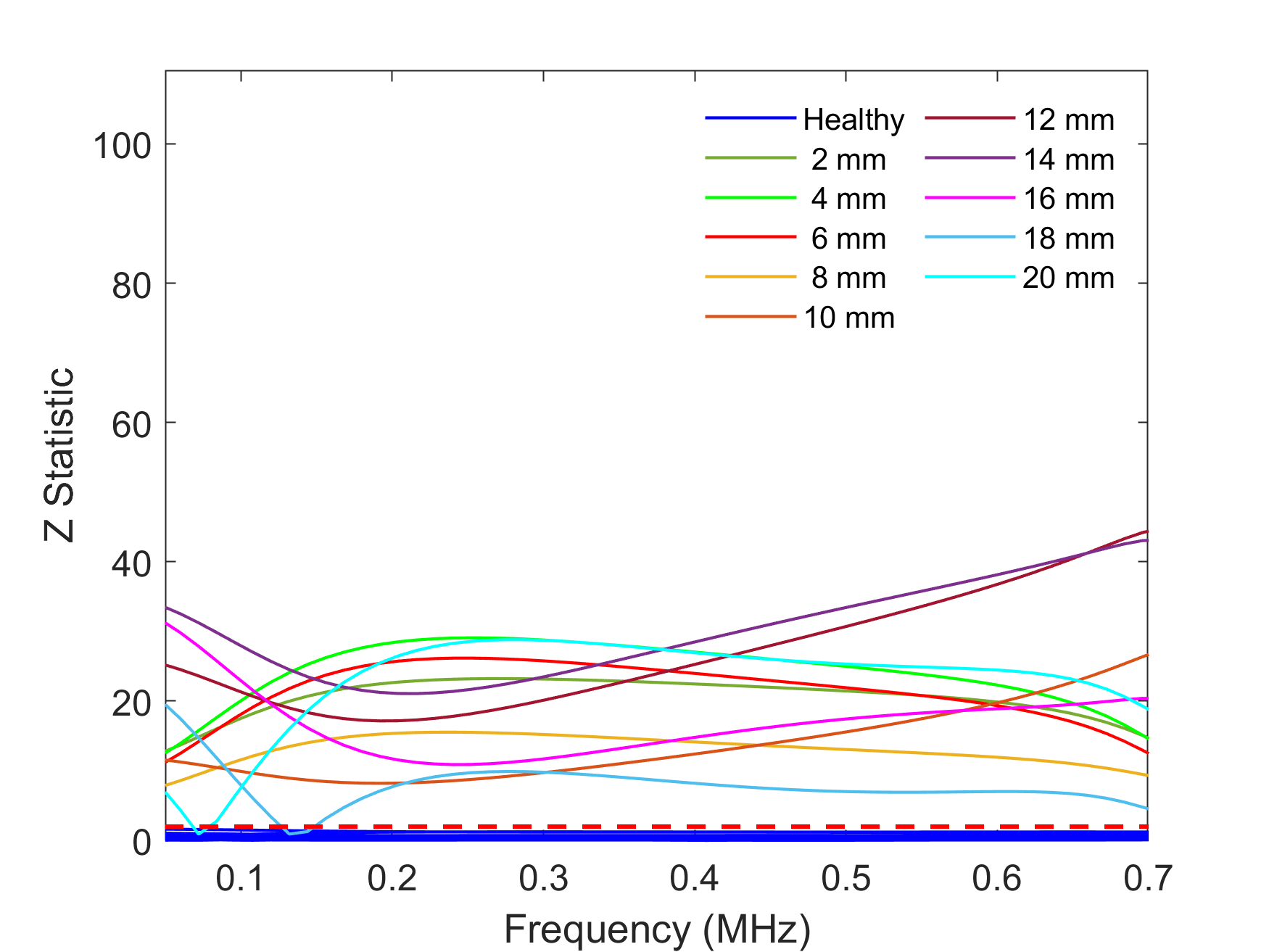}}
\put(30,295){\color{black} \large {\fontfamily{phv}\selectfont \textbf{a}}}
\put(225,295){\color{black} \large {\fontfamily{phv}\selectfont \textbf{b}}}
\put(30,135){\color{black} \large {\fontfamily{phv}\selectfont \textbf{c}}} 
\put(225,135){\color{black} \large {\fontfamily{phv}\selectfont \textbf{d}}}
\end{picture} 
\caption{Indicative results from applying the proposed NP-TS approach to the first-arrival wave packet from path 6-3 in the Al coupon under different damage sizes: (a) Welch PSD -- the red and the black dashed lines indicate the theoretical (estimation uncertainty) and the experimental  95\% confidence bounds of the healthy PSD, respectively; (b) $F$ statistic; (c) $F_m$ statistic; (d) $Z$ statistic.}
\label{fig:Al_6-3_npts1}
\end{figure}

\begin{figure}[t!]
\centering
 \begin{picture}(400,300)
\put(-3,160){\includegraphics[scale=0.5]{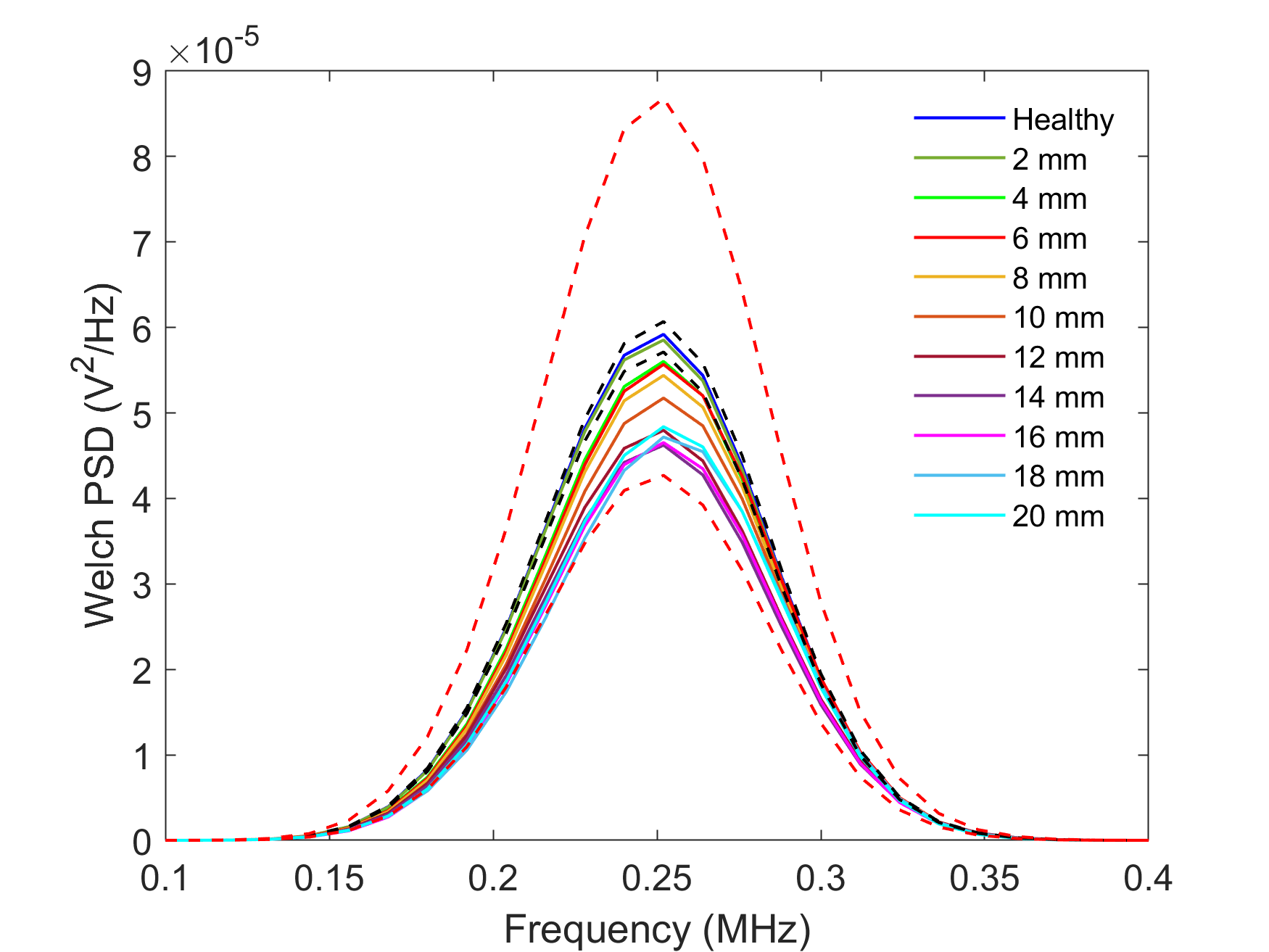}}
\put(192,160){\includegraphics[scale=0.5]{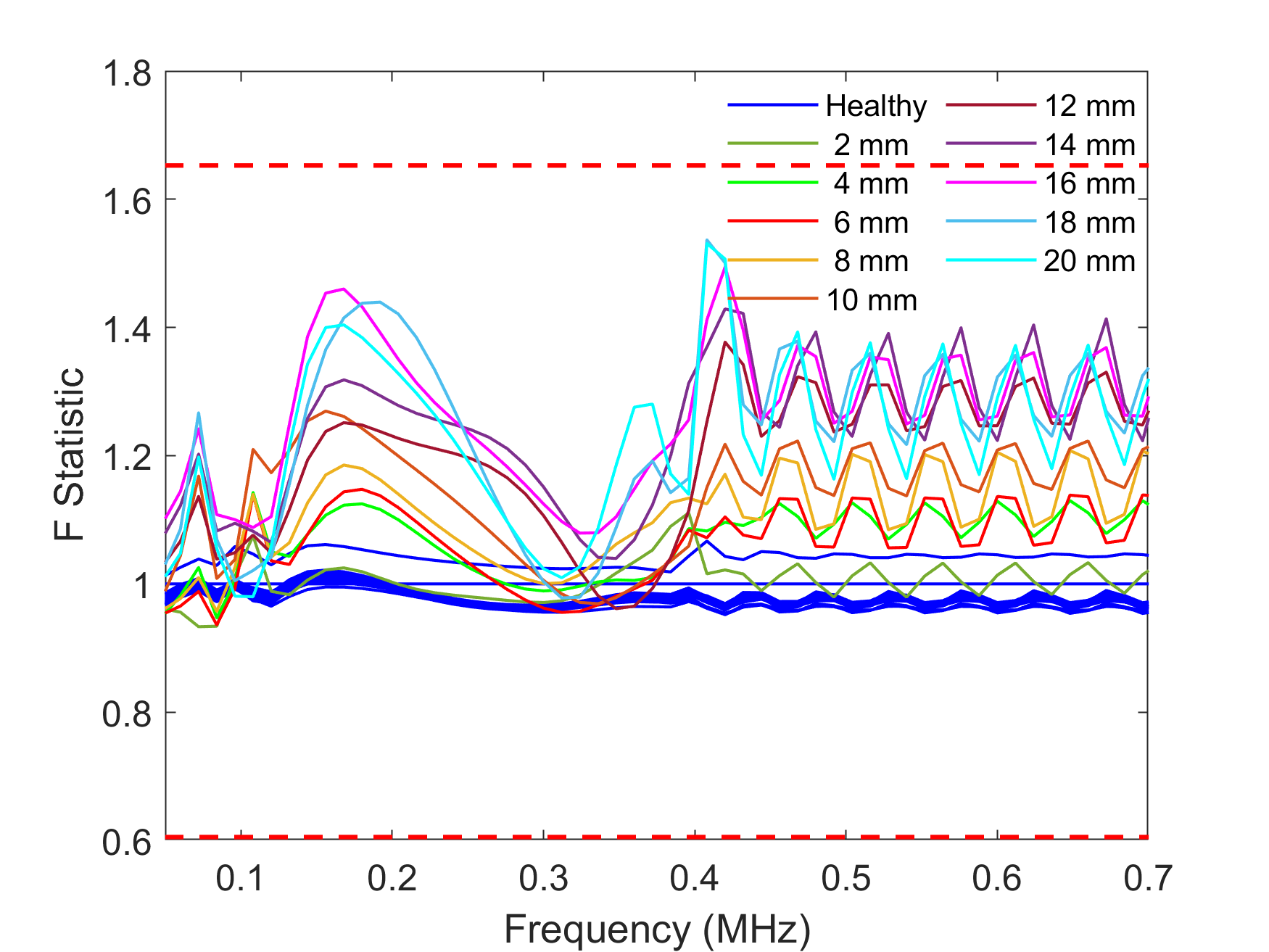}}
\put(-3,0){\includegraphics[scale=0.5]{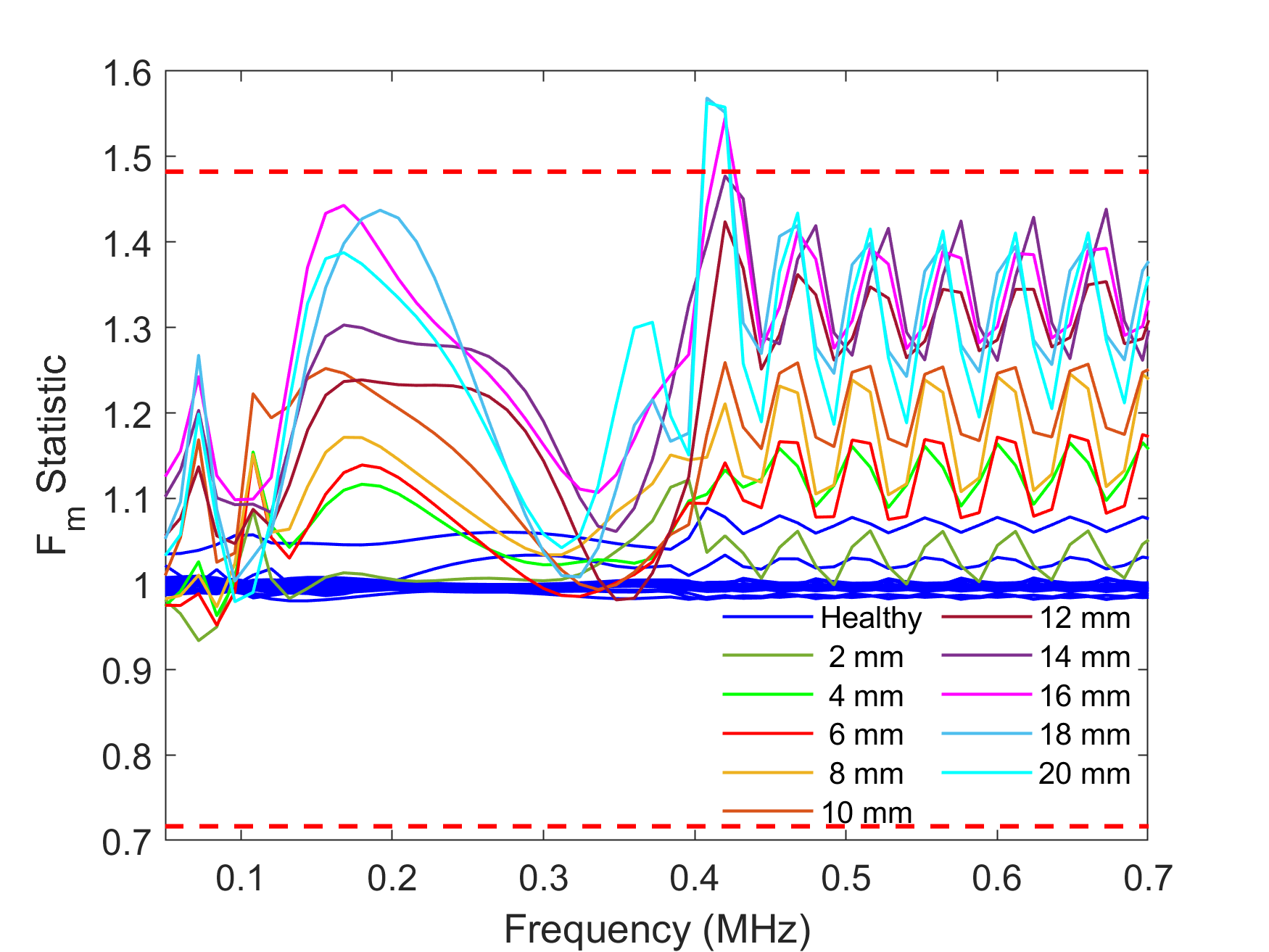}}
\put(192,0){\includegraphics[scale=0.5]{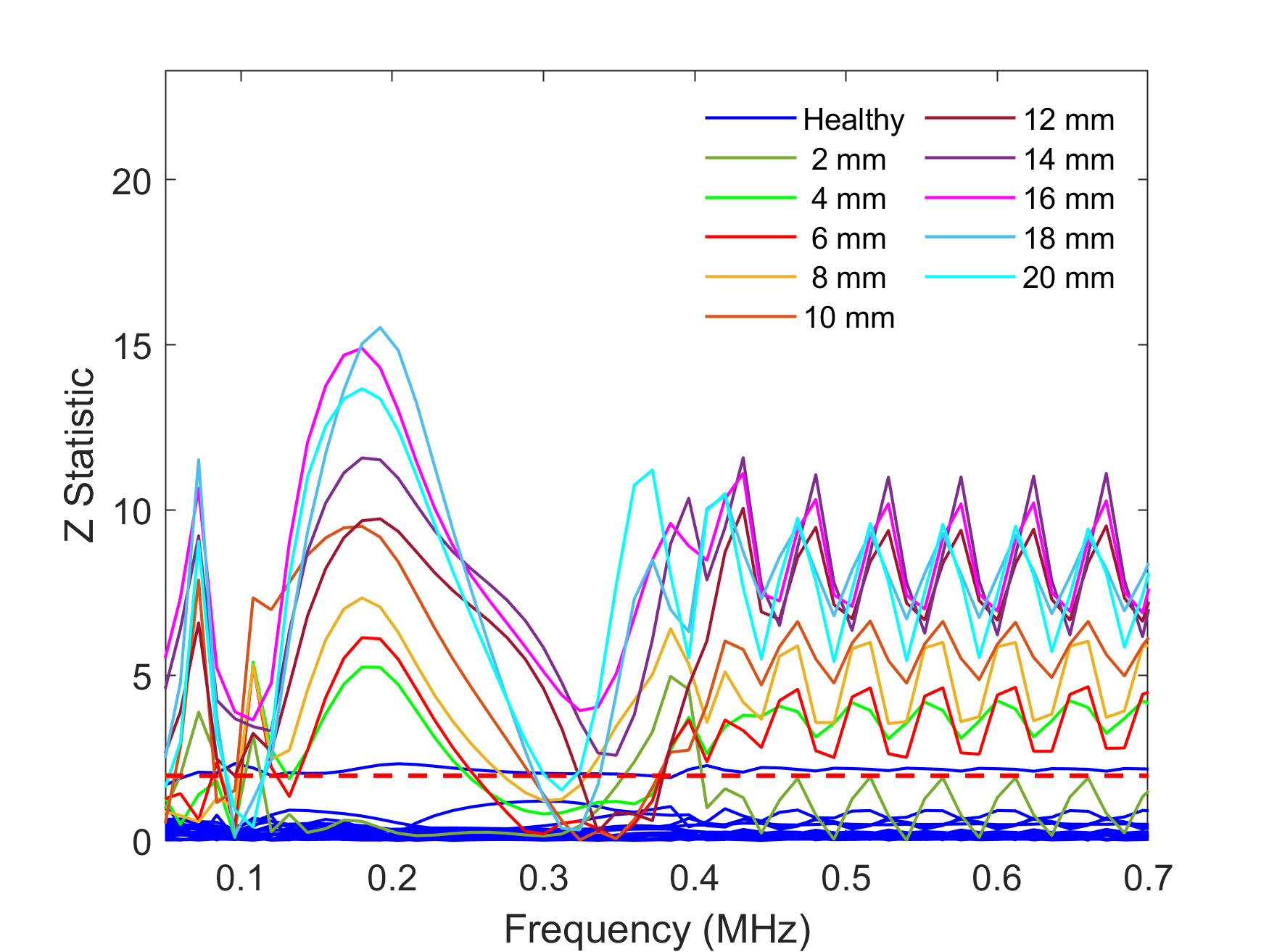}}
\put(30,295){\color{black} \large {\fontfamily{phv}\selectfont \textbf{a}}}
\put(225,295){\color{black} \large {\fontfamily{phv}\selectfont \textbf{b}}}
\put(30,135){\color{black} \large {\fontfamily{phv}\selectfont \textbf{c}}} 
\put(225,135){\color{black} \large {\fontfamily{phv}\selectfont \textbf{d}}}
\end{picture} 
\caption{Indicative results from applying the proposed NP-TS approach to the full signal from path 6-3 in the Al coupon under different damage sizes: (a) Welch PSD  -- the red and the black dashed lines indicate the theoretical (estimation uncertainty) and the experimental 95\% confidence bounds of the healthy PSD, respectively; (b) $F$ statistic; (c) $F_m$ statistic; (d) $Z$ statistic.}
\label{fig:Al_6-3_npts2}
\end{figure}

Moving on to the damage-non-intersecting path (path 6-3), Figure \ref{fig:Al_6-3_npts1} shows indicative results for a single wave packet (see Table \ref{tab:Al_6-3_FA/MF} in the Appendix for summary results). As shown in Figure \ref{fig:Al_6-3_npts1}a, using the PSD's theoretical estimation confidence intervals (red dashed lines), all damages are deemed healthy with 95\% confidence. This is attributed to the wide nature of the estimation uncertainty when the PSD  of a deterministic signal is being estimated, as is the case in this study. However, just like the DIs, all damages are detected with 95\% confidence when the experimental uncertainty is being considered. Being based on the theoretical confidence intervals, both the $F$ and $F_m$ statistics also show all damages as healthy with 95\% confidence (with the exception of the 14 mm case for the $F_m$ statistic), as shown in Figure \ref{fig:Al_6-3_npts1} panels b and c. On the other hand, the $Z$ statistic (Figure \ref{fig:Al_6-3_npts1}d) detects all damage cases with 95\% confidence because its formulation is based on the experimental uncertainty. The same trend can be observed when the first two wave packets are considered, as shown in Figure \ref{fig:Al_6-3_npts2}. A number of conclusions can be drawn from these observations when it comes to comparing the proposed statistics to the DIs. Firstly, for a notched Al coupon, the Welch PSD, the $F$ and $F_m$ statistics can be used as a preliminary step in differentiating between damage-intersecting and non-intersecting paths, in contrast to the DIs, which do not show a clear distinction. Secondly, the sensitivity of the $Z$ statistic seems to be the same as the DIs because of both being based on the experimental confidence intervals, an advantage that the $Z$ statistic has over the DI is the extraction of the confidence bounds based on the assumption of a normal distribution of the expectation of the signals' PSDs. Thus, the extracted damage detection thresholds emerge from the formulation of the SHM metric itself, and don't require prior experience with such materials and damages, or physics-based modelling. In contrast, the DIs require complex approaches in order to set accurate thresholds, and do not entail any theoretical distribution on the signals, from which thresholds can emerge naturally.

\subsection{Test Case II: Damage Detection in CFRP Plate}

\subsubsection{Test Setup, Damage Types and Data Acquisition}

\begin{figure}[t!]
\centering
\includegraphics[scale=0.31]{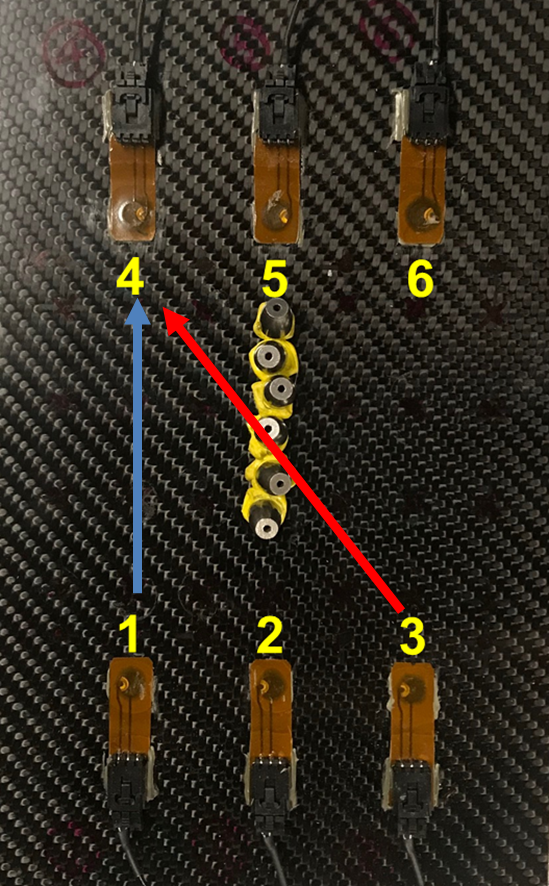}
\caption{The CFRP coupon used in this study shown here with 6 weights as simulated damage (the largest damage size of this test case). The arrows indicate the paths used in the analysis presented herein.}  
\label{fig:CFRP_coupon}
\end{figure}

The second coupon used in this study was a CFRP coupon (ACP Composites,) having the same dimensions as the Al coupon, with multiple $0/90$ unidirectional CF plies. This  coupon was also fitted with 6 single-PZT SMART Layers type PZT-5A (Acellent Technologies, Inc) as shown in Figure \ref{fig:CFRP_coupon}. Damage was simulated by attaching 1-6 three-gm weights to the surface of the coupon next to each other using tacky tape. The same actuation and data acquisition properties were used for this coupon as with the Al one. Also, similar to the case of the Al coupon, the actuation center frequency of 250 kHz was chosen for the analysis presented herein. Tables \ref{tab:CFRP_exp_info_1} and \ref{tab:CFRP_exp_info_2} summarize the experimental details for this coupon.

\begin{table}[b!]
\centering
\caption{Details of the first experimental data set for the CFRP coupon.}\label{tab:CFRP_exp_info_1}
\renewcommand{\arraystretch}{1.2}
{\footnotesize
\begin{tabular}{|ccc|} 
\hline 
Structural State & Number of Data Sets & Total Added Weight$^\dagger$ (g) \\
\hline
Healthy & 20$^{\dagger\dagger}$ & 0 \\
1 Steel weight & 1 &  $3$ \\
2 Steel weights & 1 &  $6$ \\
3 Steel weights & 1 &  $9$ \\
4 Steel weights & 1 &  $12$ \\
5 Steel weights & 1 &  $15$ \\
6 Steel weights & 1 &  $18$ \\
\hline
\multicolumn{3}{l}{{\bf Sampling Frequency:} $f_s=24$ MHz. Center frequency range: [$50:50:750$] kHz.} \\
\multicolumn{3}{l}{{\bf Number of samples} per data set $N= 8000$.}\\
\multicolumn{3}{l}{ $^\dagger$Weight of tacky tape not considered here.}\\
\multicolumn{3}{l}{ $^{\dagger\dagger}$M=20 in equation (\ref{eq:mean}).}
\end{tabular}} 
\end{table}

\begin{table}[b!]
\centering
\caption{Details of the second experimental data set for the CFRP coupon.}\label{tab:CFRP_exp_info_2}
\renewcommand{\arraystretch}{1.2}
{\footnotesize
\begin{tabular}{|ccc|} 
\hline 
Structural State & Number of Data Sets & Total Added Weight$^\dagger$ (g) \\
\hline
Healthy & 20$^{\dagger\dagger}$ & 0 \\
1 Steel weight & 20 &  $3$ \\
2 Steel weights & 20 &  $6$ \\
3 Steel weights & 20 &  $9$ \\
4 Steel weights & 20 &  $12$ \\
5 Steel weights & 20 &  $15$ \\
6 Steel weights & 20 &  $18$ \\
\hline
\multicolumn{3}{l}{{\bf Sampling Frequency:} $f_s=24$ MHz. Center frequency range: [$50:50:750$] kHz.} \\
\multicolumn{3}{l}{{\bf Number of samples} per data set $N= 8000$.}\\
\multicolumn{3}{l}{ $^\dagger$Weight of tacky tape not considered here.}\\
\multicolumn{3}{l}{ $^{\dagger\dagger}$M=20 in equation (\ref{eq:mean}).}
\end{tabular}} 
\end{table}

\subsubsection{Damage Detection Results}\label{sec:CFRP_res}

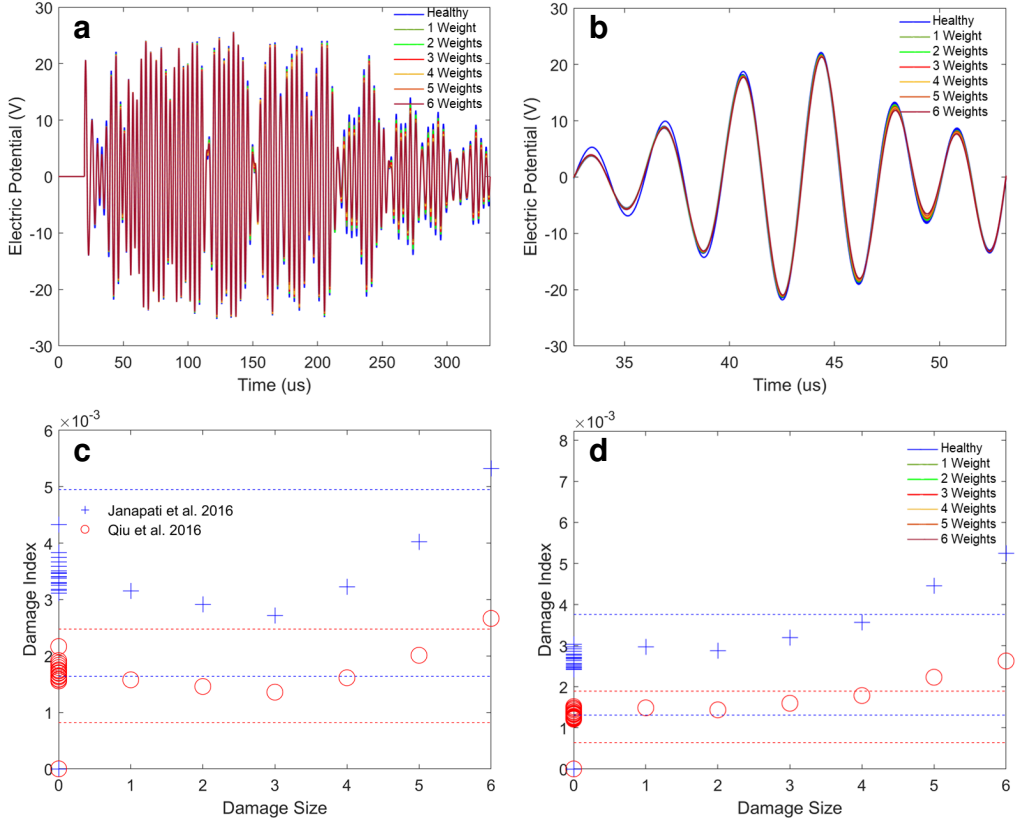
\begin{figure}[t!]
\centering
 \begin{picture}(400,300)
\put(-3,160){\includegraphics[scale=0.5]{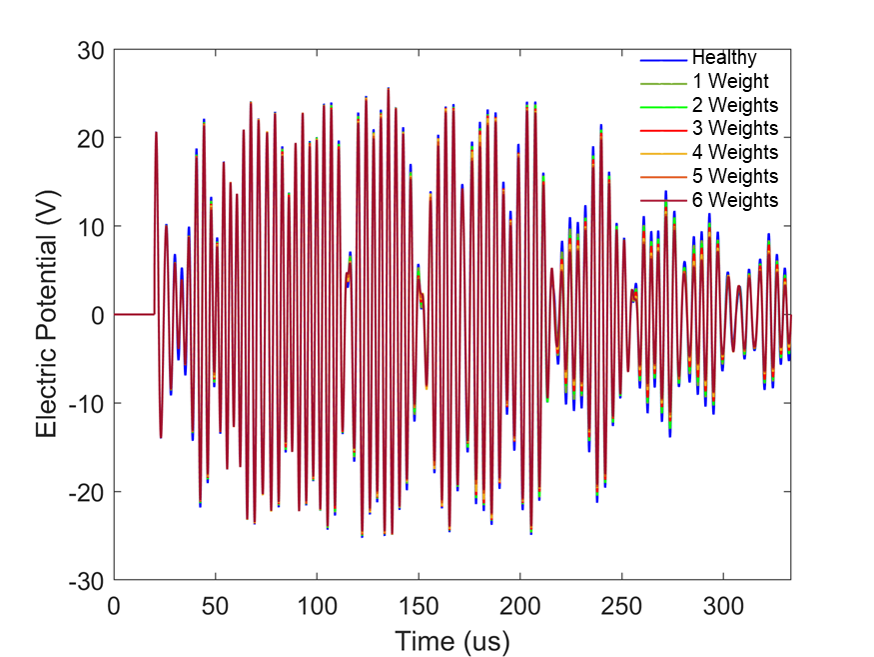}}
\put(192,160){\includegraphics[scale=0.5]{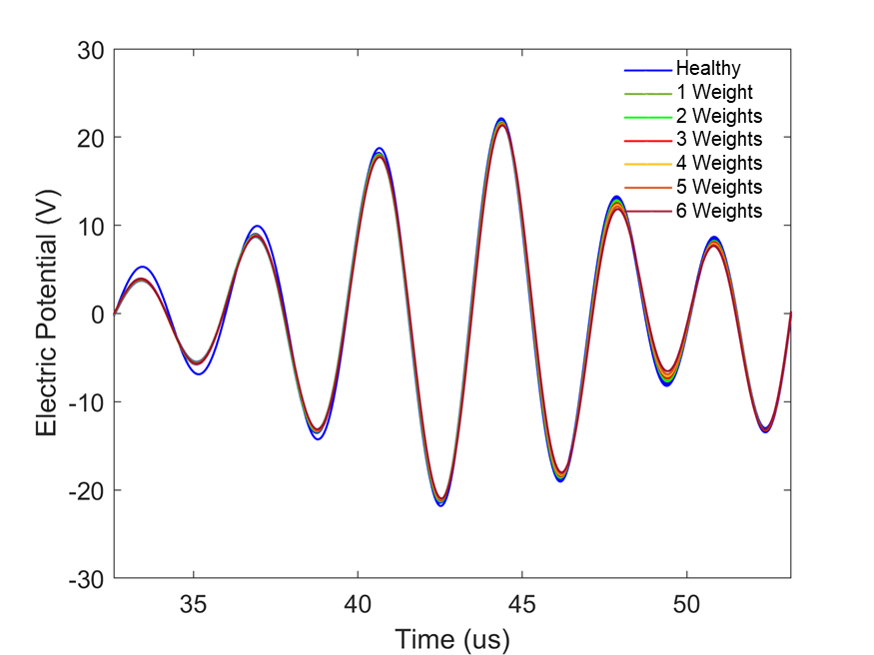}}
\put(-3,0){\includegraphics[scale=0.5]{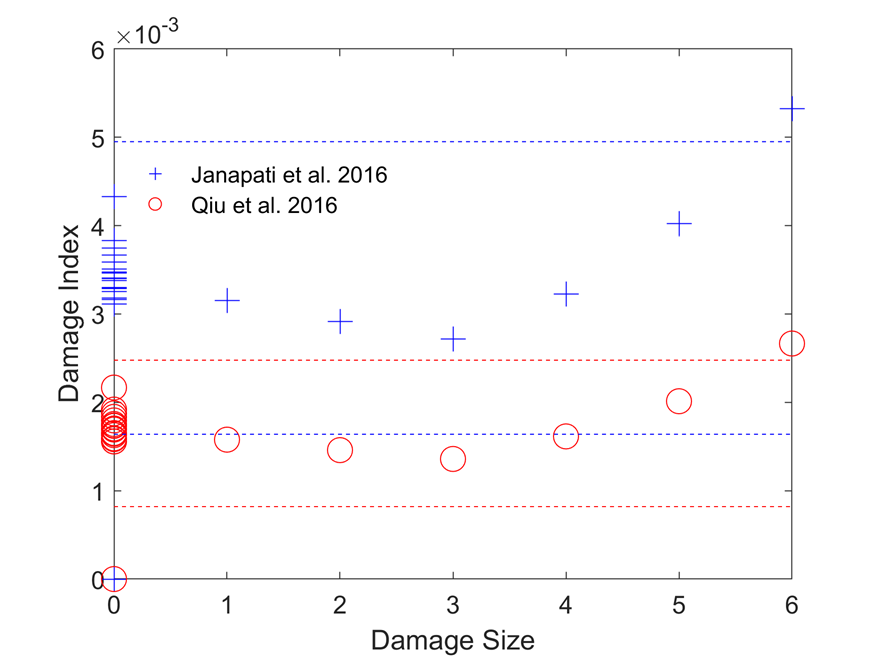}}
\put(192,0){\includegraphics[scale=0.5]{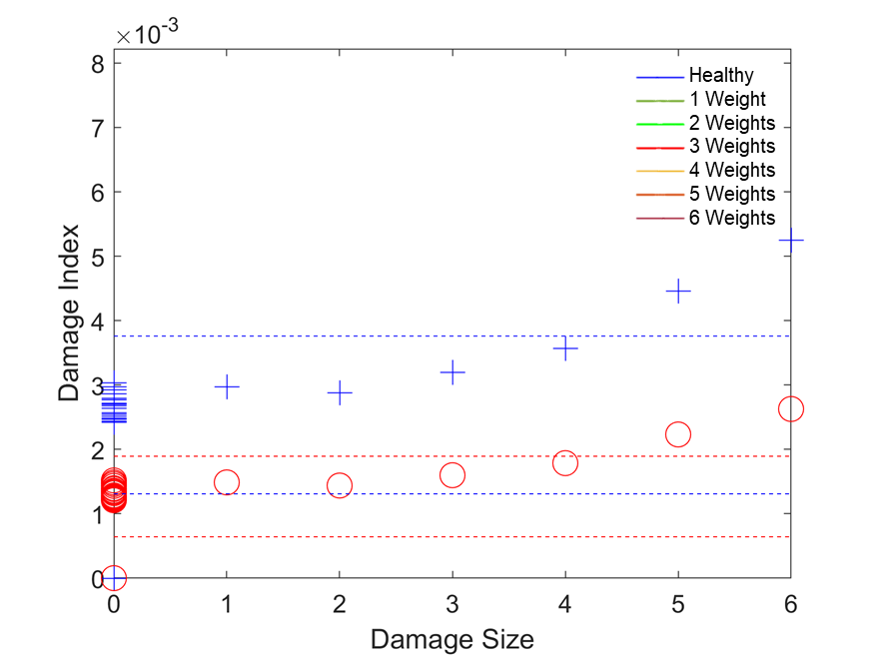}}
\put(30,295){\color{black} \large {\fontfamily{phv}\selectfont \textbf{a}}}
\put(225,295){\color{black} \large {\fontfamily{phv}\selectfont \textbf{b}}}
\put(30,135){\color{black} \large {\fontfamily{phv}\selectfont \textbf{c}}} 
\put(225,135){\color{black} \large {\fontfamily{phv}\selectfont \textbf{d}}}
\end{picture} 
\caption{Indicative signal from path 3-4 in the CFRP coupon under different simulated damage sizes (number of attached weights): (a) full signal; (b) single wave packet; (c) single-wave packet DIs -- the dashed lines designate the upper and lower $95\%$ confidence bounds for the Janapati \etal (blue) and Qiu \etal (red) DIs; (d) DIs for double the wave packet length.}  
\label{fig:CFRP_3-4_sig}
\end{figure}

\begin{figure}[t!]
\centering
 \begin{picture}(400,300)
\put(-3,160){\includegraphics[scale=0.5]{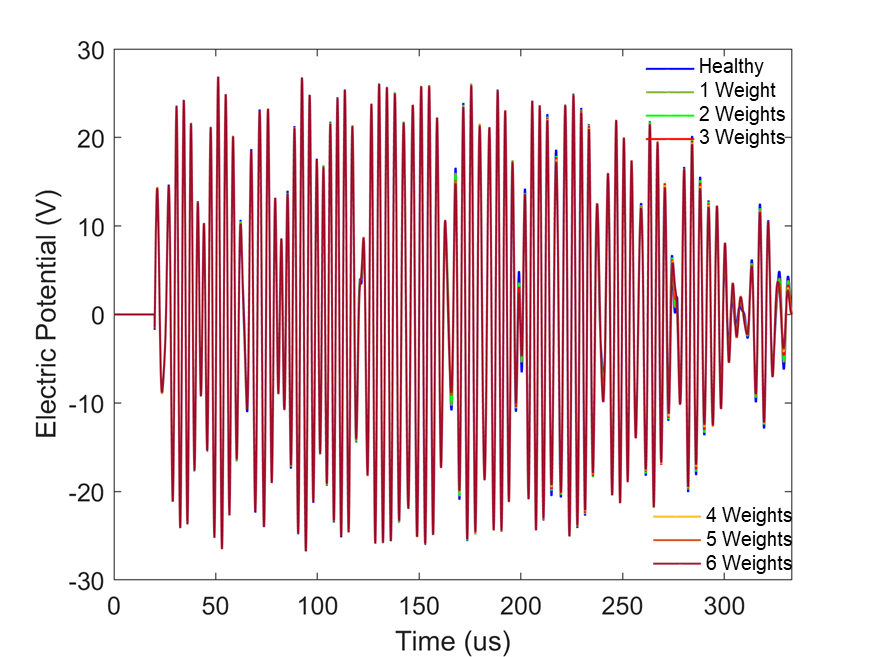}}
\put(192,160){\includegraphics[scale=0.5]{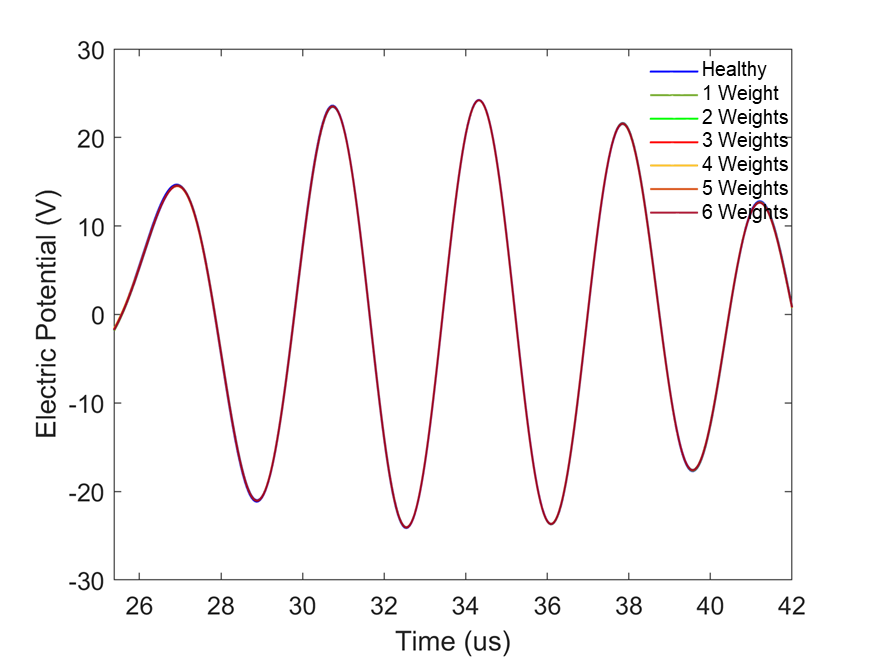}}
\put(-3,0){\includegraphics[scale=0.5]{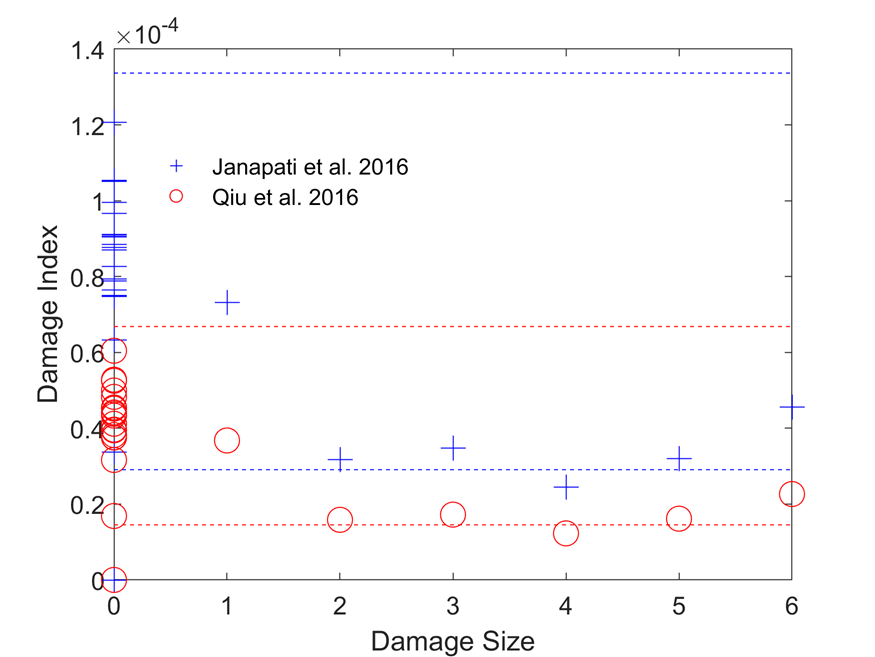}}
\put(192,0){\includegraphics[scale=0.5]{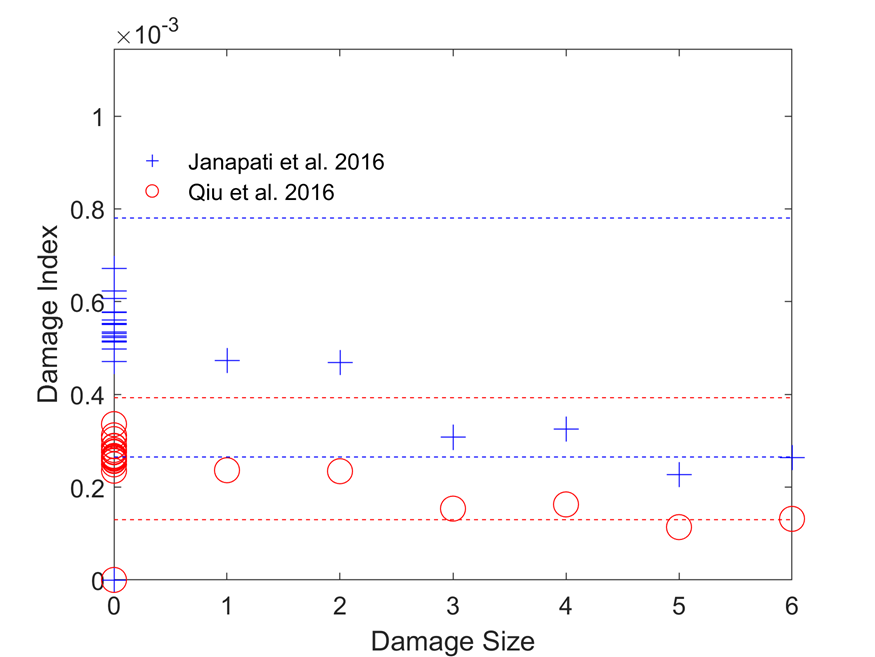}}
\put(30,295){\color{black} \large {\fontfamily{phv}\selectfont \textbf{a}}}
\put(225,295){\color{black} \large {\fontfamily{phv}\selectfont \textbf{b}}}
\put(30,135){\color{black} \large {\fontfamily{phv}\selectfont \textbf{c}}} 
\put(225,135){\color{black} \large {\fontfamily{phv}\selectfont \textbf{d}}}
\end{picture} 
\caption{Indicative signal from path 1-4 in the CFRP coupon under different simulated damage sizes (number of attached weights): (a) full signal; (b) single wave packet; (c) single wave packet DIs -- the dashed lines designate the healthy upper and lower $95\%$ confidence bounds for the Janapati \etal (blue) and Qiu \etal (red) DIs; (d)  DIs for double the wave packet length.}  
\label{fig:CFRP_1-4_sig}
\end{figure}
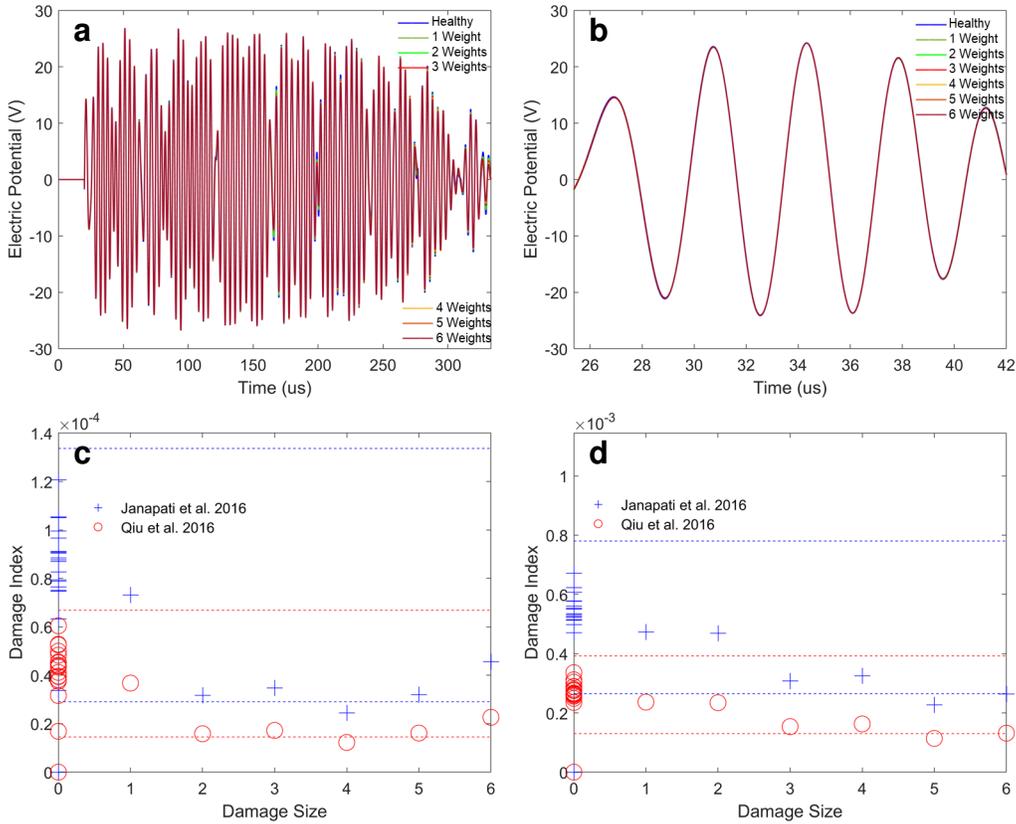

Figure \ref{fig:CFRP_3-4_sig} panels a and b present, respectively,  the signals and the first discernible wave packet, obtained at sensor 4 when sensor 3 was actuated under different damage sizes, where damage size here indicates the number of taped weights. Figure \ref{fig:CFRP_3-4_sig} panels c and d show the DIs for a single, and double wave packet lengths, respectively. As shown, because a CFRP coupon exhibits more non-linearity compared to an Al coupon, the single-wave packet DIs completely fail to follow damage evolution and can further only detect the last damage case (6 weights) within the 95\% experimental healthy confidence intervals for both DI formulations as shown in Figure \ref{fig:CFRP_3-4_sig}c. This performance is slightly enhanced when analyzing two wave packet lengths (Figure \ref{fig:CFRP_3-4_sig}d). Figure \ref{fig:CFRP_1-4_sig} shows the same 4 plots for a damage non-intersecting path (path 1-4). As shown panel c, the performance of the DIs substantially deteriorates, with a decrease in the value of both DIs with increasing simulated damage size up to 4 weights. A similar trend is observed in Figure \ref{fig:CFRP_1-4_sig}d when considering two wave packet lengths for the analysis. Furthermore, although some damage cases fall outside the 95\% confidence bounds (4 weights for single-wave packet DIs, and 5 and 6 weights for two-wave packet DIs), the general trend is a reduction in the values of the DIs, which can again be easily mistaken with changing environmental or operational conditions over an otherwise healthy component. Thus, in terms of damage detection, the DIs offer poor performance for the CFRP coupon with the simulated damage used in this study.

\begin{figure}[t!]
\centering
 \begin{picture}(400,300)
\put(-3,160){\includegraphics[scale=0.5]{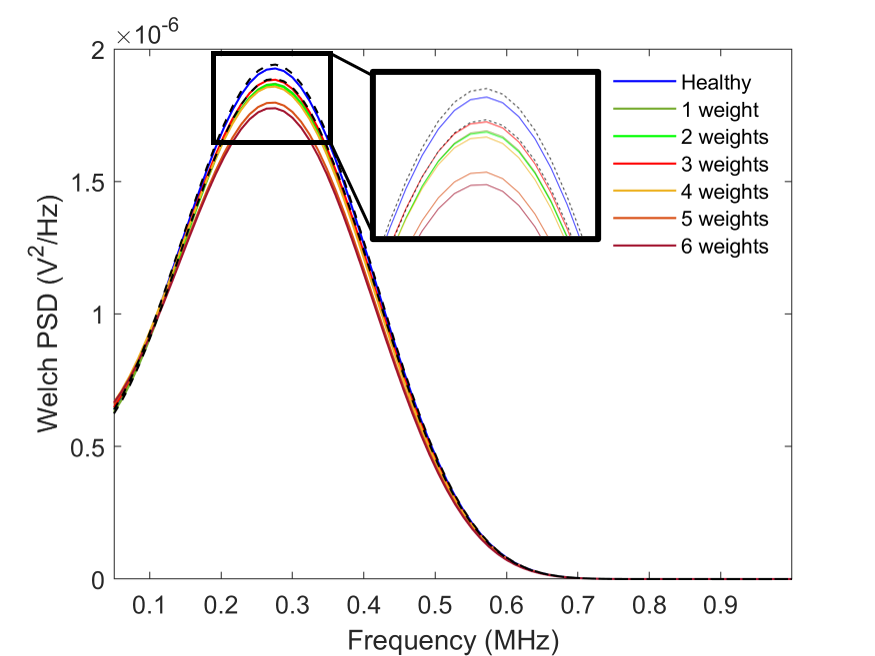}}
\put(192,160){\includegraphics[scale=0.5]{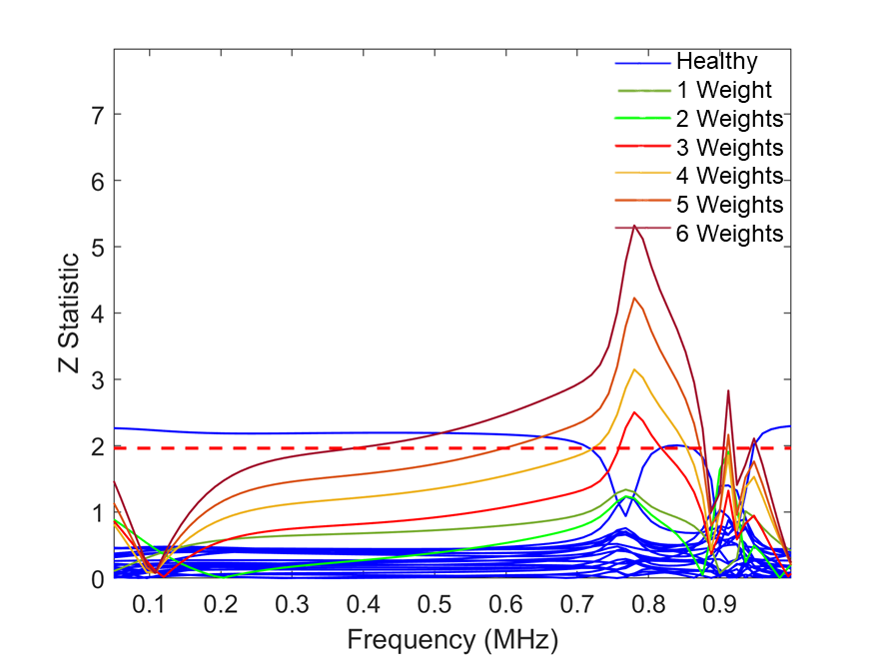}}
\put(-3,0){\includegraphics[scale=0.5]{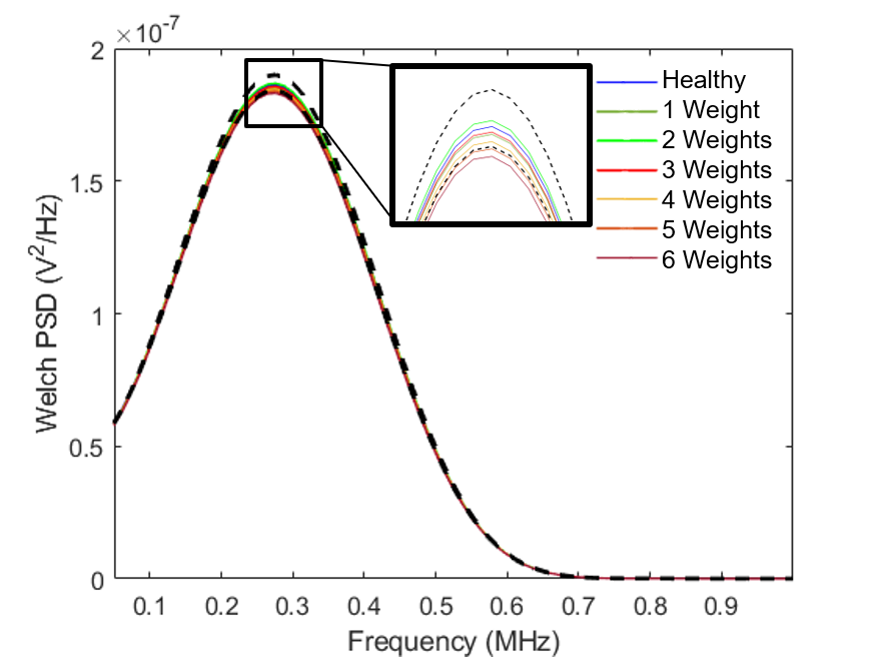}}
\put(192,0){\includegraphics[scale=0.5]{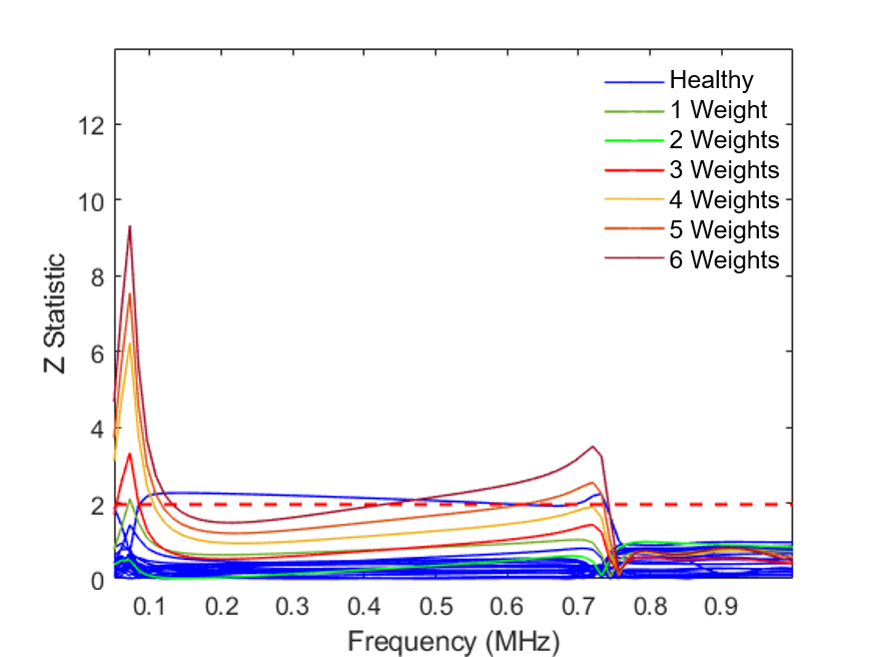}}
\put(30,295){\color{black} \large {\fontfamily{phv}\selectfont \textbf{a}}}
\put(225,295){\color{black} \large {\fontfamily{phv}\selectfont \textbf{b}}}
\put(30,135){\color{black} \large {\fontfamily{phv}\selectfont \textbf{c}}} 
\put(225,135){\color{black} \large {\fontfamily{phv}\selectfont \textbf{d}}}
\end{picture} 
\caption{Indicative results from applying the proposed NP-TS approach to the signals from path 3-4 in the CFRP coupon under different simulated damage sizes (number of attached weights): (a) Welch PSD for single wave packet -- the black dashed lines indicate the experimental 95\% confidence bounds of the healthy PSD; (b) $Z$ statistic for single wave packet; (c) Welch PSD for double the wave packet lengths; (d) $Z$ statistic for double the wave packet lengths.}
\label{fig:CFRP_3-4_npts}
\end{figure}
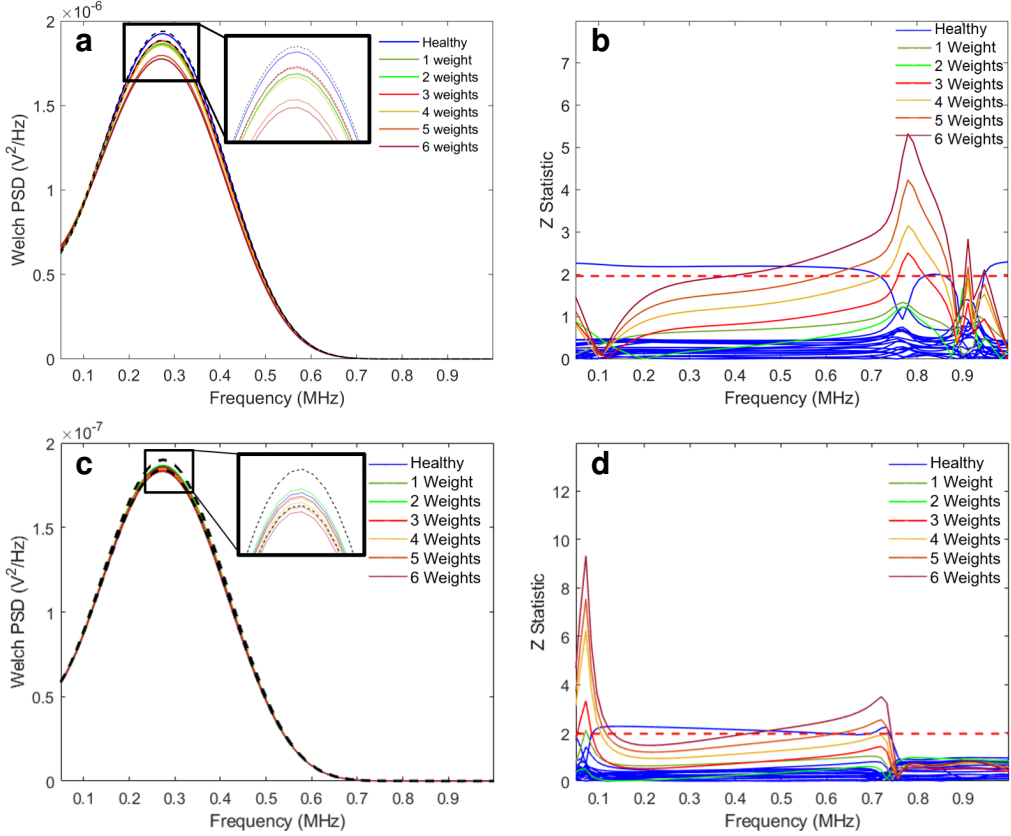

\begin{figure}[t!]
\centering
 \begin{picture}(400,300)
\put(-3,160){\includegraphics[scale=0.5]{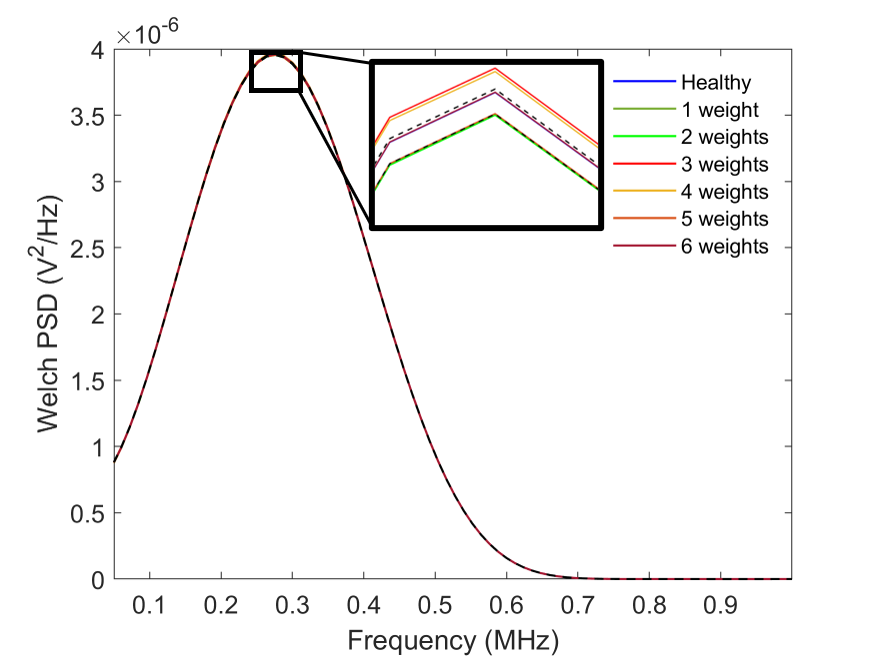}}
\put(192,160){\includegraphics[scale=0.5]{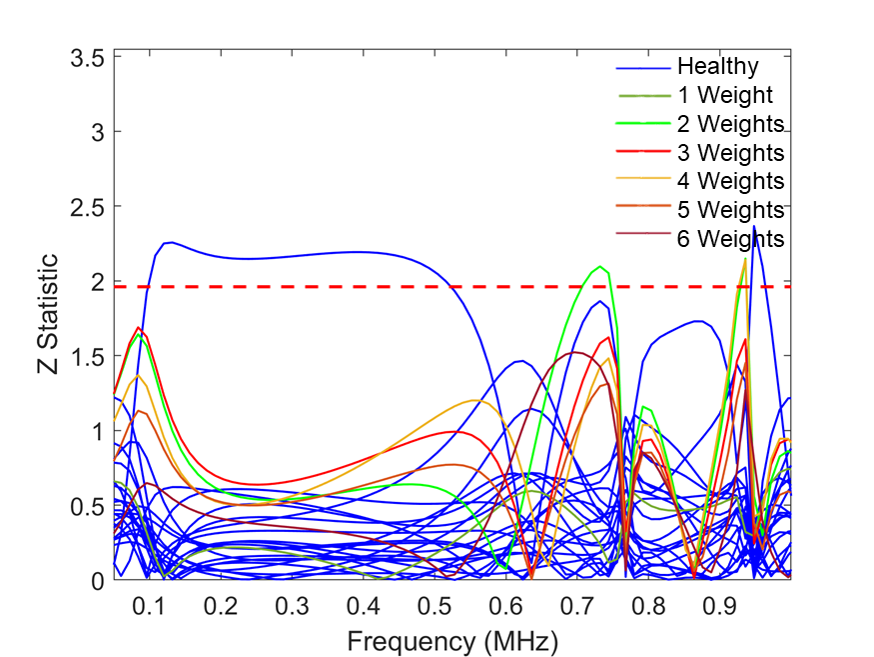}}
\put(-3,0){\includegraphics[scale=0.5]{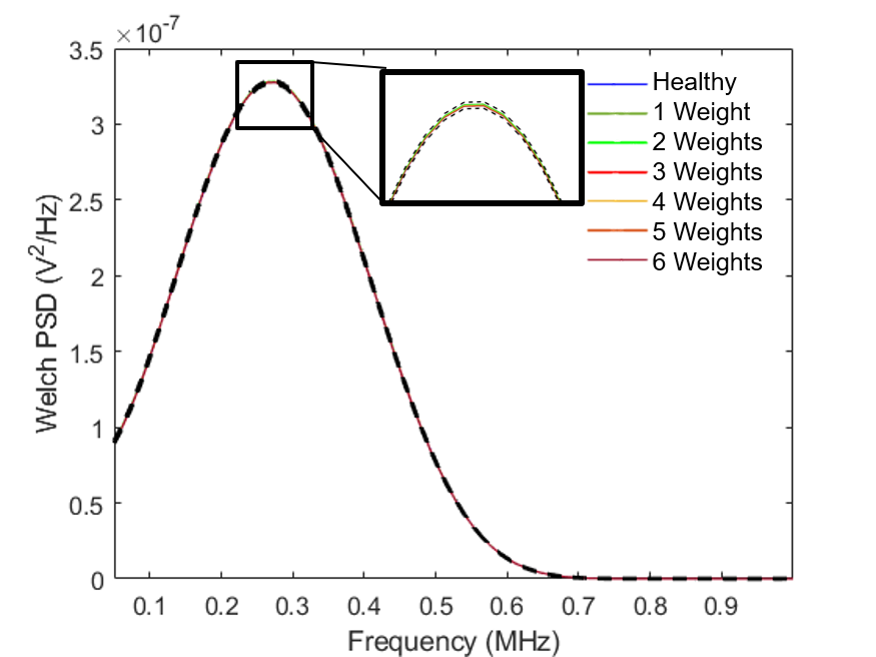}}
\put(192,0){\includegraphics[scale=0.5]{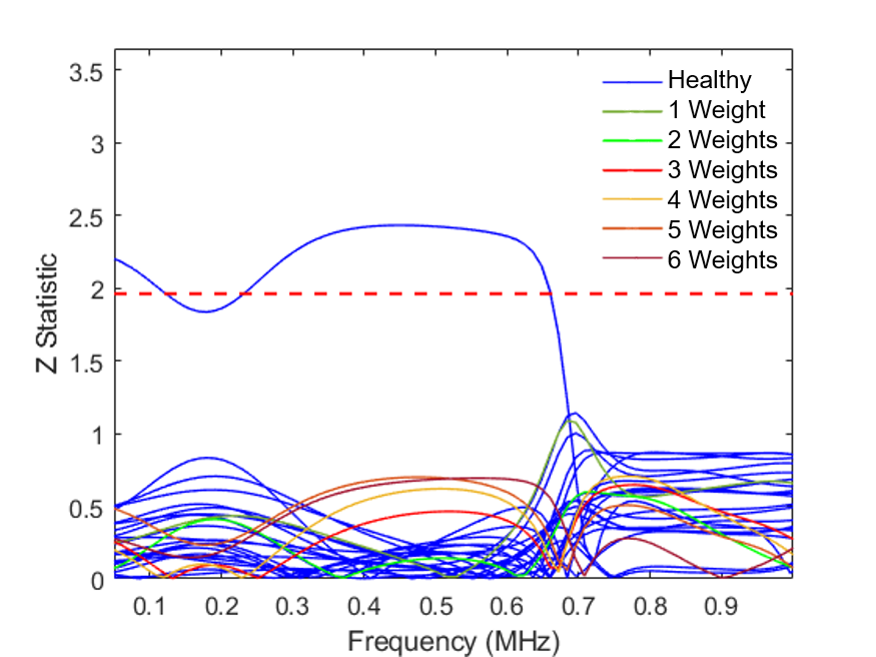}}
\put(30,295){\color{black} \large {\fontfamily{phv}\selectfont \textbf{a}}}
\put(225,295){\color{black} \large {\fontfamily{phv}\selectfont \textbf{b}}}
\put(30,135){\color{black} \large {\fontfamily{phv}\selectfont \textbf{c}}} 
\put(225,135){\color{black} \large {\fontfamily{phv}\selectfont \textbf{d}}}
\end{picture} 
\caption{Indicative results from applying the proposed NP-TS approach to the signals from path 1-4 in the CFRP coupon under different simulated damage sizes (number of attached weights): (a) Welch PSD for single wave packet -- the black dashed lines indicate the experimental 95\% confidence bounds of the healthy PSD; (b) $Z$ statistic for single wave packet; (c) Welch PSD for full signal; (d) $Z$ statistic for full signal.}
\label{fig:CFRP_1-4_npts}
\end{figure}

Figure \ref{fig:CFRP_3-4_npts}a shows the estimated Welch PSD for a single wave packet from path 3-4 under different damage cases. Note that, as mentioned, due to the nature of the actuation signal, the theoretical confidence intervals are too wide to detect any of the simulated damages, and thus they are not shown here. Figure \ref{fig:CFRP_3-4_npts}b shows the $Z$ statistic for that path, from which it can be concluded that the cases of 3-6 weights are all damage cases with 95\% confidence. Thus, the $Z$ statistic surpasses the DIs in detection performance for this damage-intersecting path. Examining a longer signal length for the analysis, it can be seen that both the Welch PSD estimate and the $Z$ statistic (Figure \ref{fig:CFRP_3-4_npts} panels c and d, respectively) show more sensitivity to damage, with the former detecting damage sizes as small as 3 weights, and the latter detecting ones as small as 2 weights. Moving onto the damage non-intersecting path (1-4), it can be seen that the Welch PSD estimates (Figure \ref{fig:CFRP_1-4_npts} panels a and c) fail to detect almost any of the damages with 95\% confidence levels. The same can be said for the $Z$ statistics, as shown in Figure \ref{fig:CFRP_1-4_npts} panels b and d) , with the exception of detecting the 2- and 4-weight cases for the single-wave packet $Z$ statistics. This reduction in sensitivity for damage-non-intersecting paths can be attributed to the effect of damage on the signal, as well as the relatively wide variability in the baseline signal amplitude, which in turn leads to widening the 95\% healthy confidence bounds.

\begin{figure}[t!]
\centering
 \begin{picture}(400,300)
\put(-3,160){\includegraphics[scale=0.5]{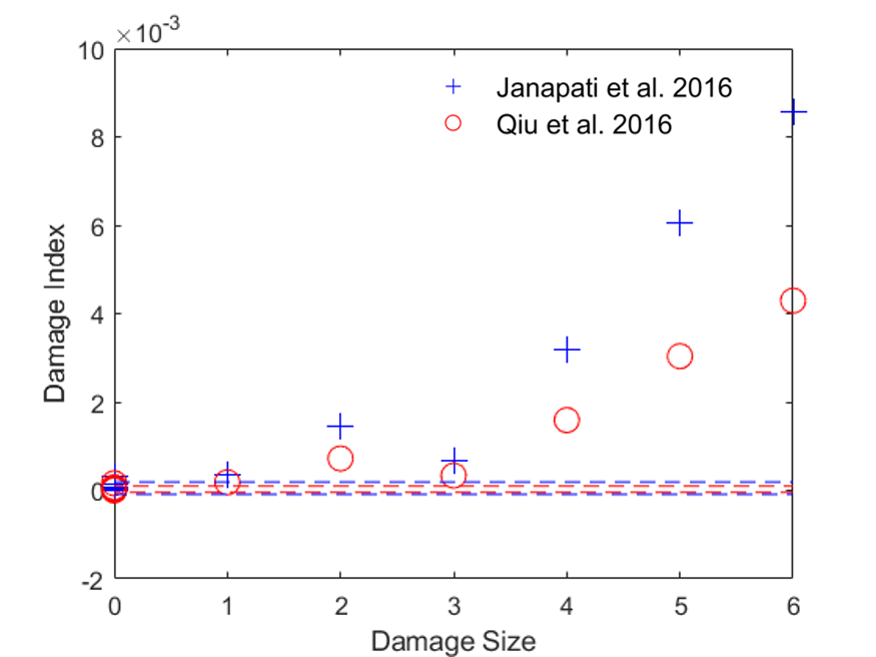}}
\put(192,160){\includegraphics[scale=0.5]{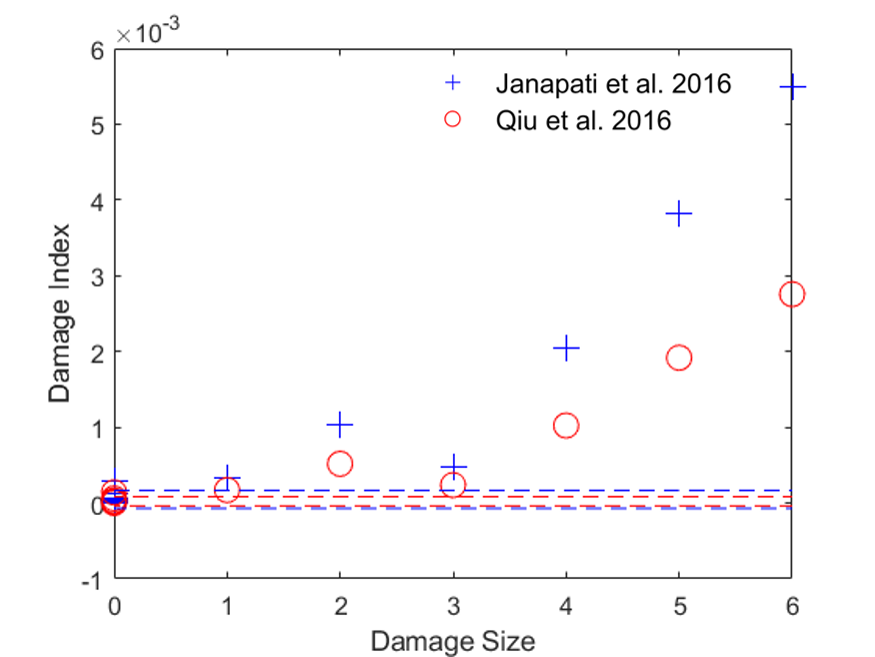}}
\put(-3,0){\includegraphics[scale=0.5]{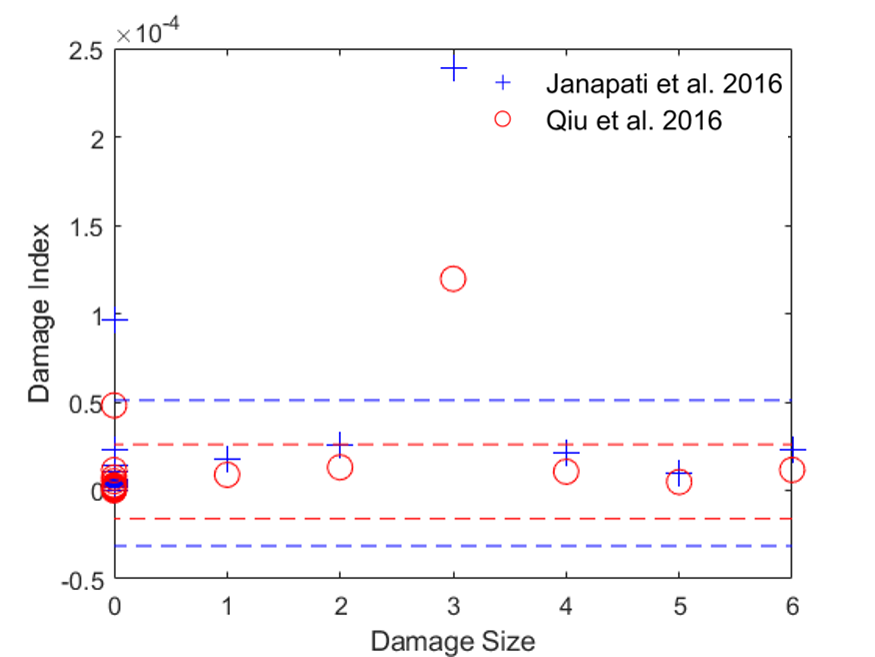}}
\put(192,0){\includegraphics[scale=0.5]{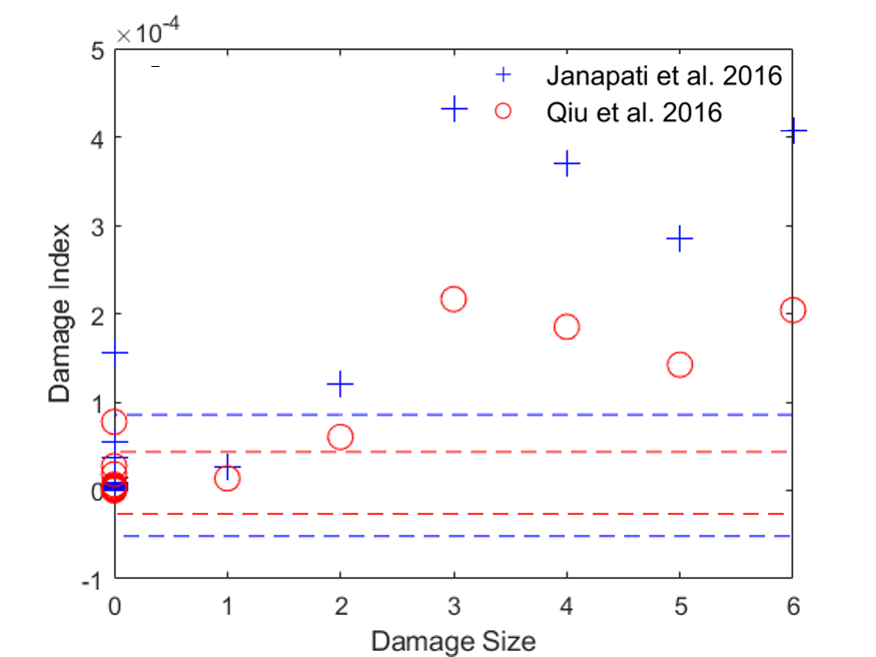}}
\put(30,295){\color{black} \large {\fontfamily{phv}\selectfont \textbf{a}}}
\put(225,295){\color{black} \large {\fontfamily{phv}\selectfont \textbf{b}}}
\put(30,135){\color{black} \large {\fontfamily{phv}\selectfont \textbf{c}}} 
\put(225,135){\color{black} \large {\fontfamily{phv}\selectfont \textbf{d}}} 
\end{picture} 
\caption{Damage Index results for the second acquired CFRP coupon data set shown in Table \ref{tab:CFRP_exp_info_2}: (a) single-wave packet DIs for path 3-4; (b) double-wave packet DIs for path 3-4; (c) single-wave packet DIs for path 1-4; (d) double-wave packet DIs for path 1-4. In all plots, the dashed red lines are the healthy $95\%$ confidence bounds for the Janapati \etal (blue) and Qiu \etal (red) DIs.}  
\label{fig:CFRP_2_DIs}
\end{figure}

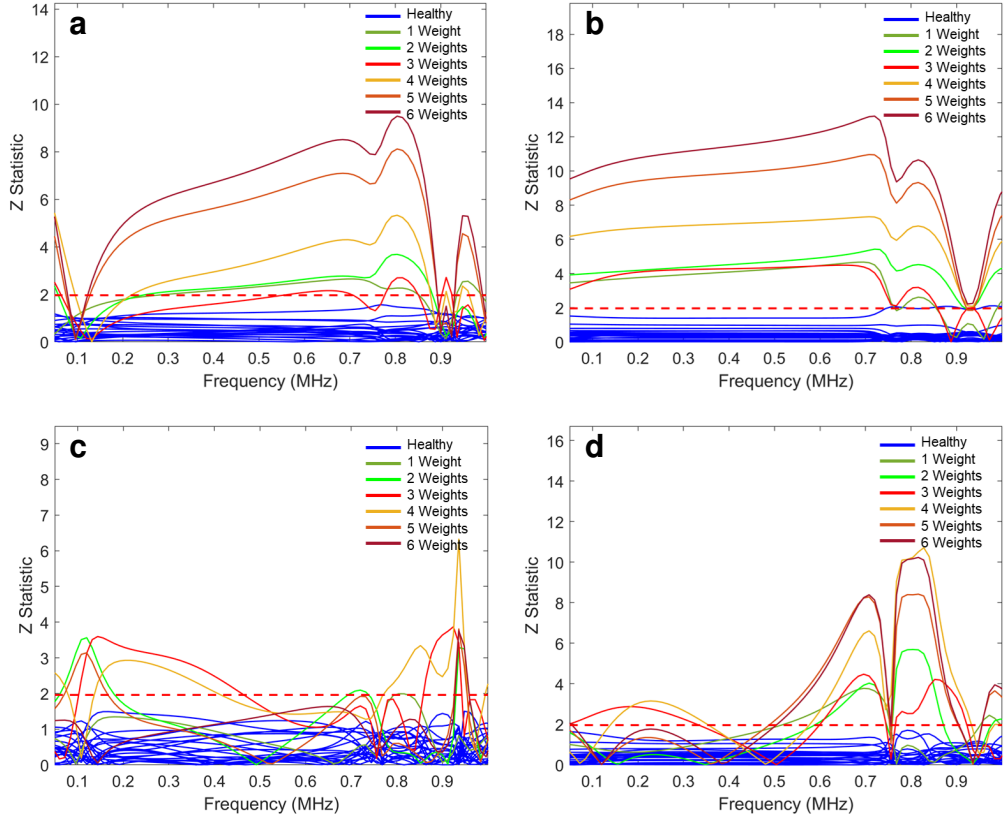
\begin{figure}[t!]
\centering
 \begin{picture}(400,300)
\put(-3,160){\includegraphics[scale=0.5]{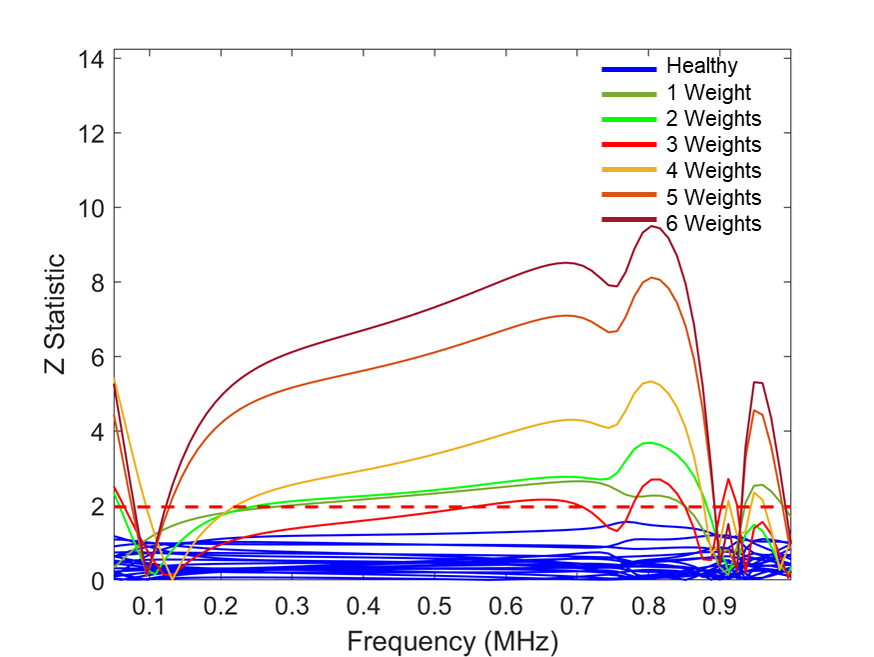}}
\put(192,160){\includegraphics[scale=0.5]{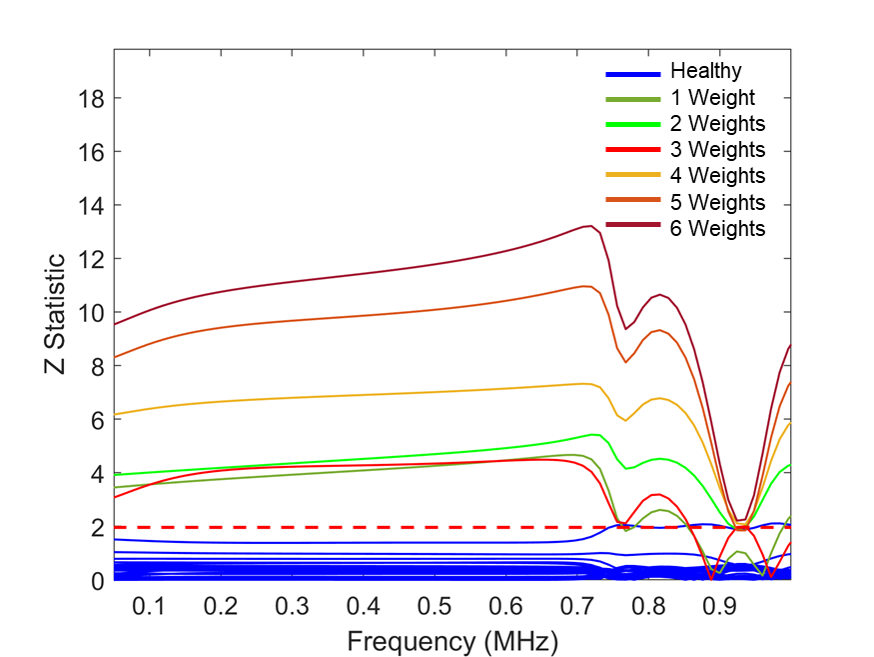}}
\put(-3,0){\includegraphics[scale=0.5]{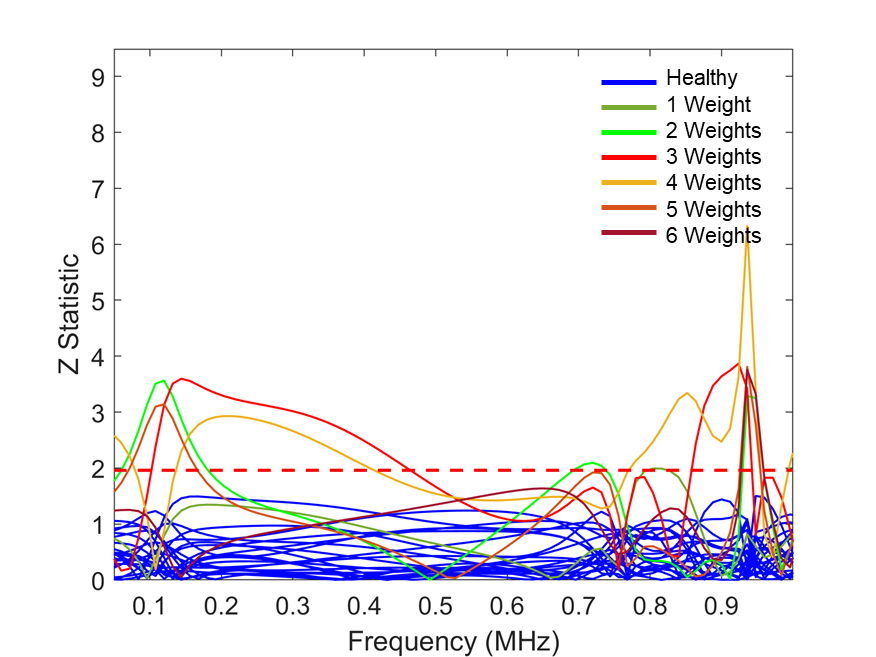}}
\put(192,0){\includegraphics[scale=0.5]{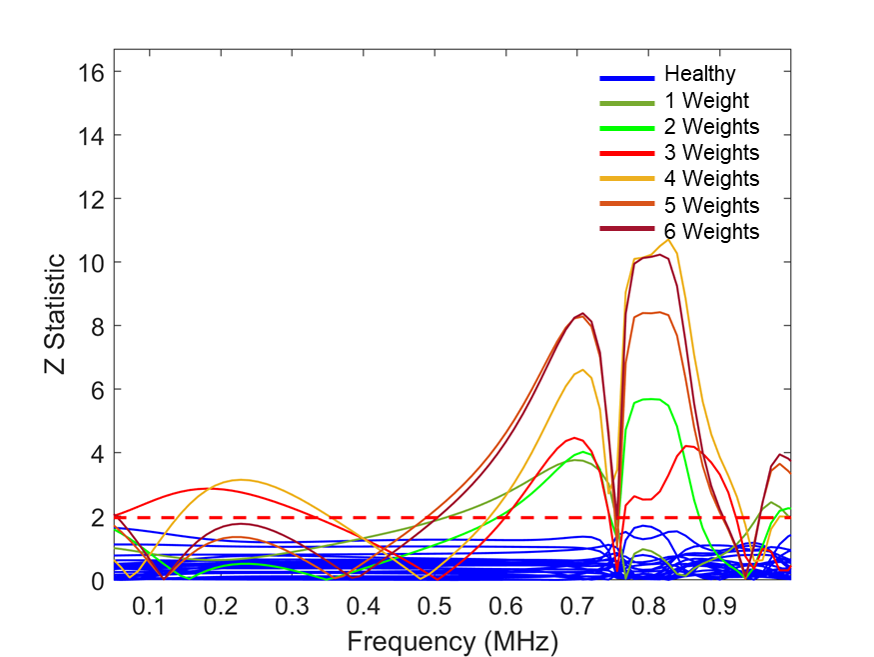}}
\put(30,295){\color{black} \large {\fontfamily{phv}\selectfont \textbf{a}}}
\put(225,295){\color{black} \large {\fontfamily{phv}\selectfont \textbf{b}}}
\put(30,135){\color{black} \large {\fontfamily{phv}\selectfont \textbf{c}}} 
\put(225,135){\color{black} \large {\fontfamily{phv}\selectfont \textbf{d}}} 
\end{picture} 
\caption{Indicative $Z$ statistic results for the second acquired CFRP coupon data set shown in Table \ref{tab:CFRP_exp_info_2}: (a) single-wave packet $Z$ statistics for path 3-4; (b) double-wave packet DIs for path 3-4; (c) single-wave packet $Z$ statistics for path 1-4; (d) double-wave packet $Z$ statistics for path 1-4. In all plots, the dashed red lines are the healthy $95\%$ confidence bounds.}  
\label{fig:CFRP_2_npts}
\end{figure}

For this reason, another experiment was taken out where the baseline data acquisition process was more restrictive (lab was empty). Table \ref{tab:CFRP_exp_info_2} presents the details of this second experiment. Figure \ref{fig:CFRP_2_DIs} shows the the DI plots for the damage-intersecting (panels a and b) and the damage-non-intersecting (panels c and d) paths of this new data set. For reference, these plots respectively correspond to the ones in Figure \ref{fig:CFRP_3-4_sig} panels c and d, and Figure \ref{fig:CFRP_1-4_sig} panels c and d. As shown, although both this and the original data sets were all acquired off of the same coupon with the same temperature setting, even in a lab environment, baseline variability between different data sets can be significant. As shown in all panels of Figure \ref{fig:CFRP_2_DIs}, the spread in the values of the healthy DIs is smaller in this new data set, which allows for good detection performance for the DIs. Exploring the $Z$ statistics (Figure \ref{fig:CFRP_2_npts}), one can observe the enhanced detection performance here too, given the narrower experimental confidence bounds in the new data set. Although both the DIs and the $Z$ statistics almost consistently follow damage size evolution for the damage-intersecting path (path 3-4), the $Z$ statistic shows better detection capability for the damage-non-intersecting path,detecting all damages with 95\% confidence.

\begin{table}[b!]
\centering
\caption{Parameters used in estimating the Welch PSD for the CFRP coupon data sets.}\label{tab:CFRP_npts_info}
\renewcommand{\arraystretch}{1.2}
{\footnotesize
\begin{tabular}{|ll|}
\hline
Segment Length & $100$ \\
Window Type & Hamming \\
Frequency Resolution & $\Delta f=12$ kHz \\
Sampling Frequency & $24$ MHz \\
\hline
\multicolumn{2}{|c|}{Single Wave Packet} \\
\hline
Data Length & $N=500$ samples ($\sim 20$ $\mu s$) \\
No of non-overlapping segments & $9$ \\
\hline
\multicolumn{2}{|c|}{Full Signal Length} \\
\hline
Data Length & $N=8000$ samples ($\sim 330$ $\mu s$) \\
No of non-overlapping segments & $159$ \\
\hline
\end{tabular}} 
\end{table}

In order to assess the performance of all 4 metrics (the DI, $F$, $F_m$, and $Z$ statistics) at different alpha (false alarm levels) i.e. at different confidence intervals, the corresponding Receiver Operating Characteristics (ROC) curves were explored for different signal lengths (also, see Tables \ref{tab:CFRP_3-4_FA/MF} and \ref{tab:CFRP_1-4_FA/MF} in the Appendix for summary results). Figure \ref{fig:CFRP_roc}a shows the ROC for the 4 metrics at alpha levels ranging from 1E-6 to 1, as applied to a single wave packet off of the damage-intersecting path 3-4. As shown, because this is a damage-intersecting path, 3 out of the 4 metrics exhibit perfect detection performance with an area under the ROC curve equal to 1. Also, although the performance of the $F_m$ statistic doesn't seem to be better than the worst statistical estimator (the dashed line), the $F$ statistic shows optimal performance as the alpha levels change, in contrast to its weak detection performance at an alpha level of 0.05, as mentioned in the discussion of Figure \ref{fig:CFRP_3-4_npts}. Moving onto two wave packets of the same path (Figure \ref{fig:CFRP_roc}b, one can observe that the $Z$ statistic outperforms all other metrics in damage detection. In addition, the $F$ statistic outperforms the DI metric, which hints on the advantages of using frequency-domain approaches and statistical hypothesis tests instead of time-domain approaches. For the damage-non-intersecting path 1-4, it can be observed that the $Z$ statistic outperforms the DI for both: a single- (Figure \ref{fig:CFRP_roc}c) and two- (Figure \ref{fig:CFRP_roc}d) wave packet lengths. Thus, it can be concluded that the $Z$ statistic emerges as the best damage detection statistic in this study for the CFRP coupon investigated herein.

\begin{figure}[t!]
\centering
 \begin{picture}(400,300)
\put(-3,160){\includegraphics[scale=0.5]{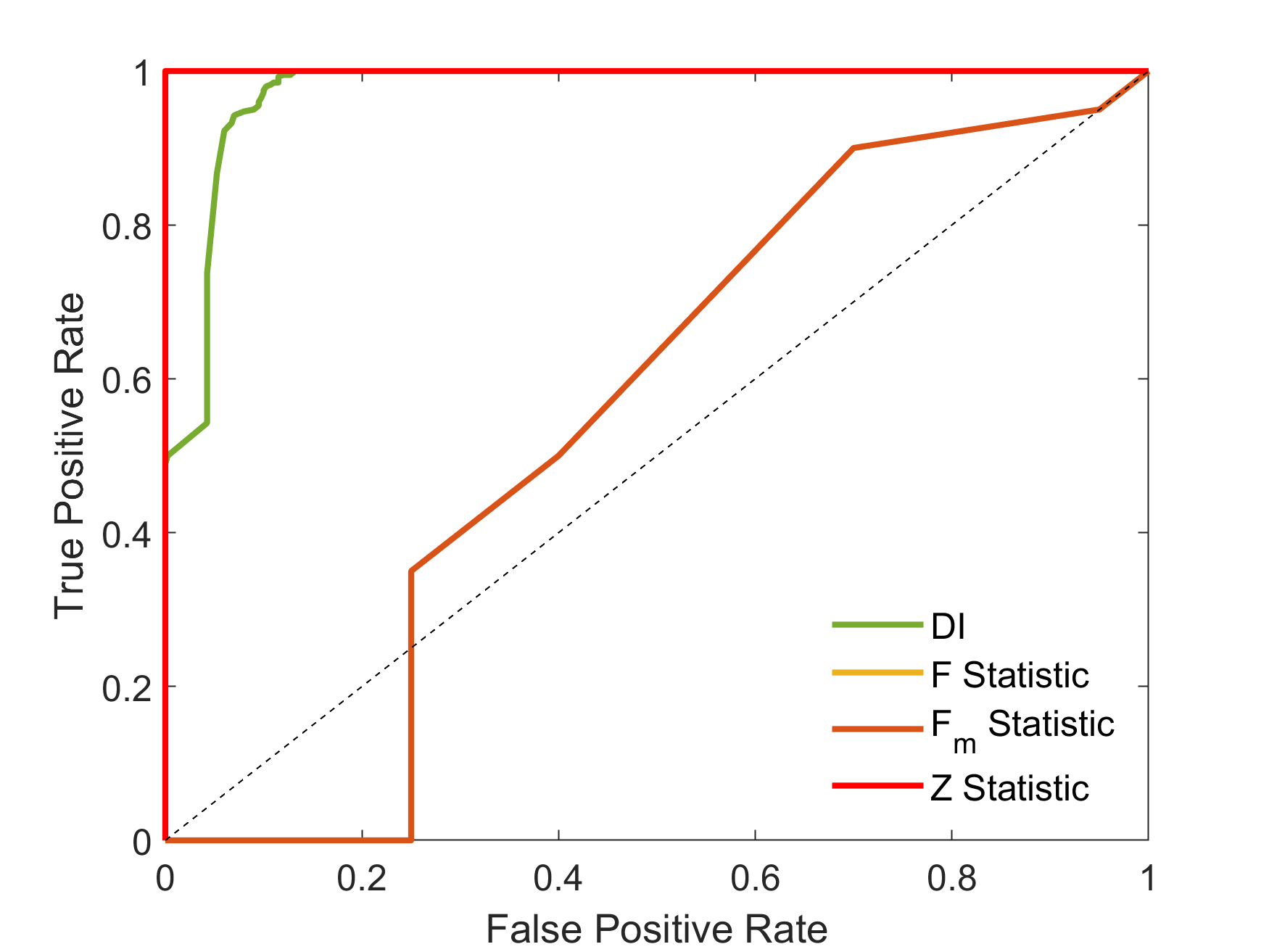}}
\put(192,160){\includegraphics[scale=0.5]{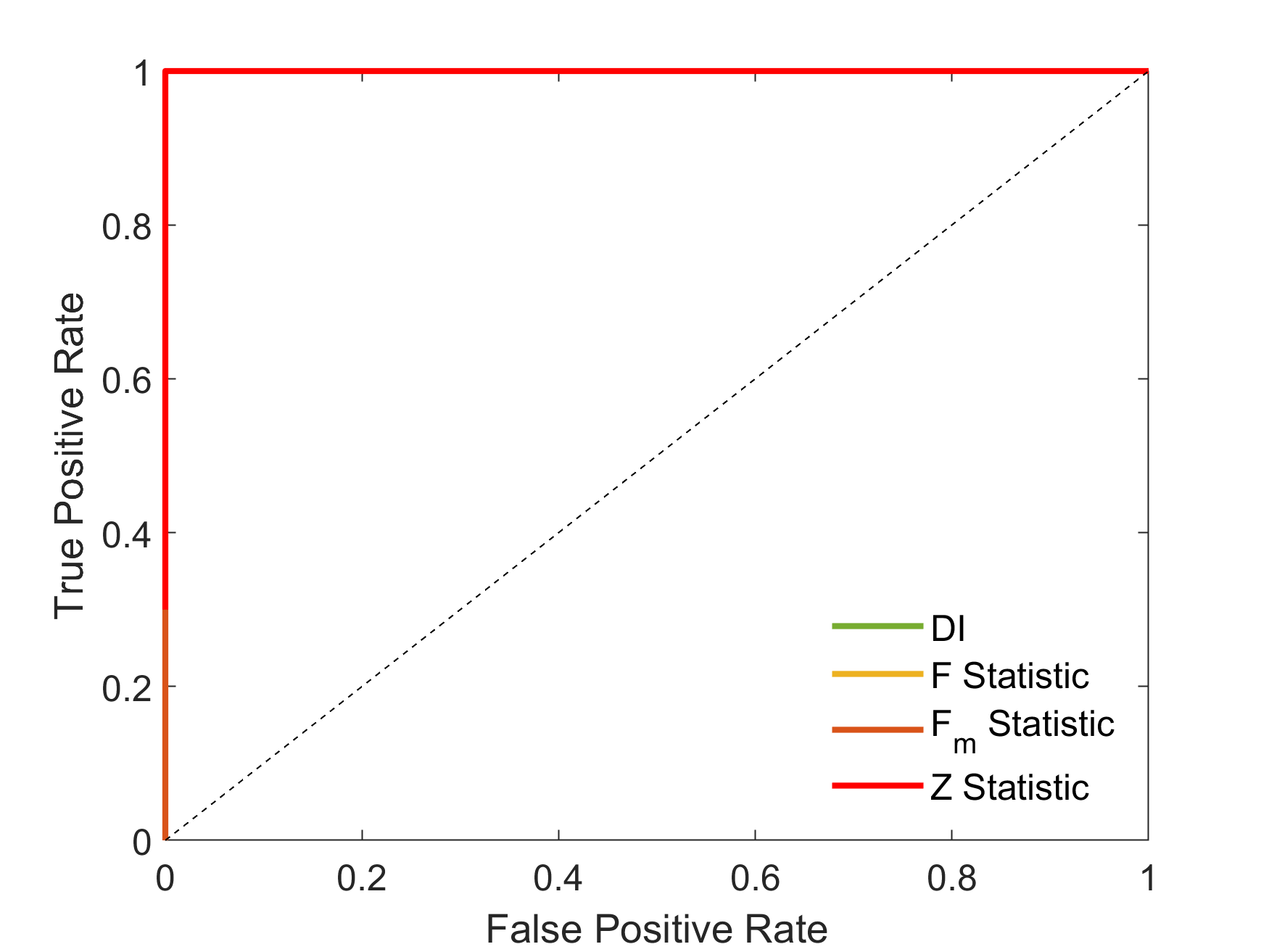}}
\put(-3,0){\includegraphics[scale=0.5]{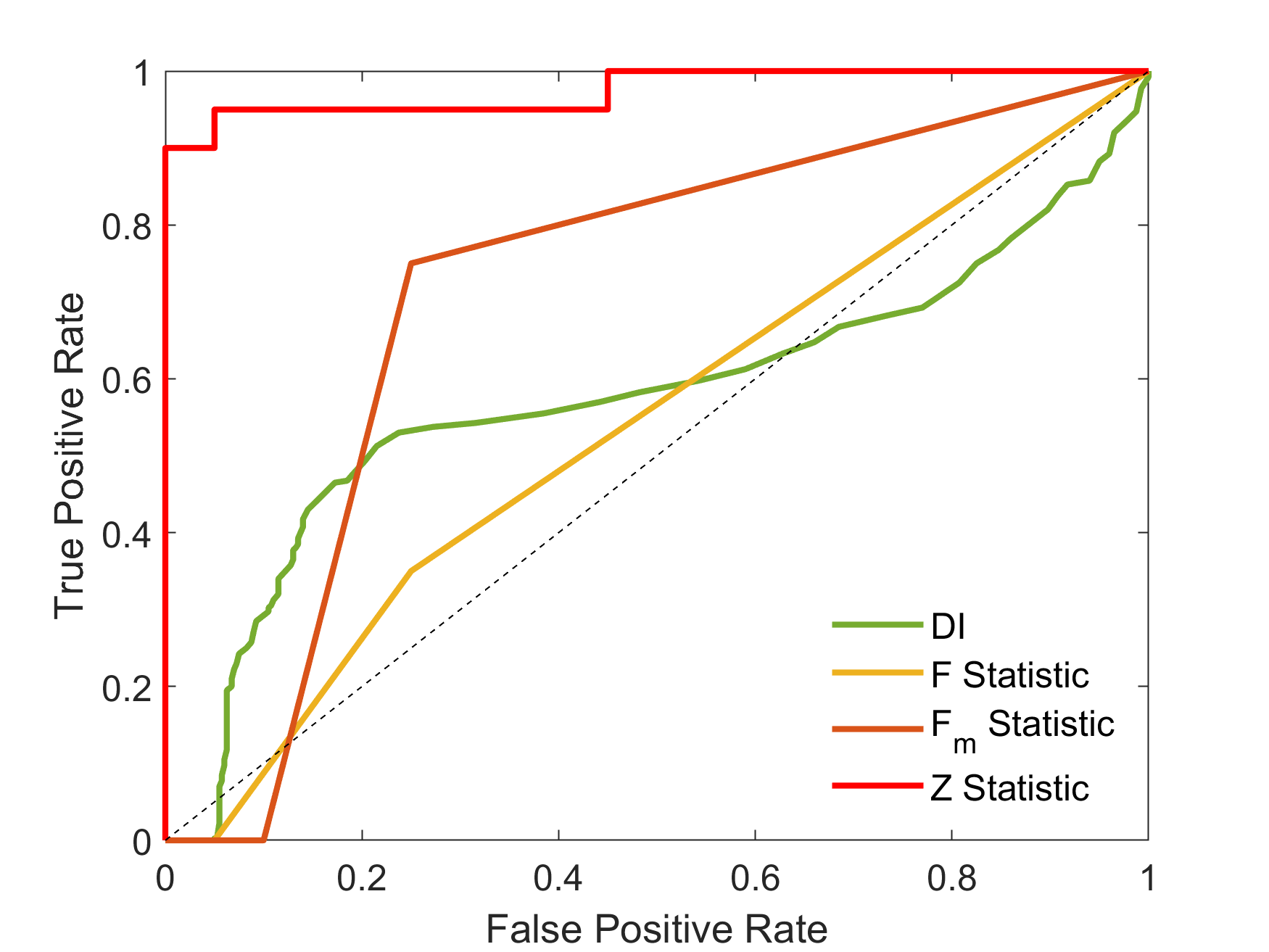}}
\put(192,0){\includegraphics[scale=0.5]{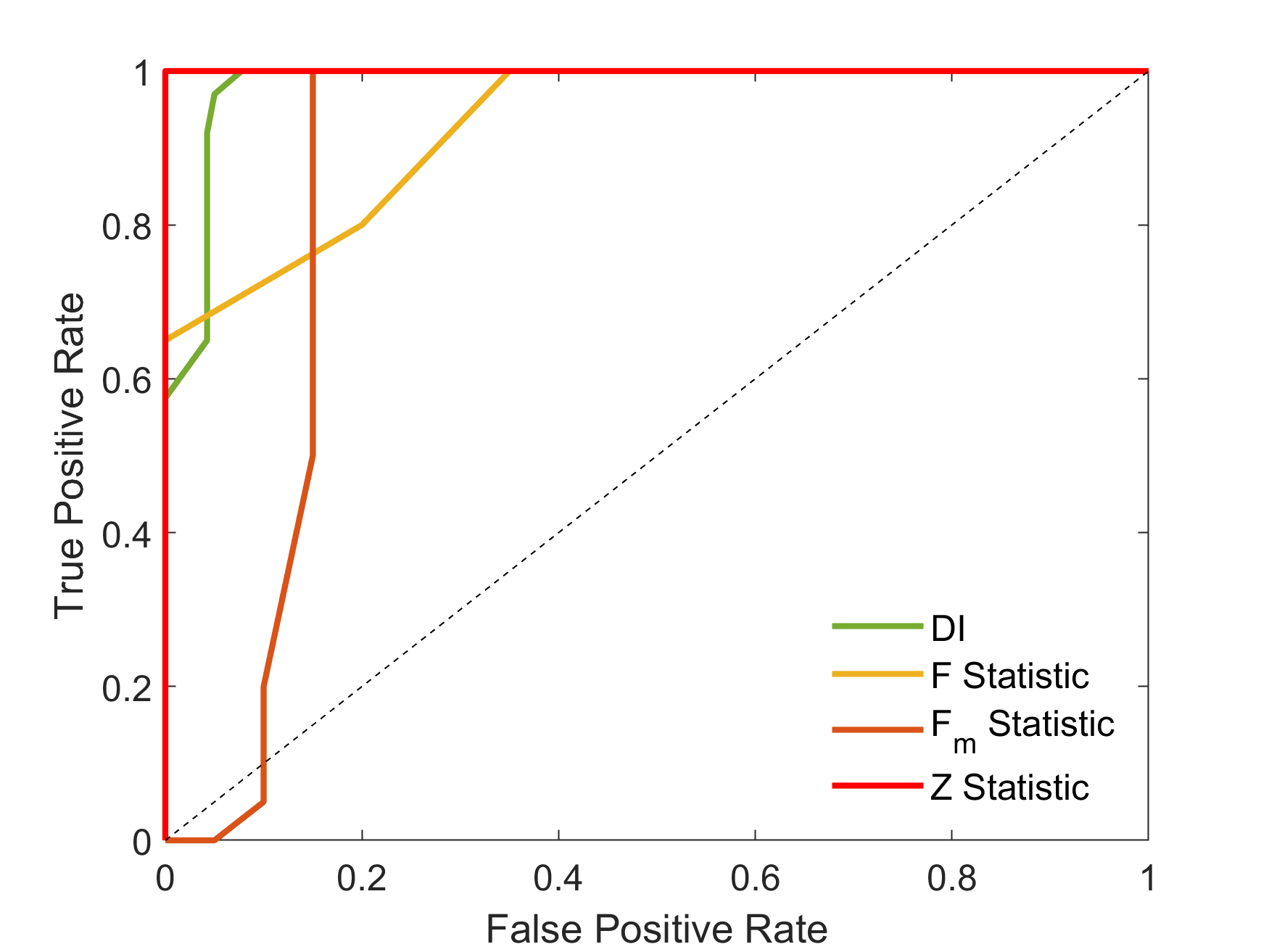}}
\put(30,295){\color{black} \large {\fontfamily{phv}\selectfont \textbf{a}}}
\put(225,295){\color{black} \large {\fontfamily{phv}\selectfont \textbf{b}}}
\put(30,135){\color{black} \large {\fontfamily{phv}\selectfont \textbf{c}}} 
\put(225,135){\color{black} \large {\fontfamily{phv}\selectfont \textbf{d}}}
\end{picture} 
\caption{Receiver Operating Characteristics (ROC) plots comparing the different damage detection methods for the new data set of the CFRP coupon under the effect of the first simulated damage (1 weight): (a) path 3-4 wave packet; (b) path 3-4 full signal; (c) path 1-4 wave packet; (d) path 1-4 full signal. In all subplots, 15 out of 20 healthy signals were used for calculating the mean in estimating the $F_m$ and the $Z$ statistics.}
\label{fig:CFRP_roc}
\end{figure}

\subsection{Test Case III: Damage Detection in the Open Guided-Waves CFRP Panel}


\subsubsection{Test Setup, Damage Types and Data Acquisition}

The third test case used in this study was the CFRP panel utilized in the Open Guided-Waves project \cite{OGW}, which had a quasi-isotropic construction with layup $[45/0/-45/90/-45/0/-45/90]_S$. The panel had the dimensions of $500 \times 500$ mm ($19.69 \times 19.69$ in), and a thickness of 2 mm ($0.079$ in). During the fabrication process of the panel, 12 PZT sensors, 5 mm ($0.2$ in) in diameter and $0.2$ mm ($0.0079$ in) in thickness, were co-bonded on the panel. To simulate damage, a 10-mm diameter, $2.35$-mm-thick ($0.0925$ in) Al disk ($0.5$ g) was consecutively attached using tacky tape on 28 different locations on the panel grouped into seven groups. Figure \ref{fig:OGW_3-12_sig_damage_intersecting}a shows a schematic of the CFRP panel, where the sensor and damage locations are shown. Also, the inset in Figure \ref{fig:OGW_3-12_sig_damage_intersecting}a shows the simulated damage on one of the locations.

Each sensor was consecutively actuated using a 5-peak tone burst signal (5-cycle Hanning-filtered sine wave) having an amplitude of $\pm 100$ V. Response signals were sampled over the remaining sensors at a sampling rate of $10$ MHz. Three sets of 20 baseline (healthy) signal realizations per sensor were recorded. After acquiring the first baseline set (the first 20 healthy signals), a single signal per sensor was recorded for each damage location, with the weight on locations D$1-11$. After that, the second baseline set was acquired (healthy signals $21-40$), followed by recording a single damage signal per sensor per weight location for locations D$12-20$. Finally, signals were recorded with the weight at locations D$21-28$ after the third baseline set was acquired (healthy signals $41-60$). This resulted in $60$ baseline realizations and $28$ damage signals per sensor (one signal for each damage location for each sensor.) Other data sets were also recorded that are not used in this study. Table \ref{tab:OGW_exp_info} summarizes the experimental details of this panel, and the readers are directed to the original study \cite{OGW} for more information on test setup. For ease of comparison of the proposed damage detection methods, only the response signals for the actuation with $260$ kHz center frequency were chosen for analysis in this study.

In the present study, signals from simulated damages in the same damage group (see Figure \ref{fig:OGW_3-12_sig_damage_intersecting}a and Table \ref{tab:OGW_exp_info}) were treated as different realizations of single damage in the vicinity of that group on the CFRP panel. In addition, for all the detection metrics in this study, each damage group was analyzed against its corresponding baseline data set only (the healthy data set immediately preceding that damage group in the data acquisition process) for accuracy. Actuator-sensor path 3-12 was used to demonstrate the performance of the different detection techniques proposed herein. As such, damage groups $2$ and $3$ (see Table \ref{tab:OGW_exp_info}) were considered as different realizations of signals for two damages intersected by the path (with the first baseline set used for comparison). On the other hand, damage groups $7$ and $8$ were treated as different realizations of signals for two damages not intersected by the signal path (with the third baseline set used for comparison).


\begin{table}[t!]
\centering
\caption{Summary of experimental details for the CFRP panel \cite{OGW}.} \label{tab:OGW_exp_info}
\renewcommand{\arraystretch}{1.2}
{\footnotesize
\begin{tabular}{|ccc|} 
\hline 
Structural State$^\dagger$ & Number of Data Sets & Data Set Label \\
\hline
\hline
Healthy (weight unattached) & 20$^{\dagger\dagger}$ & First Baseline Set \\
Weight on D1-4 & 4 & Damage Group 1 \\
Weight on D5-8 & 4 & Damage Group 2 \\
Weight on D9-11 & 3 & Damage Group 3 \\
\hline
Healthy (weight unattached) & 20$^{\dagger\dagger}$ & Second Baseline Set \\
Weight on D12 & 1 & Damage Group 4 \\
Weight on D13-16 & 4 & Damage Group 5 \\
Weight on D17-20 & 4 & Damage Group 6 \\
\hline
Healthy (weight unattached) & 20$^{\dagger\dagger}$ & Third Baseline Set \\
Weight on D21-24 & 4 & Damage Group 7 \\
Weight on D25-28 & 4 & Damage Group 8 \\
\hline
\multicolumn{3}{l}{{\bf Sampling Frequency:} $f_s=10$ MHz. Center frequency range: [$40:20:260$] kHz.} \\
\multicolumn{3}{l}{{\bf Number of samples} per data set $N= 13106$.}\\
\multicolumn{3}{l}{$^\dagger$Weight was attached to one location (e.g. D4) at a time within each damage group.}\\
\multicolumn{3}{l}{$^{\dagger\dagger}$M=20 in equation (\ref{eq:mean})}
\end{tabular}} 
\end{table}

\subsubsection{Damage Detection Results}

\begin{figure}[t!]
\centering
 \begin{picture}(400,300)
\put(30,160){\includegraphics[scale=0.35]{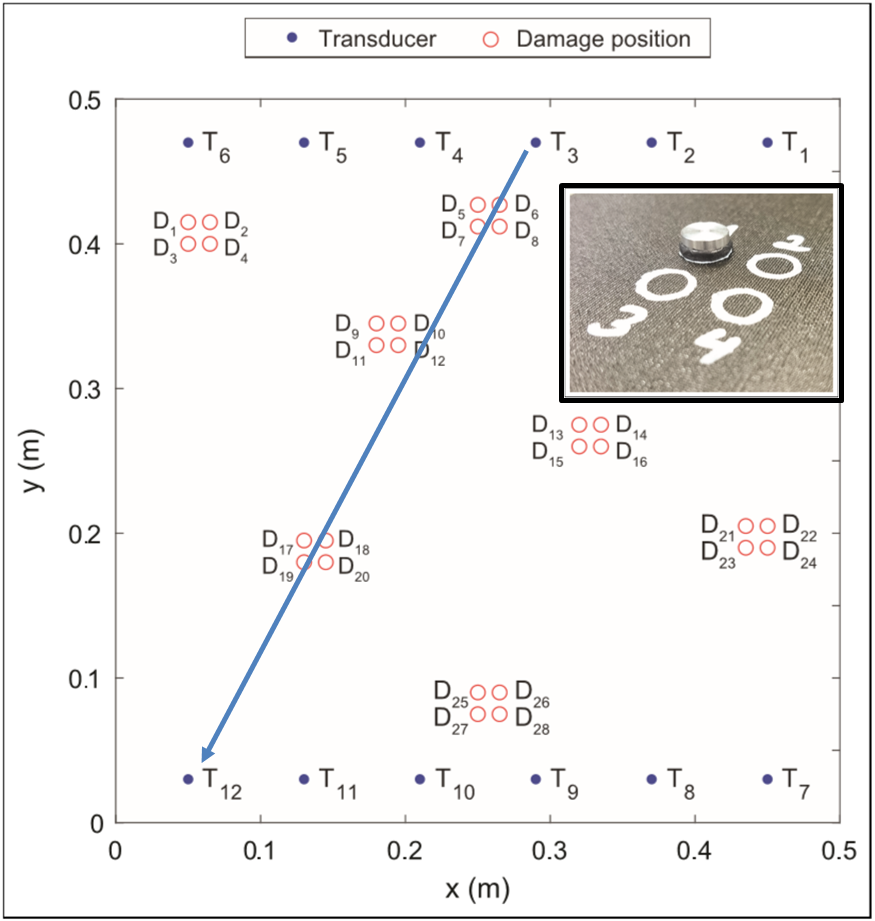}}
\put(192,160){\includegraphics[scale=0.5]{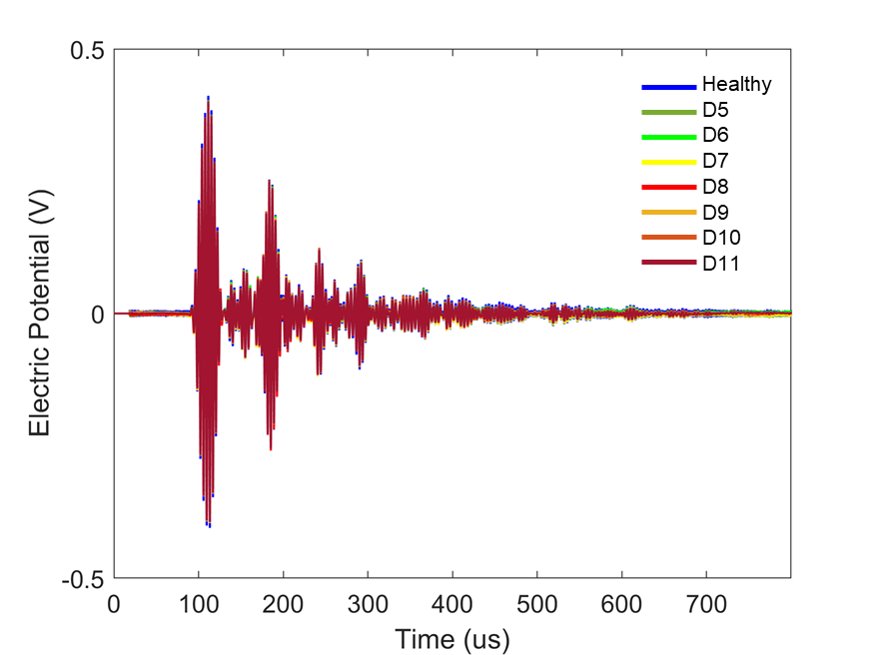}}
\put(-3,0){\includegraphics[scale=0.5]{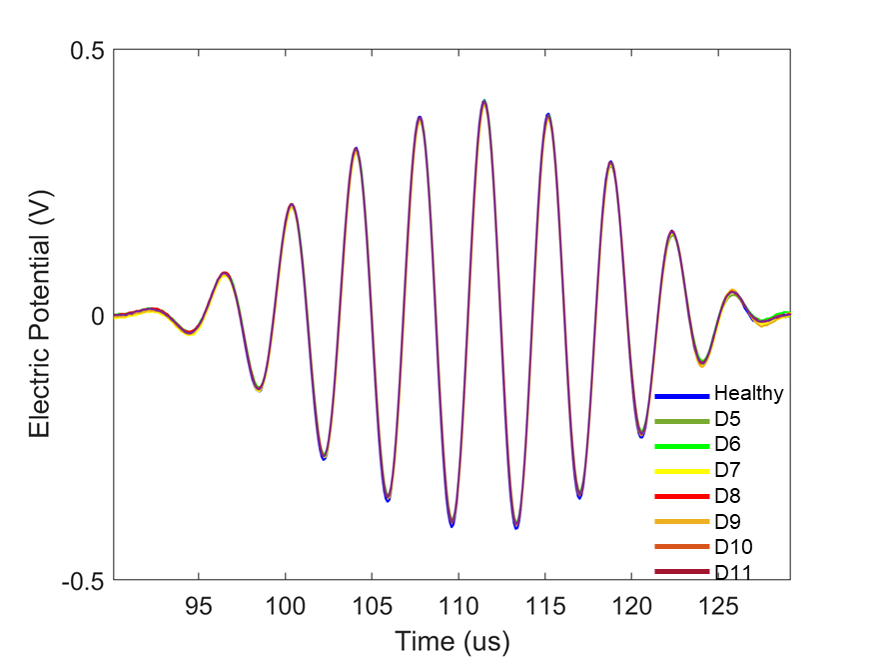}}
\put(192,0){\includegraphics[scale=0.5]{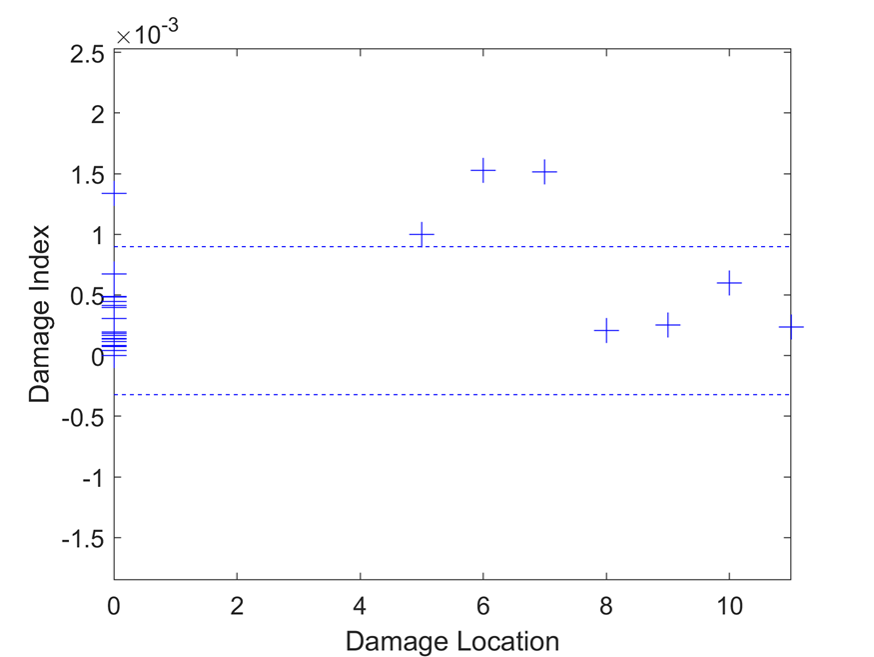}}
\put(35,290){\color{black} \large {\fontfamily{phv}\selectfont \textbf{a}}}
\put(225,290){\color{black} \large {\fontfamily{phv}\selectfont \textbf{b}}}
\put(30,130){\color{black} \large {\fontfamily{phv}\selectfont \textbf{c}}} 
\put(225,130){\color{black} \large {\fontfamily{phv}\selectfont \textbf{d}}} 
\end{picture} 
\caption{(a) A schematic of the CFRP panel used in the Open Guided Waves open source data project \cite{OGW} with all the simulated damage locations.  The inset shows a snapshot of part of the actual panel with damage location markings and the Al weight used to simulate damage on one of the locations. The arrow indicates the path used in the analysis presented herein; (b) indicative signals from path 3-12 for the healthy case, as well as when the Al weight (simulated damage) was on locations D5-10 (damage-intersecting case); (c) the first arrival wave packets; (d) the Janapati \etal DI for the first arrival wave packets -- the dashed blue lines are the upper and lower $95\%$ confidence bounds for the Janapati \etal DI as applied to the DI values of corresponding 20-baseline signal data set.}  
\label{fig:OGW_3-12_sig_damage_intersecting}
\end{figure}
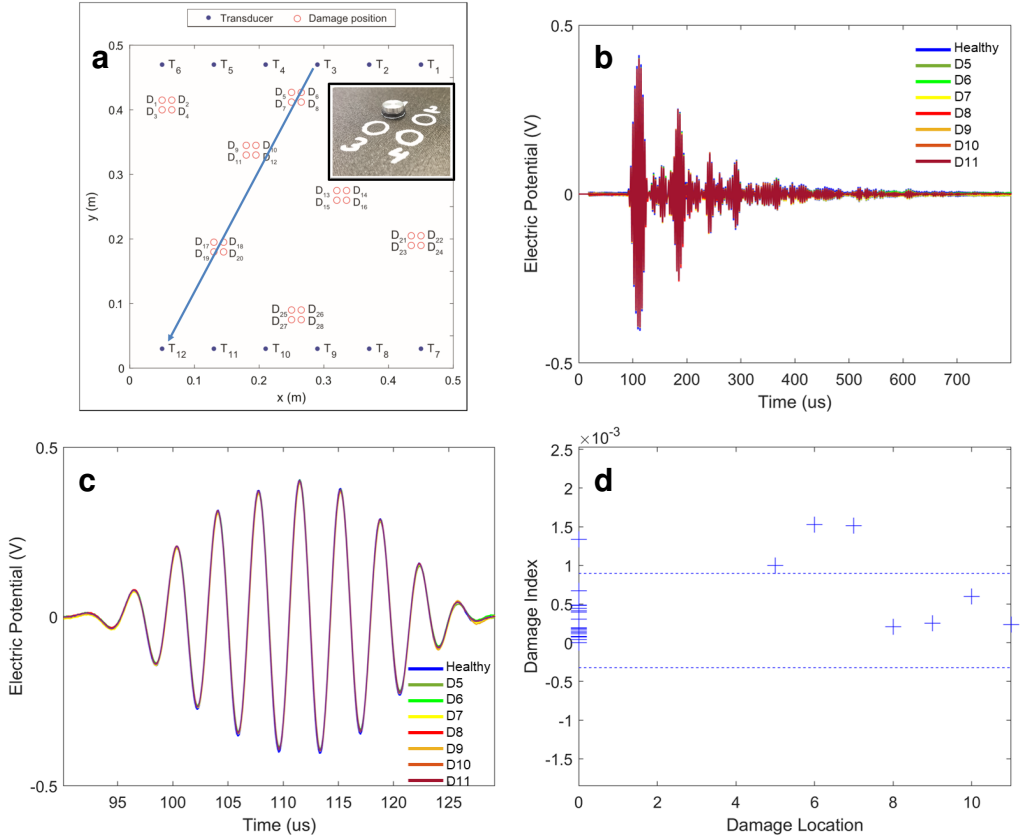

Figure \ref{fig:OGW_3-12_sig_damage_intersecting} panels b and c show the complete signal and the first-arrival wave packet, respectively, for signal path 3-12 on the CFRP panel when the Al weight was on locations 5-11 (damage-intersecting case). It is worth noting that, examining other paths (not shown here), the packet shown in panel c is actually a combination of two wave packets merged together, as can also be observed from the number of cycles in the shown packet. Even though this limits the analysis to only this single wave structure, this path was still chosen because it directly intersects (or does not intersect) almost complete damage location groups, which allows for the analysis of detection performance for both types of paths. 

\begin{figure}[t!]
\centering
 \begin{picture}(400,300)
 \put(-3,160){\includegraphics[scale=0.5]{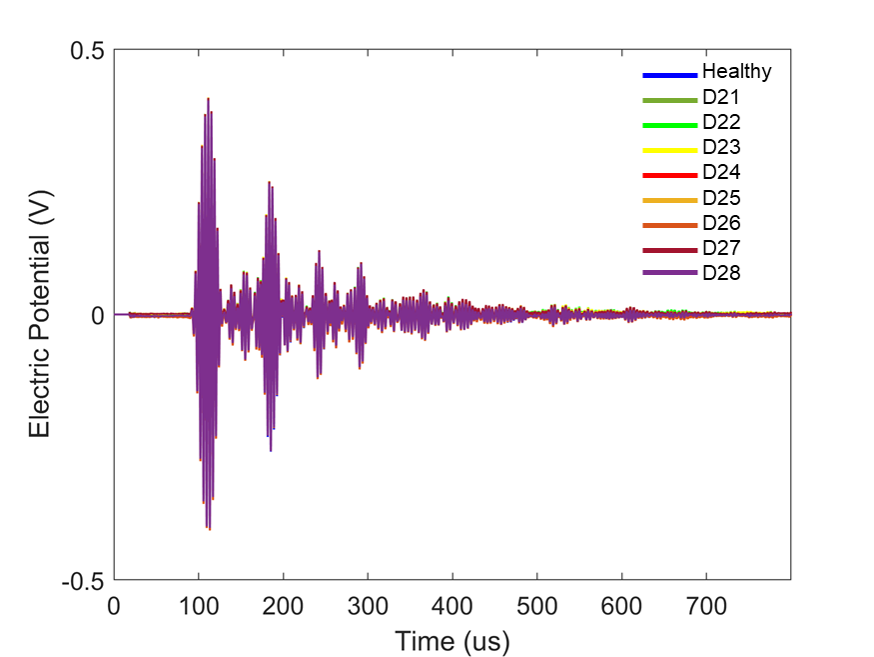}}
\put(192,160){\includegraphics[scale=0.5]{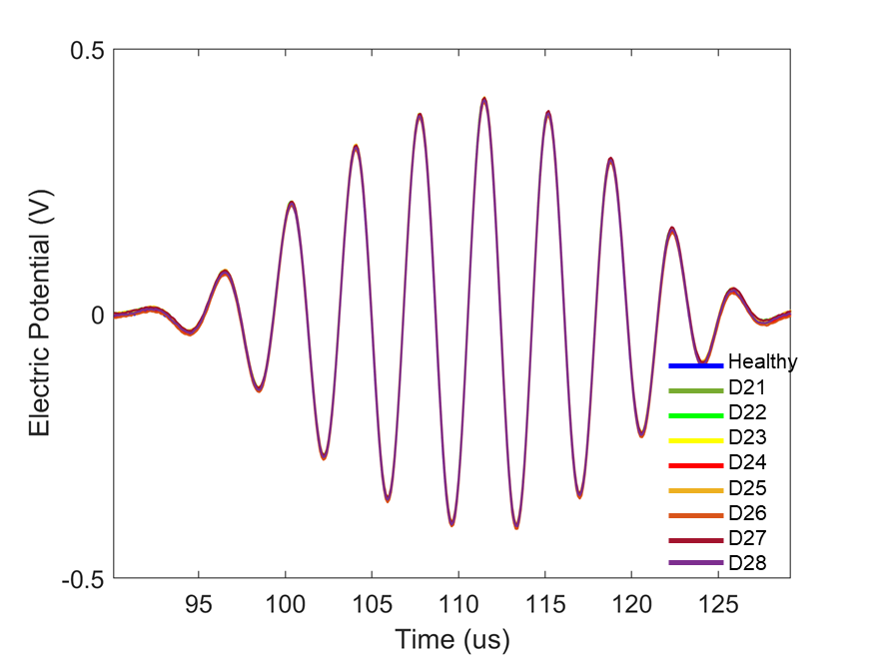}}
\put(100,0){\includegraphics[scale=0.5]{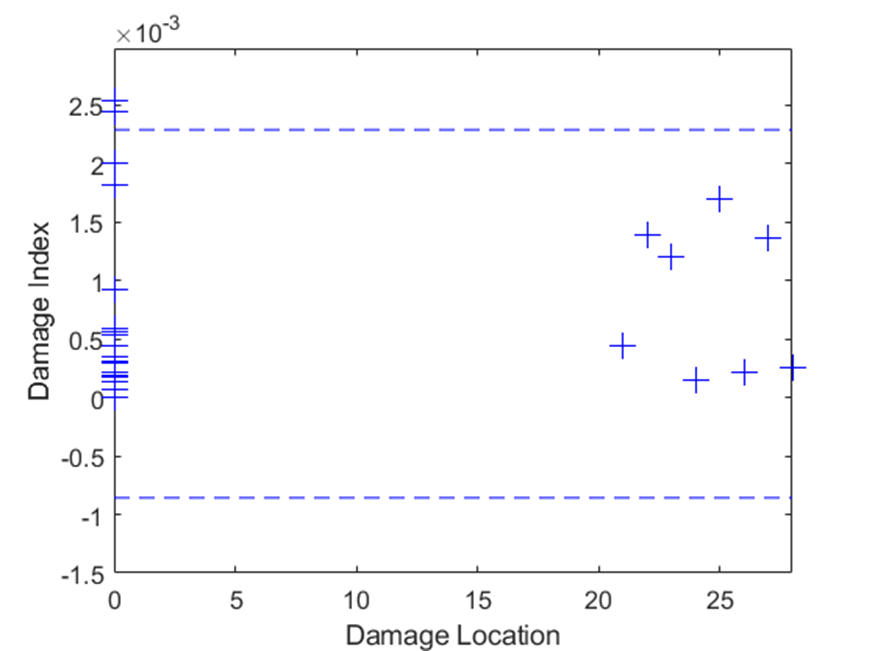}}
\put(30,295){\color{black} \large {\fontfamily{phv}\selectfont \textbf{a}}}
\put(225,295){\color{black} \large {\fontfamily{phv}\selectfont \textbf{b}}}
\put(135,135){\color{black} \large {\fontfamily{phv}\selectfont \textbf{c}}} 
\end{picture} 
\caption{(a) Indicative signals from path 3-12 for the healthy case, as well as when the Al weight (simulated damage) was on locations D21-28 (damage-non-intersecting case); (b) the first arrival wave packets; (c) the Janapati \etal DI for the first arrival wave packets - the dashed blue lines are the upper and lower $95\%$ confidence bounds for the Janapati \etal DI as applied to the DI values of corresponding 20-baseline signal data set.}  
\label{fig:OGW_3-12_sig_damage_non_intersecting}
\end{figure}

Figure \ref{fig:OGW_3-12_sig_damage_intersecting}d shows the values of the DI formulated by Janapati \etal \cite{Janapati-etal16} for the first 20 baseline signals and the signals corresponding to damage locations 5-11. As shown, within the $95$\% healthy confidence bounds (dashed blue lines), only damages at locations 5-7 are detected, while the rest of the damage cases are considered healthy with $95$\% confidence. Noting that the healthy signals in each baseline set were taken under controlled temperatures, it can be again concluded that the DI lacks robustness to uncertainties even in controlled environments, where the values of the DI still fluctuate even for the healthy case, producing wide healthy bounds that affect detection performance. Figure \ref{fig:OGW_3-12_sig_damage_non_intersecting} shows the same set of figures for the third baseline set (healthy signals 41-60), and the signals corresponding to damage locations 21-28 (damage-non-intersecting case.) As shown, the DI fails to detect any of the damage cases with $95$\% confidence.

\begin{figure}[t!]
\centering
 \begin{picture}(400,300)
\put(-3,160){\includegraphics[scale=0.5]{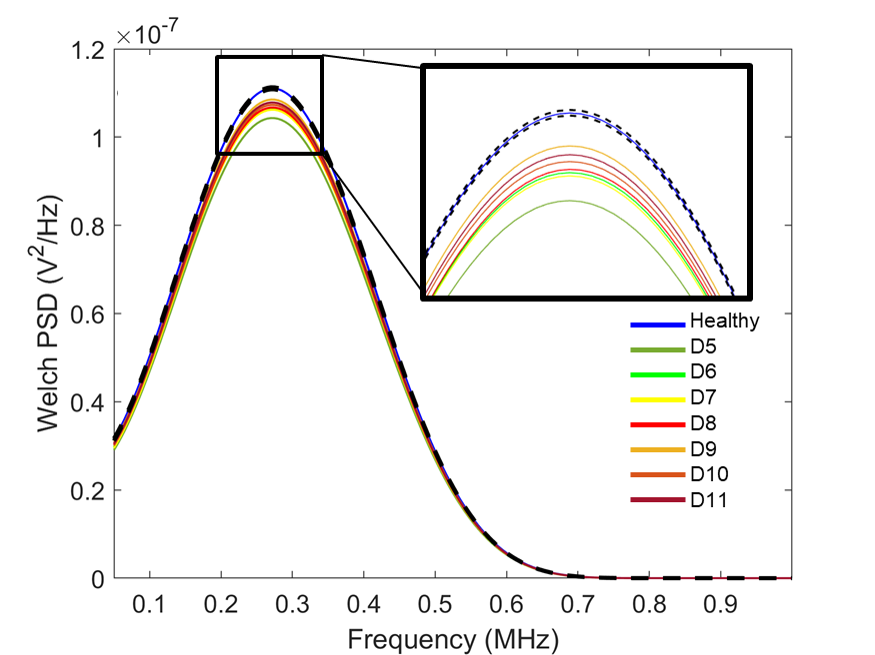}}
\put(192,160){\includegraphics[scale=0.5]{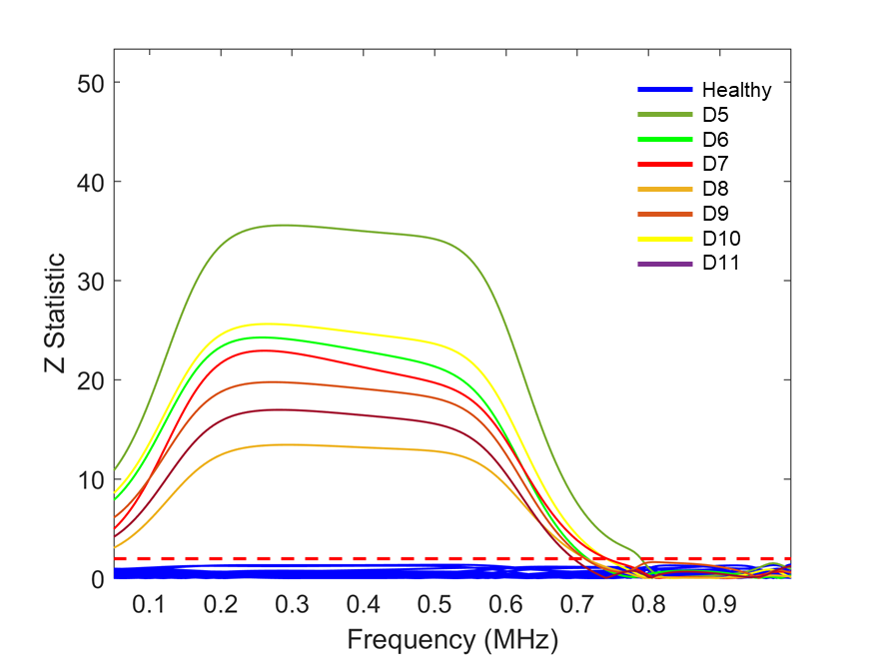}}
\put(-3,0){\includegraphics[scale=0.5]{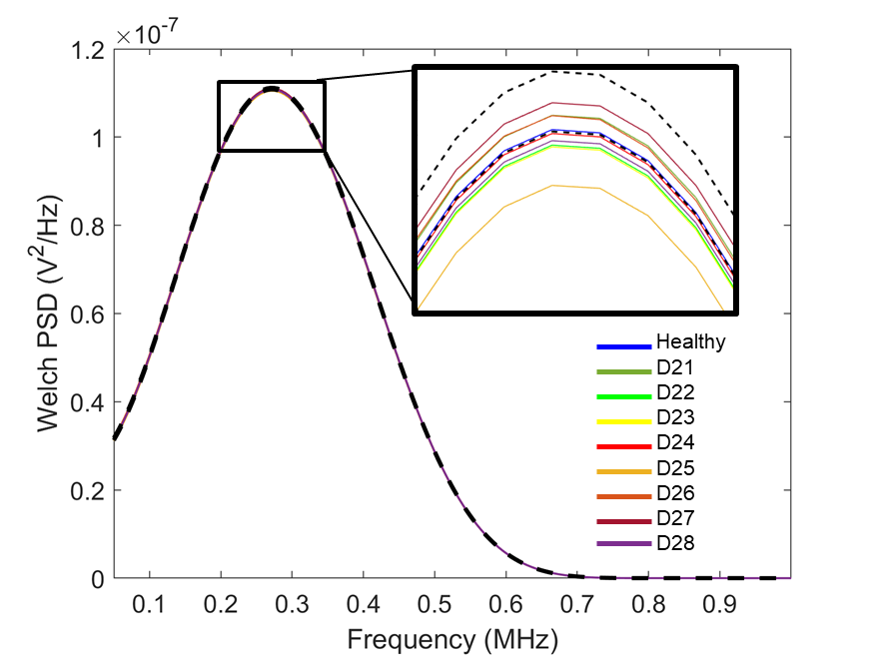}}
\put(192,0){\includegraphics[scale=0.5]{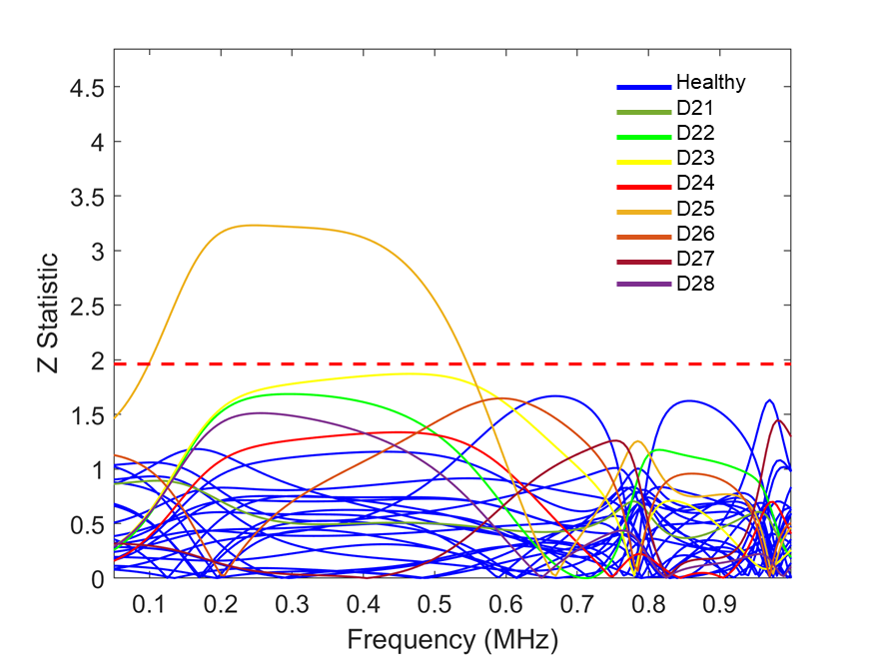}}
\put(30,290){\color{black} \large {\fontfamily{phv}\selectfont \textbf{a}}}
\put(225,290){\color{black} \large {\fontfamily{phv}\selectfont \textbf{b}}}
\put(30,130){\color{black} \large {\fontfamily{phv}\selectfont \textbf{c}}} 
\put(225,130){\color{black} \large {\fontfamily{phv}\selectfont \textbf{d}}} 
\end{picture} 
\caption{The results of applying the proposed NP-TS approach to the signals from path 3-12 in the Open Guided-Waves CFRP panel under different simulated damage locations: (a) Welch PSD for D5-10 (damage-intersecting case) - the black dashed lines indicate the theoretical (estimation uncertainty) and the experimental 95\% confidence bounds of the healthy PSD, respectively; (b) $Z$ Statistic for D5-10 (damage-intersecting case); (c) Welch PSD for D21-28 (damage-non-intersecting case); (d) $Z$ Statistic for D21-28 (damage-non-intersecting case)}
\label{fig:OGW_3-12_npts}
\end{figure}

Examining the Welch PSD estimates for the damage-intersecting case, Figure \ref{fig:OGW_3-12_npts}a shows that all damage cases are detected with $95$\% confidence. This immediately shows the advantage of this frequency-domain metric when it comes to damage detection. Figure \ref{fig:OGW_3-12_npts}b shows the $Z$ statistic for that case, where again all damage cases are detected with $95$\% confidence. Also, it can be observed that there are no false alarms in this case, whereas there was at least one false alarm event with the DI (see Figure \ref{fig:OGW_3-12_sig_damage_intersecting}c). Figure \ref{fig:OGW_3-12_npts}c presents the Welch PSD estimate for the damage-non-intersecting set of signals. As shown, at least 5 out of the 8 damage cases were detected with $95$\% accuracy. Examining the $Z$ statistic, it can be observed that only one damage case is detected with $95$\% confidence, while the rest are deemed healthy. It is worth noting that neither of the other two statistics (the $F$ and $F_m$ statistics) detected any of the damage cases with the set confidence bounds for the damage-intersecting and non-intersecting cases. Again, this called upon the exploration of the effect of different confidence intervals (manifested in different alpha false alarm levels in the statistical hypothesis tests) in order to conclusively assess the performance of the different detectors proposed herein.

\begin{table}[b!]
\centering
\caption{The Welch PSD estimation parameters for the OGW coupon data sets.} \label{tab:OGW_npts_info}
\renewcommand{\arraystretch}{1.2}
{\footnotesize
\begin{tabular}{|ll|}
\hline
Segment Length & $40$ \\
Window Type & Hamming \\
Frequency Resolution & $\Delta f=5$ kHz \\
Sampling Frequency & $10$ MHz \\
\hline
\multicolumn{2}{|c|}{Single Wave Packet} \\
\hline
Data Length & $N=360$ samples ($\sim 40$ $\mu s$) \\
No of non-overlapping segments & $9$ \\
\hline
\end{tabular}} 
\end{table}

\begin{figure}[t]
\centering
 \begin{picture}(400,300)
\put(-3,160){\includegraphics[scale=0.5]{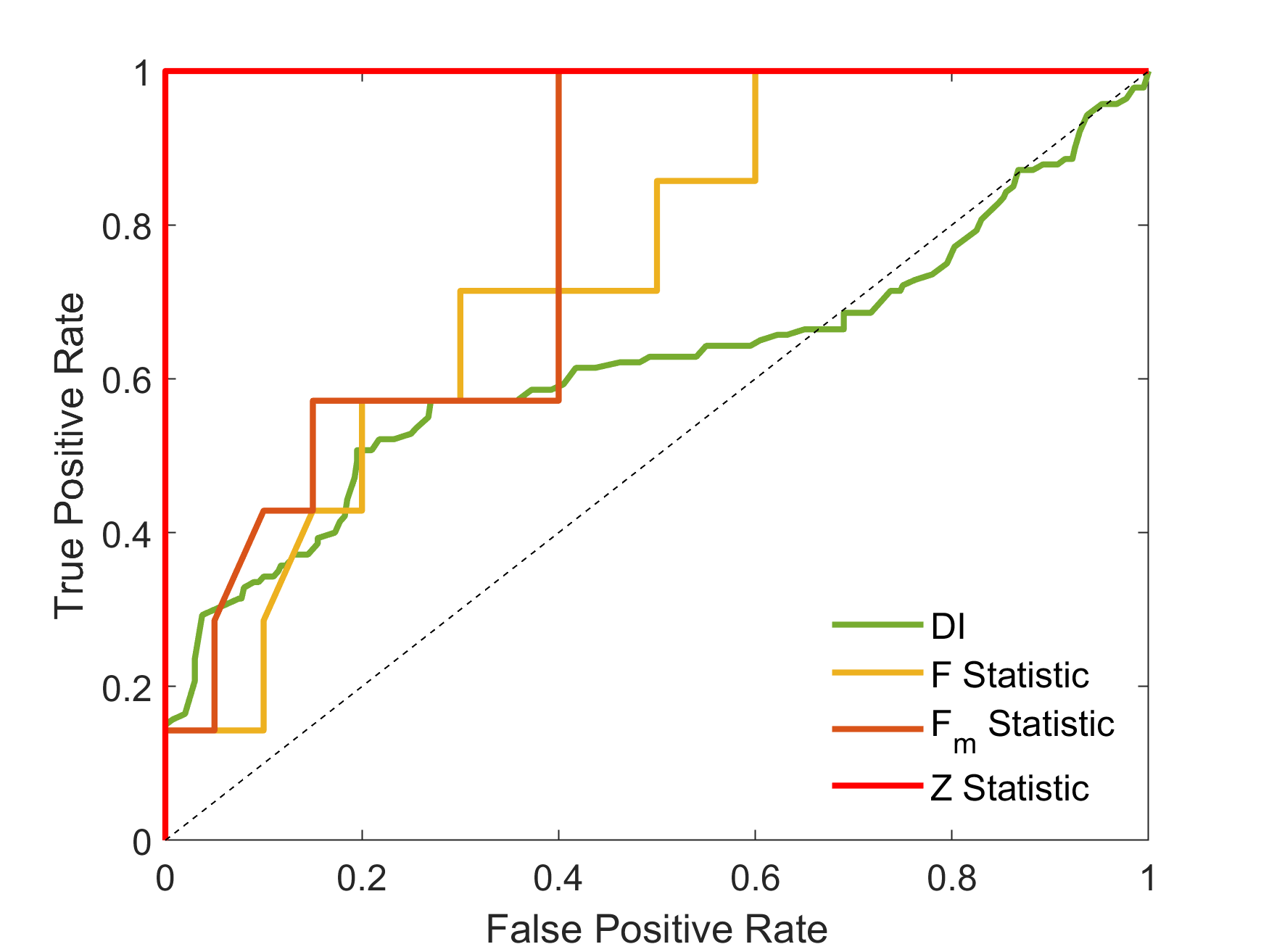}}
\put(192,160){\includegraphics[scale=0.5]{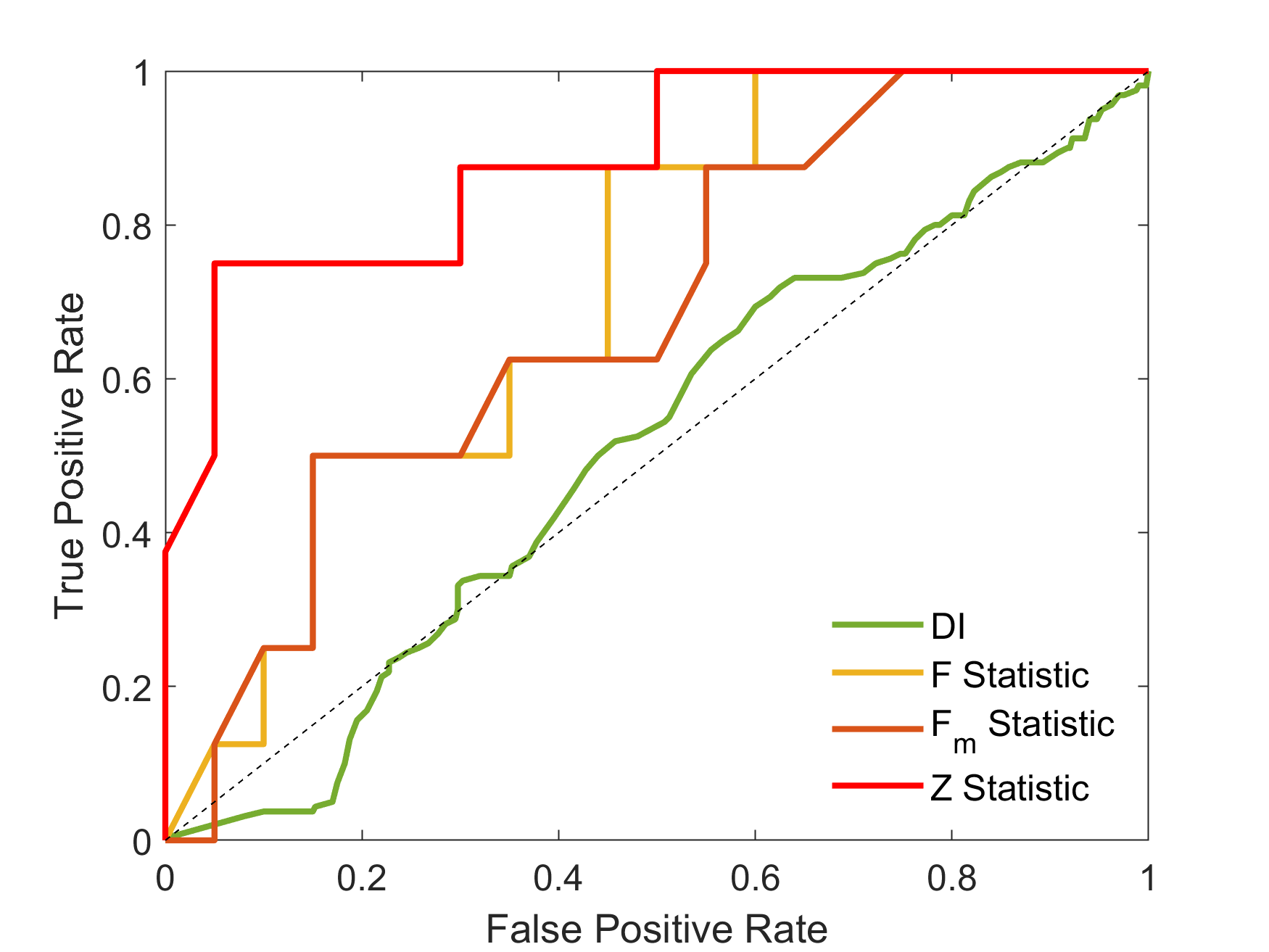}}
\put(30,290){\color{black} \large {\fontfamily{phv}\selectfont \textbf{a}}}
\put(225,290){\color{black} \large {\fontfamily{phv}\selectfont \textbf{b}}}
\end{picture} \vspace{-160pt}
\caption{ROC plots comparing the different damage detection methods for the Open Guided Waves project's CFRP panel (path 3-12): (a) the Al weight (simulated damage) on D5-11 (damage-intersecting case); (d) the Al weight (simulated damage) on D21-28 (damage-non-intersecting case). In both subplots, 15 out of 20 healthy signals were used for calculating the mean in estimating the $F_m$ and the $Z$ statistics.}
\label{fig:OGW_roc}
\end{figure}

Figure \ref{fig:OGW_roc} panels a and b show the ROC plots for the damage-intersecting case and the damage-non-intersecting case, respectively. In constructing each plot, detection statistics from all corresponding damage locations were used, and only corresponding baseline groups were considered in each case. As shown in the damage-intersecting case (panel a), the $Z$, $F$, and $F_m$ statistics all outperform the DI in overall detection performance, with larger areas under the curves. For the damage-non-intersecting case (panel b), although the $F$, $F_m$ statistics and the DI show similar performance, the $Z$ statistic surpasses all of them, even for low alpha levels (wider confidence bounds). Tables \ref{tab:OGW_3-12_damage_intersecting} and \ref{tab:OGW_3-12_damage_non_intersecting} in the Appendix also present summary detection results. All of these results show the superiority of the $Z$ statistic when it comes to damage detection.

\section{Conclusions}\label{sec:conc}

In this study, three frequency-domain damage detection metrics based on stochastic non-parametric time series representations were developed and compared with state-of-the-art damage indices as applied to three different test cases: a notched Al plate, a CFRP coupon with stacked weights, and the CFRP panel with different weight locations used in the Open Guided-Waves project \cite{OGW}. It was shown that, although the DIs can accurately detect damage and follow damage evolution in the isotropic Al coupon case, it fails to do either in the CFRP coupon for $95$\% healthy confidence bounds. In addition,  it also shows poor detection performance for the different damage cases of the CFRP panel at the same confidence levels. Examining the $F$ and $F_m$ statistics, because their detection thresholds are either solely dependent on the theoretical estimation confidence bounds of the Welch PSD estimator ($F$ statistic), or dependent on the theoretical estimation intervals with the incorporation of some experimental statistics ($F_m$ statistics), their detection performance at $95$\% confidence can, in some cases, be even worse than the DIs. However, for different confidence levels, both, especially the $F$ statistic, exhibit a detection performance more or less similar to that of the DIs, as shown in the different Receiver Operating Characteristics plots in this study. On the other hand, the $Z$ statistic outperforms all other detectors used in this study for all three test cases, for both: damage-intersecting and non-intersecting paths. In addition, it also better follows the evolution of damage for the Al and CFRP coupons in the damage-intersecting case compared to the DIs, which hints on its damage quantification capabilities.

Overall, it can be concluded from this study that, for the three test cases studied herein, methods based on frequency-domain non-parametric statistical time series models show greater sensitivity to damage, even when used to analyze damage-non-intersecting signals, compared to time-domain DI-based approaches, especially in materials exhibiting non-linearities and anisotropic behaviour such as composites. This was clearly demonstrated when constructing $95$\% healthy confidence bounds accounting for the same experimental uncertainties in both approaches. In addition, the proposed approaches show increased robustness to uncertainty with less fluctuation in the values of the metrics for the healthy test cases compared to the time-domain-based DIs. Thus, non-parametric time series representations emerge as sources of constructing accurate and robust metrics that promise enhancement in damage detection performance for SHM systems.

\section*{Acknowledgment}

This work is carried out at the Rensselaer Polytechnic Institute under the Vertical Lift Research Center of Excellence (VLRCOE) Program, grant number W911W61120012, with Dr. Mahendra Bhagwat and Dr. William Lewis as Technical Monitors. 


\bibliographystyle{aiaa} 

\bibliography{references} 

\newpage

\appendices
\appendixpage
\addappheadtotoc

\setcounter{table}{0}
\renewcommand\thetable{\Alph{section}.\arabic{table}}

\setcounter{figure}{0}
\renewcommand\thefigure{\Alph{section}.\arabic{figure}}


\section{Damage Detection Summary Results: Notched Al Coupon}


\begin{table}[h!] 
\centering
\caption{Damage detection summary results at an $\alpha$ value of $95\%$ for path 2-6 (single wave packet) in the Al plate (damage presented in units of mm).}\label{tab:Al_2-6_FA/MF}
\renewcommand{\arraystretch}{1.2}
{\footnotesize
\begin{tabular}{lccccccccccc} 
\hline 
Method & \multicolumn{1}{|c}{False} & \multicolumn{10}{|c}{Missed Damage ($\%$)} \\
\cline{3-12}
         & \multicolumn{1}{|c|}{Alarms ($\%$)} & 2 & 4 & 6 & 8 & 10 & 12 & 14 & 16 & 18 & 20 \\
\hline
DI$^{\dagger}$ \cite{Janapati-etal16}   & 6 & 0 & 0 & 0 & 0 & 0 & 0 & 0 & 0 & 0 & 0 \\
$F$ Statistic$^\dagger$           & 0 & 100 & 100 & 100 & 100 & 0 & 0 & 0 & 0 & 0 & 0 \\
$F_m$ Statistic$^{\dagger\dagger}$        & 0 & 100 & 100 & 100 & 0 & 0 & 0 & 0 & 0 & 0 & 0 \\
$Z$ Statistic$^{\dagger\dagger}$        & 0 & 0 & 0 & 0 & 0 & 0 & 0 & 0 & 0 & 0 & 0 \\
\hline
\multicolumn{12}{l}{{\bf False alarms} presented as percentage of 400 test cases for the DI and $F$ statistic,} \\
  & \multicolumn{11}{l}{and as percentage of 20 data sets for the $F_m$ and $Z$ statistics.} \\
\multicolumn{12}{l}{{\bf Missed damages} presented as percentage of 20 test cases.}  \\
\multicolumn{12}{l}{$^\dagger$ All 20 baseline data sets were used as reference signals consecutively.} \\
\multicolumn{12}{l}{$^{\dagger\dagger}$ 15 out of 20 baseline data sets were used to calculate the baseline mean.} 
\end{tabular}} 
\end{table}
\begin{table}[h!]
\centering
\caption{Damage detection summary results at an $\alpha$ value of $95\%$ for path 6-3 (single wave packet) in the Al plate (damage presented in units of mm).}\label{tab:Al_6-3_FA/MF}
\renewcommand{\arraystretch}{1.2}
{\footnotesize
\begin{tabular}{lccccccccccc} 
\hline 
Method & \multicolumn{1}{|c}{False} & \multicolumn{10}{|c}{Missed Damage ($\%$)} \\
\cline{3-12}
         & \multicolumn{1}{|c|}{Alarms ($\%$)} & 2 & 4 & 6 & 8 & 10 & 12 & 14 & 16 & 18 & 20 \\
\hline
DI$^\dagger$ \cite{Janapati-etal16}   & 5.5 & 0 & 0 & 0 & 0 & 0 & 0 & 0 & 0 & 0 & 0 \\
$F$ Statistic$^{\dagger}$           & 0 & 100 & 100 & 100 & 100 & 100 & 100 & 100 & 100 & 100 & 100 \\
$F_m$ Statistic$^{\dagger\dagger}$        & 0 & 100 & 100 & 100 & 100 & 100 & 100 & 0 & 0 & 100 & 100 \\
$Z$ Statistic$^{\dagger\dagger}$        & 0 & 0 & 0 & 0 & 0 & 0 & 0 & 0 & 0 & 0 & 0 \\
\hline
\multicolumn{12}{l}{{\bf False alarms} presented as percentage of 400 test cases for the DI and $F$ statistic,} \\
  & \multicolumn{11}{l}{and as percentage of 20 data sets for the $F_m$ and $Z$ statistics.} \\
\multicolumn{12}{l}{{\bf Missed damage} presented as percentage of 20 test cases.}  \\
\multicolumn{12}{l}{$^\dagger$ All 20 baseline data sets were used as reference signals consecutively.} \\
\multicolumn{12}{l}{$^{\dagger\dagger}$ 15 out of 20 baseline data sets were used to calculate the baseline mean.} 
\end{tabular}} 
\end{table}

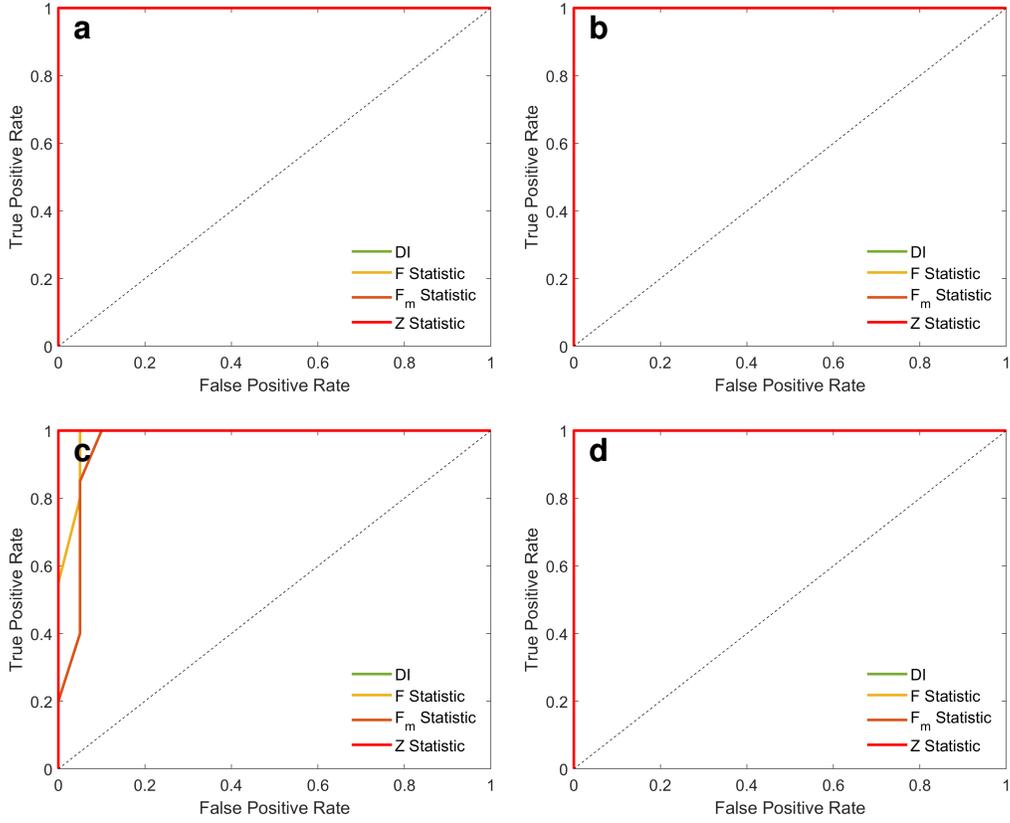
\begin{figure}[h!]
\centering
 \begin{picture}(400,300)
\put(-3,160){\includegraphics[scale=0.5]{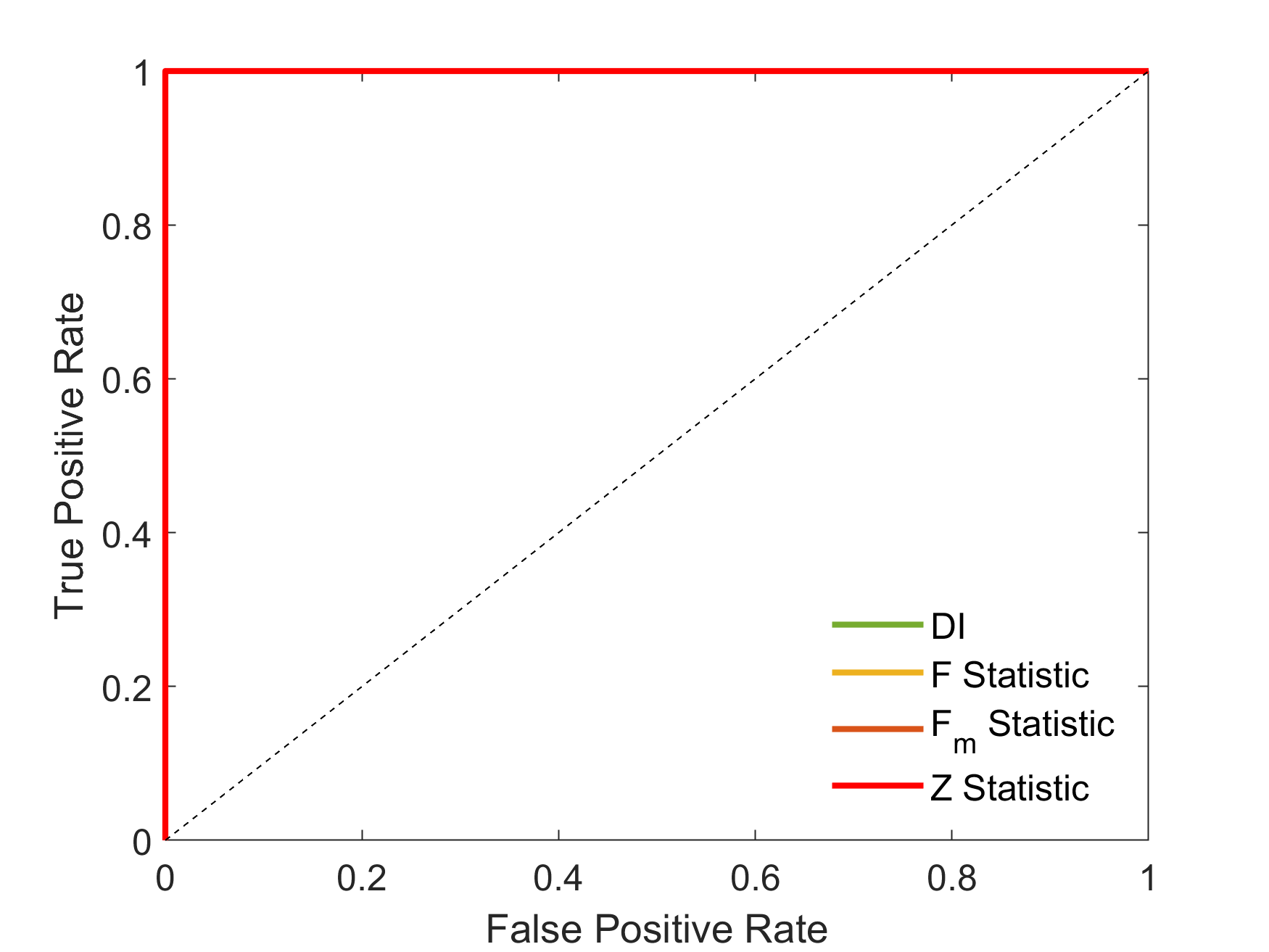}}
\put(192,160){\includegraphics[scale=0.5]{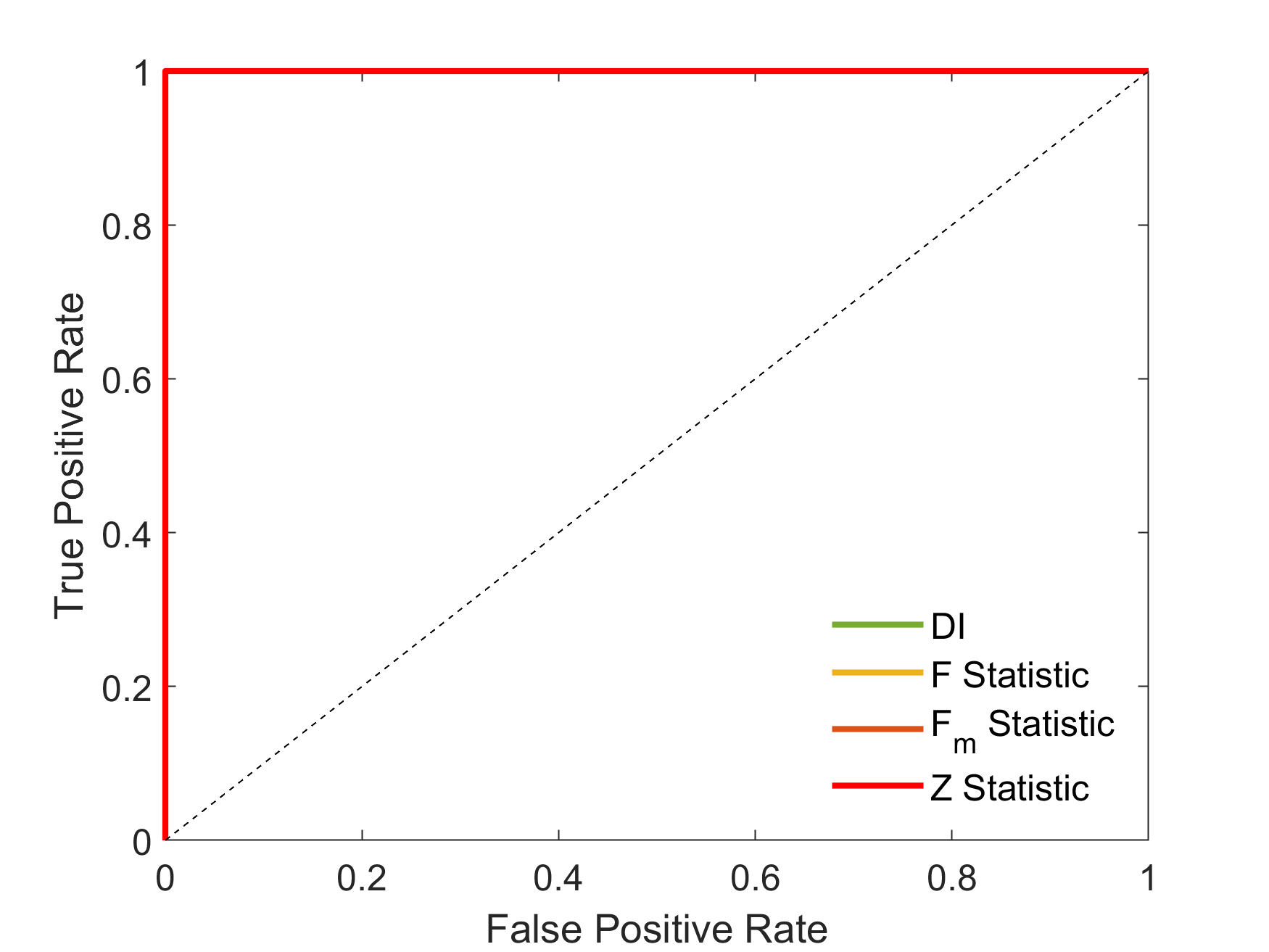}}
\put(-3,0){\includegraphics[scale=0.5]{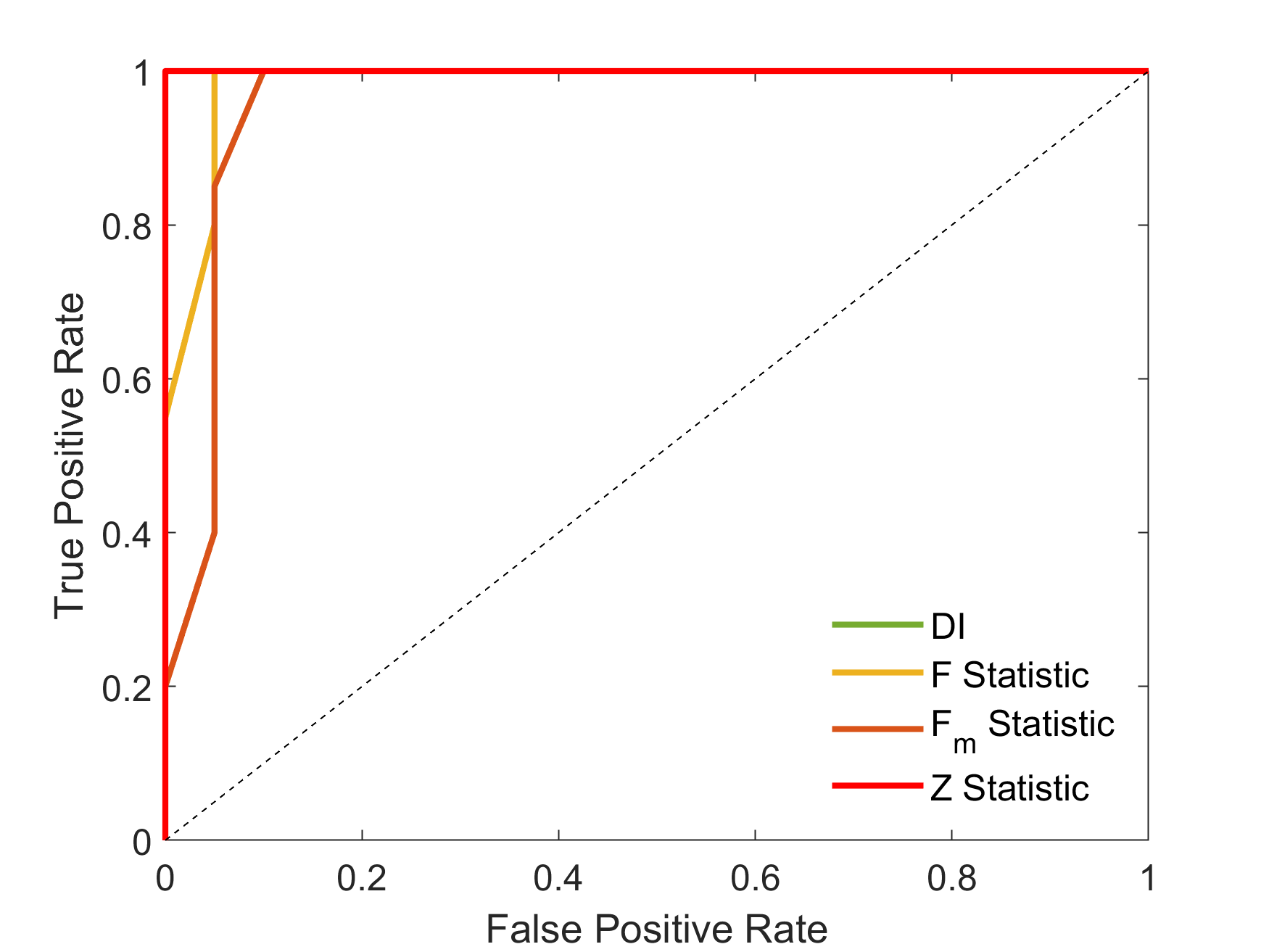}}
\put(192,0){\includegraphics[scale=0.5]{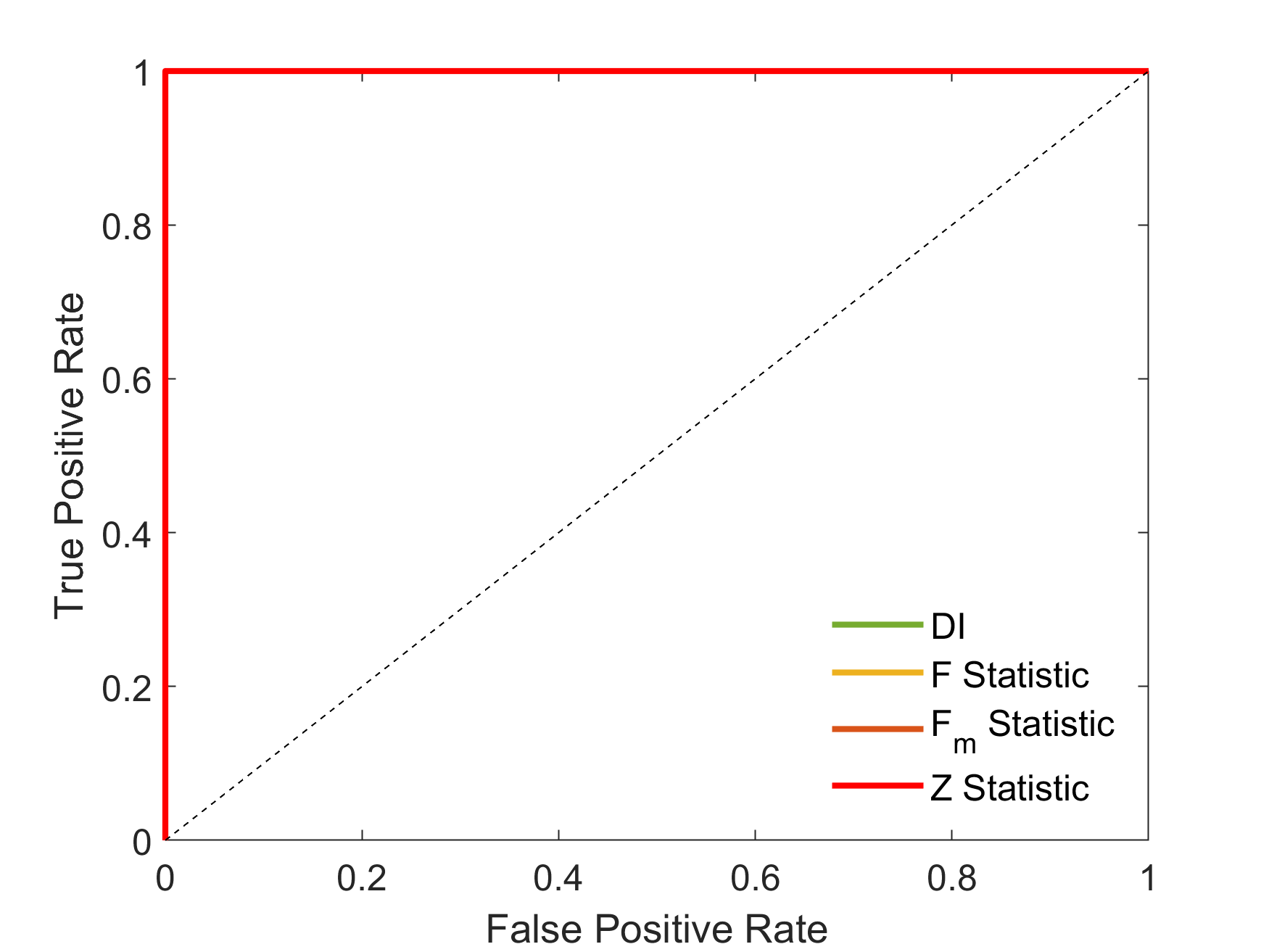}}
\put(30,295){\color{black} \large {\fontfamily{phv}\selectfont \textbf{a}}}
\put(225,295){\color{black} \large {\fontfamily{phv}\selectfont \textbf{b}}}
\put(30,135){\color{black} \large {\fontfamily{phv}\selectfont \textbf{c}}} 
\put(225,135){\color{black} \large {\fontfamily{phv}\selectfont \textbf{d}}} 
\end{picture} 
\caption{Receiver Operating Characteristics (ROC) plots comparing the different damage detection methods for the notched Al coupon with a notch size of 2 mm: (a) path 2-6 wave packet; (b) path 2-6 full signal; (c) path 6-3 wave packet; (d) path 6-3 full signal.}
\label{fig:Notch_roc_new}
\end{figure}


\section{Damage Detection Summary Results: CFRP Coupon}


\begin{table}[h!] 
\centering
\caption{Damage detection summary results at multiple $\alpha$ values for path 1-4 (single wave packet) in the CFRP plate.}\label{tab:CFRP_1-4_FA/MF}
\renewcommand{\arraystretch}{1.2}
{\footnotesize
\begin{tabular}{lccccccc} 
\hline 
Method & \multicolumn{1}{|c}{False} & \multicolumn{6}{|c}{Missed damage ($\%$)} \\
\cline{3-8}
         & \multicolumn{1}{|c|}{alarms ($\%$)} & 1 Weight & 2 Weights & 3 Weights & 4 Weights & 5 Weights & 6 Weights \\
\hline
DI$^a$$^\dagger$ \cite{Janapati-etal16}   & 5.25 &  99.75 & 84.5 & 94.5 & 99.25 & 93.75 & 53.5 \\
$F$ Statistic$^b$$^\dagger$           & 25 & 65 & 5 & 5 & 0 & 0 & 0 \\
$F_m$ Statistic$^c$$^{\dagger\dagger}$        & 25 & 40 & 10 & 60 & 0 & 0 & 0 \\
$Z$ Statistic$^a$$^{\dagger\dagger}$        & 0 & 25 & 10 & 30 & 10 & 0 & 0 \\
\hline
\multicolumn{8}{l}{{\bf False alarms} presented as percentage of 20 test cases.} \\
\multicolumn{8}{l}{{\bf Missed damages} presented as percentage of 20 test cases.}  \\
\multicolumn{8}{l}{$^a$ $\alpha = 95\%$.; $^b$ $\alpha = 1\%$; $^c$ $\alpha = 10\%$} \\
\multicolumn{8}{l}{$^\dagger$ All 20 baseline data sets were used as reference signals consecutively.} \\
\multicolumn{8}{l}{$^{\dagger\dagger}$ 15 out of 20 baseline data sets were used to calculate the baseline mean.} 
\end{tabular}} 
\end{table}

\begin{table}[h!]
\centering
\caption{Damage detection summary results at multiple $\alpha$ values for path 3-4 (single wave packet) in the CFRP plate.}\label{tab:CFRP_3-4_FA/MF}
\renewcommand{\arraystretch}{1.2}
{\footnotesize
\begin{tabular}{lccccccc} 
\hline 
Method & \multicolumn{1}{|c}{False} & \multicolumn{6}{|c}{Missed damage ($\%$)} \\
\cline{3-8}
         & \multicolumn{1}{|c|}{alarms ($\%$)} & 1 Weight & 2 Weights & 3 Weights & 4 Weights & 5 Weights & 6 Weights \\
\hline
DI$^a$$^\dagger$ \cite{Janapati-etal16}   & 5.25 &  13.25 & 0 & 0 & 0 & 0 & 0 \\
$F$ Statistic$^b$$^\dagger$           & 95 & 0 & 5 & 0 & 0 & 0 & 0 \\
$F_m$ Statistic$^c$$^{\dagger\dagger}$        & 30 & 60 & 0 & 0 & 0 & 0 & 0 \\
$Z$ Statistic$^a$$^{\dagger\dagger}$        & 0 & 0 & 0 & 0 & 0 & 0 & 0 \\
\hline
\multicolumn{8}{l}{{\bf False alarms} presented as percentage of 20 test cases.} \\
\multicolumn{8}{l}{{\bf Missed damages} presented as percentage of 20 test cases.}  \\
\multicolumn{8}{l}{$^a$ $\alpha = 95\%$.; $^b$ $\alpha = 1\%$; $^c$ $\alpha = 10\%$} \\
\multicolumn{8}{l}{$^\dagger$ All 20 baseline data sets were used as reference signals consecutively.} \\
\multicolumn{8}{l}{$^{\dagger\dagger}$ 15 out of 20 baseline data sets were used to calculate the baseline mean.}
\end{tabular}} 
\end{table}

\section{Damage Detection Summary Results: OGW CFRP Panel}


\begin{table}[h!]
\centering
\caption{Damage detection summary results at multiple $\alpha$ values for path 3-12 (damage-intersecting case) in the CFRP panel \cite{OGW}.}\label{tab:OGW_3-12_damage_intersecting}
\renewcommand{\arraystretch}{1.2}
{\footnotesize
\begin{tabular}{lccc} 
\hline 
Method & \multicolumn{1}{|c}{False} & \multicolumn{2}{|c}{Missed damage ($\%$)} \\
\cline{3-4}
         & \multicolumn{1}{|c|}{alarms ($\%$)} & D5/6/7/8 & D9/10/11 \\
\hline
DI$^a$$^\dagger$ \cite{Janapati-etal16}   & 7.5 &  51.25 & 100 \\
$F$ Statistic$^b$$^\dagger$                       & 0 & 75 & 100 \\
$F_m$ Statistic$^b$$^{\dagger\dagger}$        & 0 & 75 & 33 \\
$Z$ Statistic$^a$$^{\dagger\dagger}$        & 0 & 0 & 0 \\
\hline
\multicolumn{3}{l}{{\bf False alarms} presented as percentage of 20 test cases.} \\
\multicolumn{3}{l}{{\bf Missed damages} presented as percentage of all test cases per damage group.}  \\
\multicolumn{3}{l}{$^a$ $\alpha = 95\%$.; $^b$ $\alpha = 80\%$} \\
\multicolumn{3}{l}{$^\dagger$ All 20 baseline data sets were used as reference signals consecutively.} \\
\multicolumn{3}{l}{$^{\dagger\dagger}$ 15 out of 20 baseline data sets were used to calculate the baseline mean.}
\end{tabular}}
\end{table}

\begin{table}[h!]
\centering
\caption{Damage detection summary results at an $\alpha$ value of $95\%$ for path 3-12 (damage-non-intersecting case) in the CFRP panel \cite{OGW}.}\label{tab:OGW_3-12_damage_non_intersecting}
\renewcommand{\arraystretch}{1.2}
{\footnotesize
\begin{tabular}{lccc} 
\hline 
Method & \multicolumn{1}{|c}{False} & \multicolumn{2}{|c}{Missed damage ($\%$)} \\
\cline{3-4}
         & \multicolumn{1}{|c|}{alarms ($\%$)} & D21/22/23/24 & D25/26/27/28 \\
\hline
DI$^a$$^\dagger$ \cite{Janapati-etal16}   & 7.5 &  100 & 93.75 \\
$F$ Statistic$^b$$^\dagger$                       & 0 & 100 & 75 \\
$F_m$ Statistic$^b$$^{\dagger\dagger}$        & 0 & 100 & 75 \\
$Z$ Statistic$^a$$^{\dagger\dagger}$        & 0 & 50 & 75 \\
\hline
\multicolumn{3}{l}{{\bf False alarms} presented as percentage of 20 test cases.} \\
\multicolumn{3}{l}{{\bf Missed damages} presented as percentage of all test cases per damage group.}  \\
\multicolumn{3}{l}{$^a$ $\alpha = 95\%$.; $^b$ $\alpha = 70\%$} \\
\multicolumn{3}{l}{$^\dagger$ All 20 baseline data sets were used as reference signals consecutively.} \\
\multicolumn{3}{l}{$^{\dagger\dagger}$ 15 out of 20 baseline data sets were used to calculate the baseline mean.}
\end{tabular}} 
\end{table}

\end{document}